\pdfoutput=1

\documentclass[english, a4paper]{article}

\usepackage{titling}
\usepackage{authblk}

\usepackage[margin=1.3in]{geometry}
\usepackage[font=small, labelfont=bf, margin=0.35in]{caption}

\input{glyphtounicode}
\usepackage[pdftex]{hyperref}

\makeatletter
\newcounter{subsubsubsection}[subsubsection]

\newcommand\subsubsubsection{\@startsection{subsubsubsection}{4}{\z@}%
  {-3.25ex\@plus -1ex \@minus -.2ex}%
  {1.5ex \@plus .2ex}%
  {\normalfont\normalsize\bfseries}}
\let\subsubsubsectionmark\@gobble
\renewcommand\paragraph{\@startsection{paragraph}{5}{\z@}%
  {3.25ex \@plus1ex \@minus.2ex}%
  {-1em}%
  {\normalfont\normalsize\bfseries}}
\def\toclevel@subsubsubsection{4}
\def\toclevel@paragraph{5}
\makeatother

\usepackage{babel}
\usepackage[utf8]{inputenc}

\usepackage{enumitem}

\newlist{romanlist}{enumerate}{1}
\setlist[romanlist]{
  label={\roman*)},
  leftmargin=*,
  itemsep=0pt,
  topsep=2pt}

\newlist{pseudocode}{enumerate}{4}
\setlist[pseudocode]{
  label={\arabic*},
  ref={\arabic*},
  before=\raggedright,
  leftmargin=*,
  itemsep=2pt,
  topsep=1pt}
\setlist[pseudocode,2]{label*={.\arabic*}}
\setlist[pseudocode,3]{label*={.\arabic*}}
\setlist[pseudocode,4]{label*={.\arabic*}}

\usepackage{algorithm}

\makeatletter
\newenvironment{breakablealgorithm}
  {
   \begin{center}
     \refstepcounter{algorithm}
     \hrule height.8pt depth0pt \kern2pt
     \renewcommand{\caption}[2][\relax]{
       {\raggedright\textbf{\fname@algorithm~\thealgorithm} ##2\par}%
       \ifx\relax##1\relax
         \addcontentsline{loa}{algorithm}{\protect\numberline{\thealgorithm}##2}%
       \else
         \addcontentsline{loa}{algorithm}{\protect\numberline{\thealgorithm}##1}%
       \fi
       \kern2pt\hrule\kern2pt
     }
  }{
     \kern2pt\hrule\relax
   \end{center}
  }
\makeatother

\usepackage{graphicx}

\usepackage{amsmath}
\usepackage{amsthm}
\usepackage{amssymb}

\usepackage{thmtools}

\usepackage{multirow}
\usepackage{array}

\makeatletter
\newcommand{\thickhline}{%
    \noalign {\ifnum 0=`}\fi \hrule height 1pt
    \futurelet \reserved@a \@xhline
}
\newcolumntype{x}{@{\hskip\tabcolsep\vrule width 1pt\hskip\tabcolsep}}
\makeatother

\usepackage{cite}

\newcommand{\ket}[1]{\left|\, #1 \, \right\rangle}
\newcommand{\e}{\mathrm{e}}
\newcommand{\imag}{\mathrm{i}}
\newcommand{\ceil}[1]{\left\lceil #1 \right\rceil}
\newcommand{\round}[1]{\left\lceil #1 \right\rfloor}
\newcommand{\floor}[1]{\left\lfloor #1 \right\rfloor}

\newcommand{\ellint}{\Delta_{\ell}}
\newcommand{\offsets}{\mathcal I_{\ell}}

\newtheorem{theorem}{Theorem}
\newtheorem{lemma}{Lemma}

\newtheorem{theoremcorollary}{Corollary}[theorem]
\newtheorem{claim}{Claim}
\newtheorem{definition}{Definition}

\newtheorem{appclaim}{Claim}[section]
\newtheorem{applemma}{Lemma}[section]

\newtheoremstyle{TheoremNum}
  {\topsep}{\topsep}
  {\itshape}
  {}
  {\bfseries}
  {.}
  { }
  {\thmname{#1}\thmnote{ \bfseries #3}}
\theoremstyle{TheoremNum}

\newcommand{\refeq}[1]{(\ref{eq:#1})}

\title{Revisiting Shor's quantum algorithm for \\ computing general discrete logarithms}
\author[1,2]{\href{mailto:ekera@kth.se}{Martin Ekerå}}
\affil[1]{\small KTH Royal Institute of Technology, Stockholm, Sweden}
\affil[2]{\small Swedish NCSA, Swedish Armed Forces, Stockholm, Sweden}

\predate{\begin{center}}
\postdate{\end{center}\vspace{-3mm}}

\allowdisplaybreaks

\begin{document}
\maketitle

\begin{abstract}
  We heuristically show that Shor's quantum algorithm for the DLP achieves an expected success probability of approximately~$60$--$82\%$ in a single run when modified to enable efficient implementation with the semi-classical QFT.
  By slightly increasing the number of group operations that are evaluated quantumly and performing a single limited search in the post-processing, or by performing two limited searches in the post-processing, we rigorously prove that the modified algorithm can be made to achieve a success probability that, e.g., exceeds~$1 - 10^{-4}$ in a single run.
  Specifically, we give an exact analysis of the probability distribution induced by the algorithm and use it to prove a lower bound on the success probability, as a function of the group order~$r$, the size of the search space in the post-processing, and the number of additional group operations evaluated quantumly.
  In the limit as $r \rightarrow \infty$, we show that the lower bound can be made to tend to one.

  In analogy with our earlier work, we explain how the limited searches in the post-processing can be performed efficiently using meet-in-the-middle or random-walk techniques, and we show how slightly better tradeoffs can be achieved, compared to our earlier work, when~$r$ is known.
  We generalize our exact analysis to cover several quantum algorithms analyzed in our earlier works, including for parameter regimes not covered by our earlier analyses, yielding a unified lens on all of these algorithms.
  Furthermore, we generalize our exact analysis to the setting where information on multiple discrete logarithms is computed in each run of the algorithm so as to amortize costs.
  Last but not least, we give a novel exact simulator, which is based on our generalized exact analysis, and which hence works for all quantum algorithms covered by said analysis.
\end{abstract}

\section{Introduction}
In 1994, in a seminal work with profound cryptanalytical implications, Shor~\cite{shor1994, shor1997} introduced polynomial-time quantum algorithms for factoring integers --- via a reduction to order finding in cyclic subgroups of~$\mathbb Z_N^*$, for~$N$ the integer to be factored --- and for computing discrete logarithms in~$\mathbb F_p^*$.
The algorithms may be generalized to compute orders and discrete logarithms in any cyclic group for which the group operation may be implemented efficiently quantumly.

Shor's algorithms are important, in that the security of virtually all currently widely deployed asymmetric cryptographic schemes is underpinned by either of the two aforementioned problems:
The security of RSA~\cite{rsa} relies on the intractability of the integer factoring problem~(IFP), for instance, whereas the security of Diffie--Hellman~\cite{diffie-hellman} and DSA~\cite{fips186-dsa} relies on the intractability of the discrete logarithm problem~(DLP).\footnote{In the sense that RSA can be broken if the RSA modulus can be factored, and that Diffie--Hellman and DSA can be broken if a discrete logarithm can be computed, etc.}

In this paper, we analyze the success probability of Shor's algorithm for computing discrete logarithms, when modified to enable efficient implementation with the semi-classical quantum Fourier transform~(QFT)~\cite{griffiths-niu} and control qubit recycling~\cite{zalka, mosca-ekert, parker-plenio}.
We rigorously prove that the modified algorithm achieves a success probability that, e.g., exceeds~$1 - 10^{-4}$ in a single run, if the number of group operations that are evaluated quantumly is slightly increased compared to Shor's original algorithm and a limited search is performed in the classical post-processing, or if two limited searches are performed in the classical post-processing.

Furthermore, compared to our earlier result in~\cite{ekera-general}, we show how slightly better tradeoffs can be achieved --- between the number of runs, and the number of group operations that are evaluated quantumly in each run --- when the group order is known.
We compare our proposed approach to making tradeoffs to a related approach proposed by Kaliski~\cite{kaliski-magic-box}.

Finally, we generalize our analysis to the setting where information on multiple discrete logarithms is computed in each run of the quantum algorithm so as to amortize costs, and to several quantum algorithms analyzed in our earlier works, and we furthermore give an exact simulator which covers all these algorithms; see Sect.~\ref{section:our-contributions} for further details on our contributions.

The results in this paper may be used to develop estimates of the overall complexity of attacking schemes that rely on the computational intractability of the DLP, and to enable fair comparisons of quantum algorithms for computing discrete logarithms.
Thus far, the results have been used in a number of logical and physical cost estimates; see, e.g.,~\cite{gidney-ekera, gouzien23, cfs-ecc, babbush26, xue-covey, haner26, pappalardo26}.

\subsection{Preliminaries}
\label{section:preliminaries}
As in~\cite{ekera-hastad, ekera-pp, ekera-general}, we let~$g$ be a generator of a finite cyclic group~$\langle g \rangle$ under~$\odot$ of order~$r$, and
\begin{align*}
  x = [d] \, g = \underbrace{g \odot \cdots \odot g}_{d \text{ times}}.
\end{align*}

Given~$x$ and~$g$, the DLP is to compute the logarithm $d = \log_g(x)$.
In the general DLP $d \in [0, r) \cap \mathbb Z$, whereas~$d$ is smaller than~$r$ by some order of magnitude in the short DLP.
The order~$r$ may either be known or unknown when~$d$ is to be computed:
Both cases are interesting and merit study.
In cryptographic schemes that are based on the computational intractability of the DLP, the group~$\langle g \rangle$ is typically a subgroup of~$\mathbb F_p^*$ or an elliptic curve group.

Given~$g$, the order-finding problem~(OFP) is the problem of computing~$r$.
In what follows below, we survey quantum algorithms for both the OFP and the DLP.

\subsubsection{Notation}
\label{section:notation}
Throughout this paper, $u \text{ mod } n$ and~$\{ u \}_n$ denote~$u$ reduced modulo~$n$ constrained to the intervals~$[0, n)$ and~$[-n/2, n/2)$, respectively.
Furthermore, $\ceil{u}$, $\floor{u}$ and $\round{u}$ denote~$u$ rounded up, down and to the closest integer.
Ties are broken by requiring that $\round{u} = u - \{ u \}_1$.
Equivalently, $\round{u} = \floor{u + 1/2}$, so ties are rounded up.
Note that this rounding convention is translation invariant, in that $\round{u + n} = \round{u} + n$ holds for all $n \in \mathbb Z$ since $\{ u + n \}_1 = \{ u \}_1$.
It is not symmetric, however, in that $\round{-u} = -\round{u}$ holds if and only if~$u$ is not a half-integer.

The Euclidean norms of $u \in \mathbb C$, and of the vector~$\mathbf{u}$, are denoted~$|\, u \,|$ and~$|\, \mathbf{u} \,|$, respectively.
In heuristic arguments, $u \sim v$ denotes that~$u$ and~$v$ are of similar size.

\subsection{Earlier works}
\label{section:earlier-works}
Shor's order-finding algorithm works by inducing a periodicity in~$r$, extracting information on the period using the QFT, and classically post-processing the output by forming a quotient and expanding it in a continued fraction to recover~$r$.

Specifically, when the algorithm is implemented using traditional techniques, a control register of~$2m$ qubits\footnote{Shor~\cite{shor1997} originally introduced the order-finding algorithm to factor integers~$N$, and therefore picked the control register length~$l$ such that $N^2 \le 2^l < 2 N^2$ as a function of~$N > r$ to ensure that $2^l > r^2$.
Since we consider the OFP in arbitrary cyclic groups, we instead pick $l = 2m$ for~$m$ such that $r < 2^m$.} is first initialized to a uniform superposition of all values $a \in [0, 2^{2m}) \cap \mathbb Z$, for~$m$ an upper bound on the bit length of~$r$.
Then $[a] \, g$ is computed to a work register.
When the QFT is applied to the control register, this maps
\begin{align}
  \frac{1}{2^m}
  \sum_{a \, = \, 0}^{2^{2m} - 1}
  \ket{a, [a] \, g}
  \quad
  \xrightarrow{\text{ QFT }}
  \quad
  \frac{1}{2^{2m}}
  \sum_{a, \, j \, = \, 0}^{2^{2m} - 1}
  \e^{2\pi\imag \, aj / 2^{2m}}
  \ket{j, [a] \, g}.
  \label{eq:shor-order-finding-to-qft}
\end{align}

In practice, the generalized exponentiation $[a] \, g$ is computed using the generalized square-and-multiply algorithm:
Let $a = a_0 + 2 a_1 + 2^2 a_2 + \dots$ for~$a_i$ selected uniformly at random from $\{ 0, 1\}$ for $i \in [0, 2m) \cap \mathbb Z$.
Then $[a] \, g = [a_0 + 2 a_1 + 2^2 a_2 + \dots] \, g = [a_0] \, g \odot [2 a_1] \, g \odot [2^2 a_2] \, g \odot \, \dots$ so if the work register is initialized to the identity~$[0] \, g$, we may classically pre-compute~$[2^i] \, g$ and operate with this element on the work register conditioned on~$a_i$ for $i \in [0, 2m) \cap \mathbb Z$ to perform the exponentiation.
A total of~$2m$ group operations are then evaluated quantumly.

When each~$a_i$ is selected uniformly at random from~$\{0, 1\}$, as is the case above, the QFT may be implemented in an interleaved fashion using the semi-classical QFT~\cite{griffiths-niu} with control qubit recycling~\cite{zalka, mosca-ekert, parker-plenio}.
A single control qubit, and a few standard quantum operations, then suffice to implement the control register and the QFT.
The quantum cost of the algorithm is then dominated by the cost of performing the~$2m$ group operations.

If~\refeq{shor-order-finding-to-qft} is observed, the frequency~$j$ and $y = [e] \, g$ for $e \in [0, r) \cap \mathbb Z$ are yielded with probability
\begin{align*}
  \frac{1}{2^{4m}}
  \left|\,
  \sum_{t \, = \, 0}^{M}
  \e^{2\pi\imag \, (e + tr)j / 2^{2m}}
  \,\right|^2
  =
  \frac{1}{2^{4m}}
  \left|\,
  \sum_{t \, = \, 0}^{M}
  \e^{2\pi\imag \, rj \, t / 2^{2m}}
  \,\right|^2
  \cdot
  \underbrace{\:\left|\,
  \e^{2\pi\imag \, ej / 2^{2m}}
  \,\right|^2}_{= 1}
\end{align*}
where $M = \floor{(2^{2m} - e - 1) / r}$.
Constructive interference is expected to arise when $rj/2^{2m} \approx z$, for~$z$ an integer, as all unit vectors in the above sum then point in approximately the direction of the real axis.
To recover~$r$, Shor expands $j/2^{2m}$ in a continued fraction to attempt to find the unknown quotient $z/r \approx j / 2^{2m}$.
Shor uses that $r < 2^m$ to stop the expansion short.

If~$r$ is not prime, it may be that~$z$ and~$r$ have a common factor~$\tau = \gcd(z, r)$, in which case $z/\tau$ and $r/\tau$ will be found instead of~$z$ and~$r$.
To address this problem, Shor proposes to search for~$\tau$ in $\{ 1, \, 2, \, \ldots, \, B_\tau \}$, for~$B_\tau$ some bound, to attempt to find the order~$r$.\footnote{Throughout this paper, we use~$B_x$ to denote a bound on the variable~$x$.
This notation should hence not be taken to indicate that~$B_x$ depends on~$x$.}
This significantly increases the success probability.
Shor furthermore proposes to consider not only~$j$, but also $j \pm 1, \, \ldots, \, j \pm B_j$, for~$B_j$ some bound, to further increase the success probability.

In~\cite{ekera-success-order-finding}, Ekerå describes how the two aforementioned searches may be performed efficiently; both when using continued fractions-based post-processing, as originally proposed by Shor, and when using lattice-based post-processing.
Furthermore, Ekerå gives lower bounds on the probability of successfully recovering~$r$ from~$j$ in a single run of the quantum part of the order-finding algorithm, as a function of the search space in the classical post-processing part of the algorithm, and of the length of the control register in the quantum part of the algorithm:

Already for moderate~$r$, a high success probability that exceeds, e.g., $1 - 10^{-4}$, can be guaranteed without increasing the length of the control register compared to Shor.
Asymptotically, in the limit as~$r$ and hence $m \rightarrow \infty$, the search space can be bounded as a function of~$m$ so that the success probability tends to one without compromising the polynomial runtime.

By slightly decreasing the length of the control register, work in the classical post-processing part of the algorithm may be traded for work in the quantum part~\cite[see Lem.~4.3]{ekera-success-order-finding}:
The steep reduction in the success probability that arises when the control register length is decreased may be compensated for to some extent by searching an increased space in the post-processing, and vice versa.
Typically, one would strive to keep the control register as short as possible.

For earlier works that seek to increase the success probability of Shor's order-finding algorithm by increasing the control register length, see, e.g.,~\cite[p.~11 and App.~C]{cleve}, \cite[the paragraph with Eq.~(5.44) on p.~227]{nielsen-chuang}, or the analysis in~\cite{bourdon}.
See also the survey in~\cite[Sect.~1.5]{ekera-success-order-finding}.

\subsubsection{Tradeoffs in order finding}
In~\cite{seifert}, Seifert introduces the idea of making tradeoffs across runs in Shor's order-finding algorithm --- the basic idea being to perform $n \ge s$~runs with a $(1 + 1/s)m$-bit exponent, and to then simultaneously solve the outputs of these $n$~runs for~$r$ in the classical post-processing by generalizing Shor's continued fractions-based post-processing to higher dimensions.

Seifert's original motivation for introducing this variation of Shor's algorithm was to reduce the number of control qubits required to implement the algorithm.
In implementations where the semi-classical QFT is used and the control qubits are recycled, however, the effect is rather to reduce the circuit depth, and hence the number of group operations that need to be evaluated in each run of the algorithm, at the expense of having to run it multiple times.

In~\cite[App.~A]{ekera-general}, Ekerå describes how the quantum part of Shor's order-finding algorithm, and of Seifert's variation thereof, may be simulated classically for known~$r$.
Ekerå furthermore uses the simulator to estimate the number of runs~$n$ that are needed to solve for~$r$ with tradeoff factor~$s \ge 1$ and at least~$99\%$ success probability when using lattice-based post-processing.

\subsubsection{Factoring general integers via order finding}
Shor originally introduced the order-finding algorithm for the purpose of factoring general integers~$N$ via a classical reduction that follows from the seminal work of Miller~\cite{miller}:

Specifically, to factor an odd integer~$N$ that is not a perfect prime power\footnote{These requirements do not imply a loss of generality: Even integers and perfect powers are trivial to factor.}, Shor first selects~$g$ uniformly at random from $(1, N) \cap \mathbb Z$.
In the unlikely event that~$g$ is not coprime to~$N$, there is no need to proceed --- it suffices to return~$\gcd(g, N)$.
Otherwise, $g$ is perceived as a generator of a subgroup of~$\mathbb Z_N^*$ and its order~$r$ computed quantumly with the order-finding algorithm.

If~$r$ is even and $g^{r/2} \not\equiv -1 \:\: (\text{mod } N)$, Shor then uses that
\begin{align*}
  g^r - 1 = (g^{r/2} - 1)(g^{r/2} + 1) \equiv 0 \:\: (\text{mod } N)
\end{align*}
to split~$N$ into two non-trivial factors by computing $\gcd((g^{r/2} \text{ mod } N) \pm 1, N)$ in a classical post-processing step.
If~$r$ is odd, or if $g^{r/2} \equiv -1 \:\: (\text{mod } N)$ causing only trivial factors to be found, Shor originally proposed to simply re-run the algorithm for a new~$g$.
Shor gives a proof that the probability of having to perform a re-run for either of these two reasons is at most~$1/2$.

It is possible to do better, however:
Ekerå shows in~\cite{ekera-factor-completely} how any integer~$N$ may be completely factored with very high probability of success, that asymptotically tends to one, given the order~$r$ of a single element~$g$ selected uniformly at random from~$\mathbb Z_N^*$.

By combining~\cite{ekera-factor-completely} with~\cite{ekera-success-order-finding}, Ekerå obtains a lower bound~\cite[Sect.~4.2.1]{ekera-success-order-finding} on the probability of completely factoring any integer~$N$ successfully in a single run of Shor's order-finding algorithm.
The probability is large already for moderate~$N$, and it tends to one asymptotically.

For earlier proposals for how the success probability of the classical post-processing step may be increased, the reader is referred to the surveys in~\cite[Sect.~2]{ekera-factor-completely} and~\cite[Sects.~1.4--1.5]{ekera-success-order-finding}.

\subsubsection{Breaking RSA}
If the integer~$N$ is known to be of a special form, this fact may be leveraged:
Integers of the form $N = pq$, for~$p, q$ distinct random primes of similar length in bits, are of particular importance in cryptology, as the problem of factoring such integers --- commonly referred to as RSA integers --- underpins the security of RSA~\cite{rsa}, a widely deployed asymmetric scheme.

Ekerå and Håstad~\cite{ekera-hastad, ekera-pp, ekera-success-short-dlp} describe how RSA integers may be factored by computing a short discrete logarithm on the quantum computer, rather than by using order finding as originally proposed by Shor.
Ekerå--Håstad's quantum algorithm has a lower overall quantum cost than Shor's order-finding algorithm, in that it requires approximately~$3m/2$ group operations to be evaluated in each run when not making tradeoffs~\cite[Sect.~1.4 and App.~B.2]{ekera-success-short-dlp}, for~$m$ the bit length of~$N$, compared to~$2m$ group operations in each run of Shor's algorithm.
Furthermore, the probability of Ekerå--Håstad's quantum algorithm successfully finding the factors of the RSA integer in a single run exceeds, e.g., $1 - 10^{-4}$, when not making tradeoffs~\cite[App.~B.2]{ekera-success-short-dlp}.

If tradeoffs are made in Ekerå--Håstad's algorithm, in analogy with Seifert's~\cite{seifert} idea for making tradeoffs in Shor's order-finding algorithm, the number of group operations that need to be evaluated quantumly in each run can be further reduced to $(1/2 + 1/s) m$, for $s \ge 1$ the tradeoff factor, at the expense of performing $n \ge s$ runs.
This allows the requirements on the quantum computer to be further reduced, beyond what is possible with Seifert's tradeoffs.

Ekerå has estimated the number of runs required for a success probability that exceeds~$99\%$ to be achieved when making tradeoffs for different tradeoff factors~\cite[App.~A.2]{ekera-pp}.

\subsubsection{Computing discrete logarithms}
In analogy with the order-finding algorithm, Shor's algorithm for computing general discrete logarithms works by inducing a periodicity in the logarithm~$d$, extracting the period using the QFT, and classically post-processing the output to recover~$d$.

Specifically, when the algorithm is implemented using traditional techniques, two control registers, each of length~$m$ qubits for~$m$ the bit length of~$r$, are first initialized to a uniform superposition of all $a, b \in [0, r) \cap \mathbb Z$.\footnote{Shor~\cite{shor1994} originally described the algorithm for the full multiplicative group~$\mathbb F_p^*$ and so set $r = p - 1$.}
Then $[a] \, g \odot [-b] \, x$ is computed to a work register.
When QFTs of size~$2^m$ are applied to the two control registers, this maps
\begin{align}
  &\frac{1}{r}
  \sum_{a, \, b \, = \, 0}^{r - 1}
  \ket{a, b, [a] \, g \odot [-b] \, x}
  \quad \xrightarrow{\text{QFT}} \quad \notag \\
  &\quad\quad\frac{1}{2^{m}r}
  \sum_{a, \, b \, = \, 0}^{r - 1}
  \sum_{j, \, k \, = \, 0}^{2^{m} - 1}
  \e^{2\pi\imag \, (aj + bk) / 2^{m}}
  \ket{j, k, [a - bd] \, g}
  \label{eq:psi-shor-dlog}
\end{align}
after which~$(j, k)$ and $y = [e] \, g$ for $e \in [0, r) \cap \mathbb Z$ are observed.
The frequency pair~$(j, k)$ is then classically post-processed to find~$d$.
This classical post-processing requires~$r$ to be known.

As for the order-finding algorithm, the quantum cost of the above algorithm~\refeq{psi-shor-dlog} is dominated by the~$2m$ group operations.
However, the implementation is complicated by the fact that the individual qubits~$a_i$ and~$b_i$ in the two control registers are not selected uniformly at random from~$\{0, 1\}$, as required by the semi-classical QFT with control qubit recycling.
In this paper, we therefore analyze a modified version of Shor's algorithm for computing discrete logarithms, in which the superpositions in the control registers are uniform over~$[0, 2^m) \cap \mathbb Z$.

\subsubsection{On the DLP and our motivation}
The DLP underpins the security of virtually all widely deployed asymmetric schemes.
Notable examples include RSA\footnote{E.g.\ via the reduction to a short DLP.}~\cite{rsa}, Diffie--Hellman~\cite{diffie-hellman}, DSA~\cite{fips186-dsa, fips186-5-dsa} and virtually all forms of elliptic curve cryptography (ECC)~\cite{koblitz, miller-ecc}.
Despite this, Shor's algorithm for computing discrete logarithms has seemingly received less attention in the literature than Shor's order-finding and factoring algorithms:
In particular, there is seemingly no tight analysis of the success probability of Shor's algorithm for computing discrete logarithms, when modified as described above to facilitate efficient implementation.
A lower bound that slightly exceeds~$65\%$ is given in Mosca's PhD thesis~\cite[Cor.~19 on p.~58]{mosca-thesis}, for a slightly modified version of Shor's algorithm that admits efficient implementation (see Sect.~\ref{section:comparison-mosca} for further details).
In this work, we prove a tighter lower bound, and give heuristic estimates of the expected success probability.

Our original motivation for pursuing this line of work (besides the DLP being worthy of study in and of itself, as explained above) is the following:
Ekerå and Håstad have developed variations of Shor's algorithm for solving cryptanalytically relevant instances of the DLP.
When not making tradeoffs, it is known that a success probability that exceeds~$99\%$ may be achieved in a single run of Ekerå's algorithm for computing short discrete logarithms~\cite{ekera-modifying, ekera-pp}, in Ekerå--Håstad's algorithm for computing short discrete logarithms and factoring RSA integers with tradeoffs~\cite{ekera-hastad, ekera-pp}, and in Ekerå's algorithm for computing general discrete logarithms with tradeoffs~\cite{ekera-general}.
All of these algorithms may be simulated classically for known~$d$ and~$r$, enabling the success probability of the classical post-processing to be verified.

The latter of the three aforementioned algorithms computes general discrete logarithms, as does Shor's algorithm for the DLP.
Unlike Shor's algorithm, Ekerå's algorithm does however not require the group order to be known.
If the order is unknown, it may be computed along with the discrete logarithm in the classical post-processing at no additional quantum cost.
When computing discrete logarithms without making tradeoffs, Ekerå's algorithm requires approximately~$3m$ group operations to be evaluated quantumly, compared to~$2m$ operations in Shor's algorithm.
This is explained by the fact that Ekerå's algorithm computes both~$d$ and~$r$, whereas Shor's algorithm computes only~$d$ given~$r$.

If~$r$ is unknown, it is hence more efficient to run Ekerå's algorithm, compared to first running Shor's order-finding algorithm to find~$r$, and then running Shor's algorithm for the DLP to find~$d$ given~$r$, in terms of the overall number of group operations that are evaluated quantumly.\footnote{But not in terms of the number of group operations per run, and hence not in terms of the circuit depth.}
If~$r$ is known --- as is very often the case in practice --- and tradeoffs are not made, then Shor's algorithm is expected to outperform Ekerå's algorithm.

However, this is only true if one assumes Shor's algorithm to have a success probability that is on par with that of Ekerå's algorithm:
Without a tight analysis of the success probability of Shor's algorithm, making a proper comparison is difficult.
This fact served as one of our original motivations for writing down the analysis presented in this work.

It should be noted that Proos and Zalka~\cite[the short note in App.~A.2]{proos-zalka} discuss the success probability of Shor's algorithm.
They state that it should be possible to achieve a success probability that tends to one, either by performing a limited classical search, or by evaluating additional group operations quantumly, making reference to the situation being similar to that in Shor's order-finding algorithm.
However, no formal analysis is provided:
In particular, no concrete estimate of the success probability, as a function of the size of the search space, or of the number of additional group operations, is given.

\subsection{Our contributions}
\label{section:our-contributions}
In this work, we join an additional piece to the puzzle laid out in the previous sections:
We heuristically show that Shor's quantum algorithm~\cite{shor1994, shor1997} for the DLP achieves an expected success probability of approximately~$60$--$82\%$ in a single run when modified to enable efficient implementation with the semi-classical QFT~\cite{griffiths-niu} with control qubit recycling~\cite{zalka, mosca-ekert, parker-plenio}.

By slightly increasing the number of group operations that are evaluated quantumly and performing a single limited search in the post-processing, or by performing two limited searches in the post-processing, we rigorously prove that the modified algorithm can be made to achieve a success probability that, e.g., exceeds~$1 - 10^{-4}$ in a single run.
Specifically, we give an exact analysis of the probability distribution induced by the algorithm and use it to prove a lower bound on the success probability, as a function of the group order~$r$, the size of the search space in the post-processing, and the number of additional group operations evaluated quantumly.
In the limit as $r \rightarrow \infty$, we show that the lower bound can be made to tend to one.
This enables fair comparisons between Shor's algorithm for the DLP and other algorithms.

In analogy with our earlier work~\cite{ekera-success-short-dlp}, we explain how the limited searches in the post-processing can be performed efficiently using meet-in-the-middle or random-walk techniques, and we show how slightly better tradeoffs can be achieved, compared to our earlier work~\cite{ekera-general}, when~$r$ is known.
We generalize our exact analysis to cover several quantum algorithms analyzed in our earlier works~\cite{ekera-hastad, ekera-pp, ekera-general, ekera-success-order-finding, ekera-thesis, ekera-success-short-dlp}, including for parameter regimes not covered by our earlier analyses, yielding a unified lens on all of these algorithms.
Furthermore, we generalize our exact analysis to the setting where information on multiple discrete logarithms is computed in each run of the algorithm so as to amortize costs.
Last but not least, we give a novel exact simulator, which is based on our generalized exact analysis, and which hence works for all quantum algorithms covered by said analysis.

\subsection{Assumptions and reductions}
\label{section:assumptions-reductions}
When computing discrete logarithms in groups of known order~$r$, as is the primary focus in this paper, we may without loss of generality assume~$r$ to have no small prime factors.

To see why this is, note that we may efficiently find all small factors of~$r$ classically, and use Pohlig--Hellman decomposition~\cite{pohlig-hellman} to reduce the problem in the $r$-order group to problems in one or more small, and one large, subgroup.
We may then solve the problems in the small-order subgroups classically, leaving only the problem in the large-order subgroup with no small factors to be solved quantumly.
For this reason, $r$ is typically prime in cryptographic schemes that are based on the computational intractability of the DLP.

Furthermore, when computing general discrete logarithms in groups of known order~$r$, we may without loss of generality assume the logarithm~$d$ to be selected uniformly at random from $[0, r) \cap \mathbb Z$.
To see why this is, note that the DLP may be randomized:

Given~$g$ and $x' = [d'] \, g$ for $d' \in [0, r) \cap \mathbb Z$, we may pick an offset~$t$ uniformly at random from $[0, r) \cap \mathbb Z$, compute $x = x' \odot [t] \, g = [d' + t] \, g$, compute the randomized discrete logarithm $d = (d' + t) \text{ mod } r = \log_g x$ quantumly, and then compute $d' \equiv d - t \:\: (\text{mod } r)$ classically.

\subsubsection{Notes on recent space-saving implementation techniques}
It should be noted that Chevignard, Fouque and Schrottenloher~\cite{cfs, cfs-ecc} have fairly recently proposed a space-saving implementation technique that builds on the compression technique of May and Schlieper~\cite{ms22}.
In a nutshell, their idea is to compress the work register using approximate arithmetic while not recycling the control registers.
Our contributions are highly relevant also in the context of implementations that leverage this space-saving technique:
In such implementations, our results typically yield significant space savings.

For simplicity, however, throughout this work, we describe all algorithms under the assumption that they are implemented using traditional techniques.
The translation to other techniques should be straightforward.\footnote{It should perhaps also be noted in this context that the fact that the control registers are uniform up to powers of two, and that the QFTs are of sizes that are powers of two, and so forth, is less important when using the space-saving implementation technique of Chevignard, Fouque and Schrottenloher~\cite{cfs, cfs-ecc}, since the control registers are then not recycled.
Hence, some elements of our analysis, and of the quantum algorithms we analyze, could potentially be simplified if the implementation is restricted to use said space-saving technique.}

\subsection{Overview}
The remainder of this paper is organized as follows:

In Sect.~\ref{section:quantum-algorithm}, we describe the quantum algorithm that upon input of~$g$ and~$x = [d] \, g$ computes a frequency pair~$(j, k)$.
In Sect.~\ref{section:analysis-probability-distribution}, we derive a closed-form expression for the probability of observing~$(j, k)$.
In Sect.~\ref{section:capturing-probability-distribution}, we introduce the notion of $B_{\eta}$-$B_{\Delta}$-good pairs~$(j, k)$, and analyze the probability of the algorithm yielding such a pair as a function of~$B_{\eta}$, $B_{\Delta}$ and other relevant parameters.
In Sect.~\ref{section:simulating-quantum-algorithm}, we explain how the quantum algorithm may be simulated classically for known~$d$ and~$r$.
In Sect.~\ref{section:post-processing}, we describe how $B_{\eta}$-$B_{\Delta}$-good pairs~$(j, k)$ may be post-processed classically to recover the logarithm~$d$ given~$r$, both when making and when not making tradeoffs.
We discuss generalizations in Sect.~\ref{section:generalizations}, and summarize and conclude the paper in Sect.~\ref{section:summary-conclusion}.

App.~\ref{appendix:multiple-logarithms} contains generalizations to computing information on multiple discrete logarithms in each run of the quantum algorithm so as to amortize costs.
App.~\ref{appendix:tabulated-results} contains tabulated results.
App.~\ref{appendix:supporting-lemmas-claims-proofs} contains supporting lemmas, claims and proofs.

\section{The quantum algorithm}
\label{section:quantum-algorithm}
Given~$g$ and $x = [d] \, g$, the quantum algorithm that we analyze henceforth initializes two control registers, of length $m + \varsigma$ and $\ell \le m + \varsigma$ qubits, respectively, to uniform superpositions of all $a \in [0, 2^{m+\varsigma}) \cap \mathbb Z$ and $b \in [0, 2^\ell) \cap \mathbb Z$, respectively, and computes $[a] \, g \odot [-b] \, x$ to a work register.

Applying QFTs of size~$2^{m+\varsigma}$ and~$2^\ell$, respectively, to the two control registers then maps
\begin{align}
  \:&\frac{1}{\sqrt{2^{m+\varsigma+\ell}}}
  \sum_{a \, = \, 0}^{2^{m+\varsigma}-1}
  \sum_{b \, = \, 0}^{2^{\ell}-1}
  \ket{a, b, [a] \, g \odot [-b] \, x}
  \quad \xrightarrow{\text{QFT}} \quad \notag \\
  &\quad\quad \frac{1}{2^{m+\varsigma+\ell}}
  \sum_{a, \, j \, = \, 0}^{2^{m+\varsigma}-1}
  \sum_{b, \, k \, = \, 0}^{2^{\ell}-1}
  \e^{\, 2 \pi \imag \, (a j + 2^{m+\varsigma-\ell} b k) / 2^{m+\varsigma}}
  \ket{j, k, [a\, - \, bd] \, g}. \label{eq:psi}
\end{align}

Compared to Shor's original algorithm~\refeq{psi-shor-dlog}, that induces a uniform superposition of~$r$ values in the first two registers before applying the QFTs, the above algorithm~\refeq{psi} induces a uniform superposition of~$2^{m+\varsigma}$ and~$2^{\ell}$ values, respectively.
This enables it to employ the semi-classical QFT~\cite{griffiths-niu} with control qubit recycling~\cite{zalka, mosca-ekert, parker-plenio} by interleaving the computational steps.

To compute general discrete logarithms~$d$ in groups of known order~$r$, as is our primary focus, we let~$m$ be the bit length of~$r$, and~$\ell$ and~$\varsigma$ be non-negative integers.
As we shall soon see in the analysis in Sect.~\ref{section:analysis-probability-distribution}, the parameters $\ell$ and~$\varsigma$ then correspond roughly to the number of bits of information that are computed on~$d$ and~$r$, respectively, in each run.\footnote{Hence, one could consider renaming~$\ell$ and $\varsigma$ to~$\ell_d$ and~$\ell_r$, respectively, were it not for the fact that they both appear in super- and subscripts making such notation hard to read.}
It follows that we may, e.g., select $\ell \sim m/s$, for~$s$ a tradeoff factor, to compute~$d$ given~$r$ in~$\sim s$ runs.

\subsection{Notes on generalizations}
\label{section:quantum-algorithm-generalizations}
Although not the primary focus of this paper, it should be noted that the parameters~$m$, $\varsigma$ and~$\ell$ may be selected to make the above algorithm~\refeq{psi} equivalent to Ekerå's variation~\cite{ekera-general} of Shor's algorithm for computing general discrete logarithms~$d$, and optionally the group order~$r$, in groups of unknown order, by letting~$m$ be an upper bound on the bit length of~$r$, and $\varsigma = \ell$.

Furthermore, the parameters may be selected to make the above algorithm~\refeq{psi} equivalent to Ekerå--Håstad's algorithm~\cite{ekera-hastad, ekera-pp, ekera-success-short-dlp} for computing short discrete logarithms~$d$ in groups of unknown order, by letting~$m$ be an upper bound on the bit length of~$d$, and $\varsigma = \ell$.

Finally, the parameters may be selected to make the above algorithm~\refeq{psi} equivalent to Shor's original order-finding algorithm~\cite{shor1994, shor1997}, which computes the group order~$r$, and to Seifert's variation~\cite{seifert} thereof, as described and analyzed in~\cite{ekera-success-order-finding} and \cite[App.~A]{ekera-general}, by letting~$m$ be an upper bound on the bit length of~$r$, $\varsigma \sim m / s$ for~$s$ a tradeoff factor, and $\ell = 0$.

In this sense, the above algorithm~\refeq{psi}, and --- as we shall see in what follows --- our exact analysis in Sect.~\ref{section:analysis-probability-distribution} of the probability distribution it induces, as well as our exact simulator for it in Sect.~\ref{section:simulating-quantum-algorithm-without-failures}, also cover the algorithm variations in our previous works~\cite{ekera-hastad, ekera-pp, ekera-success-short-dlp, ekera-general, ekera-success-order-finding}.
We elaborate on such generalizations and exemplify their usefulness in Sect.~\ref{section:generalizations}.

\section{Analysis of the probability distribution}
\label{section:analysis-probability-distribution}
When~\refeq{psi} is observed, it collapses to~$(j, k)$ and $y = [e] \, g$ for $e \in [0, r) \cap \mathbb Z$ with probability
\begin{align}
  \label{eq:probability}
  \frac{1}{2^{2(m + \varsigma + \ell)}}
  \left|\,
  \sum_{(a, b)}
  \exp{\left[ \frac{2 \pi \imag}{2^{m + \varsigma}} (aj + 2^{m + \varsigma - \ell} bk) \right]}
  \,\right|^2
\end{align}
where~$(a, b)$ runs over all $a \in [0, 2^{m + \varsigma}) \cap \mathbb Z$ and $b \in [0, 2^{\ell}) \cap \mathbb Z$ such that $e \equiv a - bd \:\: (\text{mod } r)$.

\subsection{Simplifying the expression}
\label{section:analysis-probability-distribution-simplifying}
To simplify~\refeq{probability}, we use the fact that
\begin{align*}
  \frac{1}{r}
  \sum_{\mu \, = \, 0}^{r-1}
  \exp
  \left[
    \frac{2 \pi \imag}{r} (e - (a - bd)) \, \mu
  \right]
  =
  \left\{
    \begin{array}{cl}
      1 & \text{if } e \equiv a - bd \:\: (\text{mod } r), \\
      0 & \text{otherwise,}
    \end{array}
  \right.
\end{align*}
yielding the equivalence
\begin{align}
  \refeq{probability}
  &=
  \frac{1}{2^{2(m + \varsigma + \ell)} r^2}
  \left|\,
  \sum_{\mu \, = \, 0}^{r-1}
  \sum_{a \, = \, 0}^{2^{m + \varsigma} - 1}
  \sum_{b \, = \, 0}^{2^{\ell} - 1}
  \exp
  \left[
    \frac{2 \pi \imag}{r} (e - (a - bd)) \mu
  \right]
  \exp{\left[ \frac{2 \pi \imag}{2^{m + \varsigma}} (aj + 2^{m + \varsigma - \ell} bk) \right]}
  \,\right|^2 \notag \\
  &=
  \frac{1}{r^2}
  \left|\,
  \sum_{\mu \, = \, 0}^{r-1}
  \exp
  \left[
    \frac{2 \pi \imag}{r} e \mu
  \right]
  A_\mu(j) \, B_\mu(k)
  \,\right|^2
  \label{eq:probability2}
\end{align}
where we have introduced
\begin{align}
  A_\mu(j)
  &=
  \frac{1}{2^{m + \varsigma}}
  \sum_{a \, = \, 0}^{2^{m + \varsigma} - 1}
  \exp
  \left[
    2 \pi \imag
    \left(
      \frac{j}{2^{m + \varsigma}}
      -
      \frac{\mu}{r}
    \right)
    a
  \right], \label{eq:Amu} \\
  B_\mu(k)
  &=
  \frac{1}{2^{\ell}}
  \sum_{b \, = \, 0}^{2^{\ell} - 1}
  \exp
  \left[
    2 \pi \imag
    \left(
      \frac{k}{2^{\ell}}
      +
      \frac{d \mu}{r}
    \right)
    b
  \right].
  \label{eq:Bmu}
\end{align}

Summing~\refeq{probability}, or equivalently~\refeq{probability2}, over all $e \in [0, r) \cap \mathbb Z$ hence yields
\begin{align}
  &\frac{1}{2^{2(m + \varsigma + \ell)}}
  \sum_{e \, = \, 0}^{r-1}
  \left|\,
  \sum_{(a, b)}
  \exp{\left[ \frac{2 \pi \imag}{2^{m + \varsigma}} (aj + 2^{m + \varsigma - \ell} bk) \right]}
  \,\right|^2 \notag \\
  &\quad\:\: = \frac{1}{r^2}
  \sum_{e \, = \, 0}^{r-1}
  \left|\,
  \sum_{\mu \, = \, 0}^{r-1}
  \exp
  \left[
    \frac{2 \pi \imag}{r} e \mu
  \right]
  A_\mu(j) \, B_\mu(k)
  \,\right|^2 \notag \\
  &\quad\:\: = \frac{1}{r^2}
  \sum_{\mu, \, \mu' \, = \, 0}^{r-1}
  A_\mu(j) \, \overline{A_{\mu'}(j)} \,
  B_\mu(k) \, \overline{B_{\mu'}(k)} \,
  \sum_{e \, = \, 0}^{r-1}
  \exp
  \left[
    \frac{2 \pi \imag}{r} (\mu - \mu') e
  \right] \notag \\
  &\quad\:\: = \frac{1}{r}
  \sum_{\mu \, = \, 0}^{r-1}
  \left|\,
    A_\mu(j) \, B_\mu(k)
  \,\right|^2
  \label{eq:probability-sum}
\end{align}
where we have used that
\begin{align*}
  \sum_{e \, = \, 0}^{r-1}
  \exp
  \left[
    \frac{2 \pi \imag}{r} (\mu - \mu') e
  \right]
  &=
  \left\{
    \begin{array}{cl}
      r & \text{if } \mu \equiv \mu' \:\: (\text{mod } r), \\
      0 & \text{otherwise.}
    \end{array}
  \right.
\end{align*}

\subsection{Introducing arguments and angles}
\label{section:arguments-angles}
In analogy with the analyses in our earlier works~\cite{ekera-hastad, ekera-pp, ekera-general}, let
\begin{align}
  \alpha_r = \{ rj \}_{2^{m+\varsigma}},
  \:\quad\:
  \alpha_d = \{ dj + 2^{m+\varsigma-\ell} k \}_{2^{m+\varsigma}},
  \:\quad\:
  \theta_r = \frac{2 \pi \alpha_r}{2^{m+\varsigma}},
  \:\quad\:
  \theta_d = \frac{2 \pi \alpha_d}{2^{m+\varsigma}},
  \label{eq:arguments-angles}
\end{align}
where we recall that~$\{ u \}_n$ denotes~$u$ reduced modulo~$n$ constrained to $[-n/2, n/2)$.

Furthermore, let
\begin{align}
  \phi_u
  =
  \left\{
  \frac{2 \pi}{2^{m+\varsigma}}
  \left(
    \alpha_d
    -
    \frac{d}{r} \left( \alpha_r - 2^{m+\varsigma} u \right)
  \right)
  \right\}_{2\pi}
  =
  \left\{
    \theta_d
    -
    \frac{d}{r} \left( \theta_r - 2 \pi u \right)
  \right\}_{2\pi}
  \label{eq:phi-eta}
\end{align}
and let $z = \round{rj / 2^{m + \varsigma}}$, which implies $rj = 2^{m+\varsigma} z + \{rj\}_{2^{m + \varsigma}}$.
Then
\begin{align*}
  A_\mu(j)
  &=
  \frac{1}{2^{m + \varsigma}}
  \sum_{a \, = \, 0}^{2^{m + \varsigma} - 1}
  \exp
  \left[
    2 \pi \imag
    \left(
      \frac{j}{2^{m + \varsigma}}
      -
      \frac{\mu}{r}
    \right)
    a
  \right] \\
  &=
  \frac{1}{2^{m + \varsigma}}
  \sum_{a \, = \, 0}^{2^{m + \varsigma} - 1}
  \exp
  \left[
    \frac{2 \pi \imag}{2^{m + \varsigma} r}
    \left(
      rj
      -
      2^{m + \varsigma} \mu
    \right)
    a
  \right] \\
  &=
  \frac{1}{2^{m + \varsigma}}
  \sum_{a \, = \, 0}^{2^{m + \varsigma} - 1}
  \exp
  \left[
    \frac{2 \pi \imag}{2^{m + \varsigma} r}
    \left(
      \{ rj \}_{2^{m+\varsigma}}
      -
      2^{m + \varsigma} (\mu - z)
    \right)
    a
  \right] \\
  &=
  \frac{1}{2^{m + \varsigma}}
  \sum_{a \, = \, 0}^{2^{m + \varsigma} - 1}
  \e^{\imag (\theta_r - 2 \pi (\mu - z)) a / r},
\end{align*}
and furthermore
\begin{align*}
  B_\mu(k)
  &=
  \frac{1}{2^{\ell}}
  \sum_{b \, = \, 0}^{2^{\ell} - 1}
  \exp
  \left[
    2 \pi \imag
    \left(
      \frac{k}{2^{\ell}}
      +
      \frac{d \mu}{r}
    \right)
    b
  \right] \\
  &=
  \frac{1}{2^{\ell}}
  \sum_{b \, = \, 0}^{2^{\ell} - 1}
  \exp
  \left[
   \frac{2 \pi \imag}{2^{m+\varsigma}}
    \left(
      2^{m + \varsigma - \ell} k
      +
      \frac{d}{r} 2^{m+\varsigma} \mu
    \right)
    b
  \right] \\
  &=
  \frac{1}{2^{\ell}}
  \sum_{b \, = \, 0}^{2^{\ell} - 1}
  \exp
  \left[
    \frac{2 \pi \imag}{2^{m+\varsigma}}
    \left(
      dj + 2^{m + \varsigma - \ell} k - dj
      +
      \frac{d}{r} 2^{m+\varsigma} \mu
    \right)
    b
  \right] \\
  &=
  \frac{1}{2^{\ell}}
  \sum_{b \, = \, 0}^{2^{\ell} - 1}
  \exp
  \left[
    \frac{2 \pi \imag}{2^{m+\varsigma}}
    \left(
      dj + 2^{m + \varsigma - \ell} k
      -
      \frac{d}{r} (rj - 2^{m+\varsigma} \mu)
    \right)
    b
  \right] \\
  &=
  \frac{1}{2^{\ell}}
  \sum_{b \, = \, 0}^{2^{\ell} - 1}
  \exp
  \left[
    \frac{2 \pi \imag}{2^{m+\varsigma}}
    \left(
      \{ dj + 2^{m + \varsigma - \ell} k \}_{2^{m + \varsigma}}
      -
      \frac{d}{r} (2^{m+\varsigma} z + \{rj\}_{2^{m + \varsigma}} - 2^{m+\varsigma} \mu)
    \right)
    b
  \right] \\
  &=
  \frac{1}{2^{\ell}}
  \sum_{b \, = \, 0}^{2^{\ell} - 1}
  \exp
  \left[
    \frac{2 \pi \imag}{2^{m+\varsigma}}
    \left(
      \alpha_d
      -
      \frac{d}{r} (\alpha_r - 2^{m+\varsigma} (\mu - z))
    \right)
    b
  \right] \\
  &=
  \frac{1}{2^{\ell}}
  \sum_{b \, = \, 0}^{2^{\ell} - 1}
  \exp
  \left[
    \imag
    \left(
      \theta_d
      -
      \frac{d}{r} (\theta_r - 2 \pi (\mu - z))
    \right)
    b
  \right]
  =
  \frac{1}{2^{\ell}}
  \sum_{b \, = \, 0}^{2^{\ell} - 1}
  \e^{\imag \phi_{\mu - z} b},
\end{align*}
which implies that
\begin{align}
  \refeq{probability-sum}
  &=
  \frac{1}{r}
  \sum_{\mu \, = \, 0}^{r-1}
  \left|\,
    A_\mu(j) \, B_\mu(k)
  \,\right|^2
  =
  \frac{1}{2^{2(m + \varsigma + \ell)} r}
  \sum_{\mu \, = \, 0}^{r-1}
  \left|\,
    \sum_{a \, = \, 0}^{2^{m + \varsigma} - 1}
    \e^{\imag (\theta_r - 2 \pi (\mu - z)) a / r}
    \sum_{b \, = \, 0}^{2^{\ell} - 1}
    \e^{\imag \phi_{\mu - z} b}
  \,\right|^2 \notag \\
  &=
  \frac{1}{2^{2(m + \varsigma + \ell)} r}
  \sum_{\eta \, = \, -\floor{r/2}}^{\ceil{r/2} - 1}
  \left|\,
    \sum_{a \, = \, 0}^{2^{m + \varsigma} - 1}
    \e^{\imag (\theta_r - 2 \pi \eta) a / r}
    \sum_{b \, = \, 0}^{2^{\ell} - 1}
    \e^{\imag \phi_{\eta} b}
  \,\right|^2
  \label{eq:probability-sum2}
\end{align}
where we have let $\eta = \{ \mu - z \}_r$ by using that $\e^{\imag (\theta_r - 2 \pi \eta) a / r}$ and $\e^{\imag \phi_{\eta} b}$ are both $r$-periodic in~$\eta$.

Our rationale for introducing~$\eta$ is that it is natural to think of~$\eta$ as a peak index; $\eta = 0$ is the index of the main peak, $\eta = \pm 1$ are the indices of the two secondary peaks, etc.

\subsection{Deriving closed-form expressions}
\label{section:closed-form-expressions}
It follows from the above derivation that the probability of observing~$(j, k)$ and $y = [e] \, g$, summed over all $e \in [0, r) \cap \mathbb Z$, is given by
\begin{align}
  \refeq{probability-sum2}
  &=
  \sum_{\eta \, = \, -\floor{r/2}}^{\ceil{r/2} - 1}
  \underbrace{\frac{1}{2^{2(m + \varsigma)} r}
  \left|\,
    \sum_{a \, = \, 0}^{2^{m + \varsigma} - 1}
    \e^{\imag (\theta_r - 2 \pi \eta) a / r}
  \,\right|^2}_{F_\eta(\theta_r)}
  \cdot
  \underbrace{\frac{1}{2^{2 \ell}}
  \left|\,
    \sum_{b \, = \, 0}^{2^{\ell} - 1}
    \e^{\imag \phi_{\eta} b}
  \,\right|^2}_{h(\phi_{\eta})}
  \label{eq:probability-sum3}
\end{align}
where the two functions~$F_{\eta}(\theta_r)$ and~$h(\phi_{\eta})$ introduced above may be placed on closed form, as
\begin{align}
  \label{eq:F}
  F_{\eta}(\theta_r)
  =
  \frac{1}{2^{2(m + \varsigma)} r}
  \frac{1 - \cos(2^{m + \varsigma} (\theta_r - 2 \pi \eta) / r)}{1 - \cos((\theta_r - 2 \pi \eta) / r)}
  \quad \text{ for } \theta_r \neq 2 \pi \eta \quad \text{ and } \quad
  F_0(0) = \frac{1}{r},
\end{align}
where we use that $|\, \theta_r \,| \le \pi$ implies that $\theta_r = 2 \pi \eta$ only if $\eta = \theta_r = 0$, and as
\begin{align}
  \label{eq:h}
  h(\phi_{\eta}) =
  \frac{1}{2^{2\ell}}
  \frac{\cos(\phi_{\eta} \, 2^{\ell}) - 1}{\cos(\phi_{\eta}) - 1}
  \quad \text{ for } \phi_{\eta} \neq 0 \quad \text{ and } \quad
  h(0) = 1.
\end{align}

The derivation in Sects.~\ref{section:quantum-algorithm}--\ref{section:closed-form-expressions} is summarized and formalized in the below lemma:

\begin{lemma}
  \label{lemma:probability-distribution}
  The probability of observing~$(j, k)$ and $y = [e] \, g$, summed over all $e \in [0, r) \cap \mathbb Z$, in a run of the quantum algorithm in Sect.~\ref{section:quantum-algorithm}, is
  \begin{align}
    \frac{1}{r}
    \sum_{\mu \, = \, 0}^{r - 1}
    |\, A_{\mu}(j) \, B_{\mu}(k) \,|^2
    =
    \sum_{\eta \, = \, -\floor{r/2}}^{\ceil{r/2} - 1}
    F_{\eta}(\theta_r)
    \,
    h(\phi_{\eta}),
    \label{eq:probability-sum-lemma-equated}
  \end{align}
  where the term for~$\eta$ on the right-hand side is equal to the term for $\mu = (z + \eta) \text{ mod } r$ on the left-hand side, for $z = \round{rj / 2^{m+\varsigma}}$, for~$A_\mu(j)$ and~$B_\mu(k)$ as in~\refeq{Amu} and~\refeq{Bmu}, respectively, and for~$F_{\eta}(\theta_r)$ and~$h(\phi_{\eta})$ as in~\refeq{probability-sum3}, where~$\theta_r$ and $\phi_{\eta}$ are as in~\refeq{arguments-angles} and~\refeq{phi-eta}, respectively.
  More specifically, it holds for $\eta = \{ \mu - z \}_r$ that $|\, A_\mu(j) \,|^2 / r = F_{\eta}(\theta_r)$ and $|\, B_\mu(k) \,|^2 = h(\phi_{\eta})$.
\end{lemma}
\begin{proof}
  As shown in Sects.~\ref{section:quantum-algorithm}--\ref{section:closed-form-expressions}, it holds that $\refeq{probability} = \refeq{probability2}$ is the probability of observing~$(j, k)$ and $y = [e] \, g$ for $e \in [0, r) \cap \mathbb Z$.
  Summing~\refeq{probability2} over all $e \in [0, r) \cap \mathbb Z$ yields $\refeq{probability-sum} = \refeq{probability-sum2} = \refeq{probability-sum3}$.
  The expressions equated in~\refeq{probability-sum-lemma-equated} in the lemma are~\refeq{probability-sum} and~\refeq{probability-sum3}, and so that part follows.
  The termwise equalities follow from the expressions for~$A_\mu(j)$ and~$B_\mu(k)$, which are $r$-periodic in $\mu - z$, and from the definitions of~$F_{\eta}(\theta_r)$ and~$h(\phi_{\eta})$, and so the lemma follows.
\end{proof}

\subsubsection{Equivalent closed-form representation}
When $\theta_r \neq 2 \pi \eta$, it holds that
\begin{align}
  F_{\eta}(\theta_r)
  &=
  \frac{1}{2^{2(m + \varsigma)} r}
  \frac{1 - \cos(2^{m + \varsigma} (\theta_r - 2 \pi \eta) / r)}{1 - \cos((\theta_r - 2 \pi \eta) / r)} \notag \\
  &=
  \frac{1}{2^{2(m + \varsigma)} r}
  \sum_{t \, = \, -\infty}^{\infty}
  \frac{1 - \cos(2^{m + \varsigma} (\theta_r - 2 \pi \eta) / r)}{((\theta_r - 2 \pi (\eta + t r))/r)^2/2} \notag \\
  &=
  \frac{r}{2^{2(m + \varsigma)}}
  \sum_{t \, = \, -\infty}^{\infty}
  \frac{2(1 - \cos(2^{m + \varsigma} (\theta_r - 2 \pi (\eta + t r)) / r))}{(\theta_r - 2 \pi (\eta + t r))^2}
  \label{eq:equivalent-F}
\end{align}
as may be seen by expanding the denominator $1 - \cos((\theta_r - 2 \pi \eta) / r)$ in a partial fraction, and by observing that the numerator is $r$-periodic in~$\eta$.

By substituting~\refeq{equivalent-F} into~\refeq{probability-sum3}, we obtain the below equivalent closed-form expression
\begin{align}
  \refeq{probability-sum3}
  &=
  \sum_{\eta \, = \, -\floor{r/2}}^{\ceil{r/2} - 1}
  \underbrace{\frac{1}{2^{2(m + \varsigma)} r}
  \left|\,
    \sum_{a \, = \, 0}^{2^{m + \varsigma} - 1}
    \e^{\imag (\theta_r - 2 \pi \eta) a / r}
  \,\right|^2}_{F_\eta(\theta_r)}
  \cdot
  \underbrace{\frac{1}{2^{2 \ell}}
  \left|\,
    \sum_{b \, = \, 0}^{2^{\ell} - 1}
    \e^{\imag \phi_{\eta} b}
  \,\right|^2}_{h(\phi_{\eta})} \notag \\
  &=
  \sum_{\eta \, = \, -\floor{r/2}}^{\ceil{r/2} - 1}
  \frac{r}{2^{2(m + \varsigma)}}
  \sum_{t \, = \, -\infty}^{\infty}
  \frac{2(1 - \cos(2^{m + \varsigma} (\theta_r - 2 \pi (\eta + t r)) / r))}{(\theta_r - 2 \pi (\eta + t r))^2}
  \,
  h(\phi_{\eta}) \notag \\
  &=
  \sum_{\eta \, = \, -\infty}^{\infty}
  \underbrace{\frac{r}{2^{2(m + \varsigma)}}
  \frac{2(1 - \cos(2^{m + \varsigma} (\theta_r - 2 \pi \eta) / r))}{(\theta_r - 2 \pi \eta)^2}}_{f_\eta(\theta_r)}
  \,
  h(\phi_{\eta}),
  \label{eq:probability-sum4}
\end{align}
when $\theta_r \neq 2 \pi \eta$, where we have used that $h(\phi_{\eta})$ is $r$-periodic in $\eta$.

Note furthermore that the fact that $|\, \theta_r \,| \le \pi$ implies that $\theta_r = 2 \pi \eta$ only if $\eta = \theta_r = 0$, in which case we define $f_{0}(0) = 1/r$ for completeness.
The equality between~\refeq{probability-sum3} and the last expression in~\refeq{probability-sum4} then holds also when $\theta_r = 0$, as~\refeq{equivalent-F} holds for $\eta \neq 0$, as $F_0(0) = 1/r = f_0(0)$, and as $f_{\eta}(0) = 0$ for all $\eta \in r \mathbb Z \setminus \{ 0 \}$.

The above derivation is summarized and formalized in the below lemma:

\begin{lemma}
  \label{lemma:probability-distribution-f}
  The probability of observing~$(j, k)$ and $y = [e] \, g$, summed over all $e \in [0, r) \cap \mathbb Z$, in a run of the quantum algorithm in Sect.~\ref{section:quantum-algorithm}, is
  \begin{align}
    \sum_{\eta \, = \, -\infty}^{\infty}
    f_{\eta}(\theta_r)
    \,
    h(\phi_{\eta}),
    \label{eq:probability-sum-lemma-equated-f}
  \end{align}
  where~$f_{\eta}(\theta_r)$ and~$h(\phi_{\eta})$ are as in~\refeq{probability-sum4} and~\refeq{probability-sum3}, respectively, with $f_0(0) = 1/r$, and~$\theta_r$ and $\phi_{\eta}$ are as in~\refeq{arguments-angles} and~\refeq{phi-eta}, respectively.
\end{lemma}
\begin{proof}
  The lemma follows from Lem.~\ref{lemma:probability-distribution}, by substituting~\refeq{equivalent-F} into~\refeq{probability-sum-lemma-equated} as explained above.
\end{proof}

\subsubsection{Truncated closed-form expressions}
It follows from the above derivation that the probability of observing a pair $(j, k)$ that is associated with a peak with index ${\eta \in [-B_\eta, B_\eta] \cap \mathbb Z}$ can be lower bounded as follows:

\begin{lemma}
  \label{lemma:probability-distribution-truncated}
  The probability of observing a pair~$(j, k)$ that is associated with a peak with index ${\eta \in [-B_\eta, B_\eta] \cap \mathbb Z}$ and $y = [e] \, g$, summed over all $e \in [0, r) \cap \mathbb Z$, in a run of the quantum algorithm in Sect.~\ref{section:quantum-algorithm}, is
  \begin{align}
    \sum_{\eta \, = \, -B_\eta}^{B_\eta}
    F_\eta(\theta_r) \, h(\phi_\eta)
    \ge
    \sum_{\eta \, = \, -B_\eta}^{B_\eta}
    f_\eta(\theta_r) \, h(\phi_\eta),
    \label{eq:probability-partial-sum}
  \end{align}
  where $0 \le B_\eta \le \ceil{r/2}-1 = \floor{(r-1)/2} \le \floor{r/2}$, $F_{\eta}(\theta_r)$ and~$h(\phi_{\eta})$ are as in~\refeq{probability-sum3}, $f_{\eta}(\theta_r)$ is as in~\refeq{probability-sum4}, with $f_0(0) = 1/r$, and~$\theta_r$ and $\phi_{\eta}$ are as in~\refeq{arguments-angles} and~\refeq{phi-eta}, respectively.
\end{lemma}
\begin{proof}
  The left-hand side of~\refeq{probability-partial-sum} is the right-hand side of~\refeq{probability-sum-lemma-equated} in Lem.~\ref{lemma:probability-distribution} truncated to the peaks with indices~$\eta \in [-B_\eta, B_\eta] \cap \mathbb Z$.
  The inequality follows from the fact that $f_\eta(\theta_r) \le F_\eta(\theta_r)$:
  If $\theta_r \neq 2 \pi \eta$, then $F_\eta(\theta_r)$ is by~\refeq{equivalent-F} a sum of non-negative terms, the term for $t = 0$ of which is~$f_\eta(\theta_r)$.
  Otherwise, $\eta = \theta_r = 0$, and $F_0(0) = f_0(0) = 1/r$, and so the lemma follows.
\end{proof}

\section{Capturing the probability distribution}
\label{section:capturing-probability-distribution}
In what follows below, we describe how the probability distribution induced by the quantum algorithm may be captured by summing up the parts of the distribution where the probability mass is expected to be concentrated due to constructive interference arising.

\subsection{The notion of \texorpdfstring{$B_{\eta}$-$B_{\Delta}$}{B\_eta-B\_Delta}-good pairs~\texorpdfstring{$(j, k)$}{(j, k)}}
\label{section:notion-good-pair}
For each~$\eta$ and~$j$, there is an optimal value~$k_{\eta, 0}$ of~$k$ that minimizes~$|\, \phi_{\eta} \,|$, as given by
\begin{align}
  k_{\eta, 0}
  &=
  \round{\frac{1}{2^{m + \varsigma - \ell}} \left(-dj + \frac{d}{r} \left( \{ rj \}_{2^{m+\varsigma}} - 2^{m + \varsigma} \eta \right) \right)} \text{ $\mathrm{mod}$ } 2^\ell \label{eq:k0} \\
  &=
  \left(
    \frac{1}{2^{m + \varsigma - \ell}} \left(-dj + \frac{d}{r} \left( \{ rj \}_{2^{m+\varsigma}} - 2^{m + \varsigma} \eta \right) \right) - \delta_{\eta}
  \right)
  +
  2^\ell \nu_\eta \notag
\end{align}
for some~$\delta_{\eta} \in [-\frac{1}{2}, \frac{1}{2})$ and some integer~$\nu_\eta$.
More specifically,
\begin{align}
  \delta_\eta
  =
  \left\{
    \frac{1}{2^{m + \varsigma - \ell}} \left(-dj + \frac{d}{r} \left( \{ rj \}_{2^{m+\varsigma}} - 2^{m + \varsigma} \eta \right) \right)
  \right\}_1.
  \label{eq:delta-eta}
\end{align}

For $k = k_{\eta, \Delta} = (k_{\eta, 0} + \Delta) \text{ mod } 2^\ell$, for $\Delta \in [-2^{\ell-1}, 2^{\ell-1}) \cap \mathbb Z$, it then holds that
\begin{align}
  \phi_{\eta}
  &=
  \refeq{phi-eta}
  =
  \left\{
    \theta_d - \frac{d}{r} \left( \theta_r - 2 \pi \eta \right)
  \right\}_{2\pi}
  =
  \frac{2\pi}{2^{m+\varsigma}}
  \left\{
    \alpha_d - \frac{d}{r} \left( \alpha_r - 2^{m+\varsigma} \eta \right)
  \right\}_{2^{m+\varsigma}} \notag \\
  &=
  \frac{2\pi}{2^{m+\varsigma}}
  \left\{
    dj + 2^{m+\varsigma-\ell} k - \frac{d}{r} \left( \{ rj \}_{2^{m+\varsigma}} - 2^{m+\varsigma} \eta \right)
  \right\}_{2^{m+\varsigma}} \notag \\
  &=
  \frac{2\pi}{2^{m+\varsigma}}
  \left\{
    2^{m+\varsigma-\ell} \left( \Delta - \delta_{\eta} \right)
  \right\}_{2^{m+\varsigma}}
  =
  \frac{2\pi}{2^{\ell}}
  \left\{
    \Delta - \delta_{\eta}
  \right\}_{2^{\ell}},
  \label{eq:phi-eta-extended}
\end{align}
which is close to zero for small~$\Delta$, leading to constructive interference arising.

It is convenient therefore to introduce the following notion of a $B_{\eta}$-$B_{\Delta}$-good pair~$(j, k)$:
\begin{definition}
A pair~$(j, k)$ is said to be $B_{\eta}$-$B_{\Delta}$-good if
\begin{align*}
  \phi_{\eta}
  &=
  \frac{2\pi}{2^{m+\varsigma}}
  \left\{
    dj + 2^{m+\varsigma-\ell} k - \frac{d}{r} \left( \{ rj \}_{2^{m+\varsigma}} - 2^{m+\varsigma} \eta \right)
  \right\}_{2^{m+\varsigma}}
  =
  \frac{2\pi}{2^{\ell}}
  \left\{
  \Delta - \delta_{\eta}
  \right\}_{2^\ell}
\end{align*}
for~$\eta$ and~$\Delta$ integers such that $|\, \eta \,| \le B_{\eta}$ and $\Delta \in \offsets(B_{\Delta})$ for $B_{\Delta} \in [0, 2^{\ell-1}] \cap \mathbb Z$, and for some $\delta_{\eta} \in [-\frac{1}{2}, \frac{1}{2})$, where $\offsets(B) = [-B, B] \cap [-2^{\ell-1}, 2^{\ell-1}) \cap \mathbb Z$ for real $B \ge 0$.
\end{definition}

\subsection{The distribution of arguments~\texorpdfstring{$\alpha_r$}{alpha\_r}}
\label{section:distribution-alpha-r}
Before proceeding to capture the distribution, we need to understand how the arguments~$\alpha_r$ are distributed on $[-2^{m+\varsigma-1}, 2^{m+\varsigma-1}) \cap \mathbb Z$ as a function of~$r$.
In analogy with~\cite[App.~A.3]{ekera-general}:

\begin{claim}
  \label{claim:alpha-r}
  There are~$2^{\kappa_r}$ distinct $j \in [0, 2^{m+\varsigma}) \cap \mathbb Z$ that yield a given argument $\alpha_r = \{ rj \}_{2^{m+\varsigma}}$ for~$2^{\kappa_r}$ the greatest power of two to divide both~$r$ and~$2^{m+\varsigma}$.
  It follows that there are~$2^{m+\varsigma-\kappa_r}$ distinct arguments~$\alpha_r$, that are all multiples of~$2^{\kappa_r}$, and that all occur with multiplicity~$2^{\kappa_r}$.
  \begin{proof}
    It is easy to see that the frequencies $j + 2^{m + \varsigma - \kappa_r} t_r \text{ mod } 2^{m+\varsigma}$ for $t_r \in [0, 2^{\kappa_r}) \cap \mathbb Z$ yield the same argument $\alpha_r = \{ rj \}_{2^{m+\varsigma}}$, and that all arguments must be multiples of~$2^{\kappa_r}$, since the modulus when computing~$\alpha_r$ is~$2^{m+\varsigma}$.
    Conversely, if two frequencies~$j$ and~$j'$ yield the same argument, then $r (j - j') \equiv 0 \:\: (\text{mod } 2^{m+\varsigma})$, which implies that $j \equiv j' \:\: (\text{mod } 2^{m+\varsigma-\kappa_r})$, since $r / 2^{\kappa_r}$ is odd if $\kappa_r < m + \varsigma$ (and trivially if $\kappa_r = m + \varsigma$).
    It follows that exactly~$2^{\kappa_r}$ frequencies yield each argument, and so the claim follows.
  \end{proof}
\end{claim}

In what follows, we shall assume that~$r$ is not a multiple of~$2^{m+\varsigma}$, in which case~$2^{\kappa_r}$ is the greatest power of two to divide~$r$.
This slightly simplifies the analysis.
Note that the situation where~$r$ is a multiple of~$2^{m+\varsigma}$ cannot arise when computing general discrete logarithms in groups of known order~$r$, since~$m$ is then the bit length of~$r$.
When computing short discrete logarithms, it can in theory arise, but not if~$r$ is prime as is typically the case; see Sect.~\ref{section:assumptions-reductions}.

A notable special case is Ekerå--Håstad's algorithm~\cite{ekera-hastad, ekera-pp} for factoring $N = pq$, where $p, q$ are two distinct random primes of similar bit length, by means of computing a short discrete logarithm in a cyclic $r$-order subgroup of~$\mathbb Z_N^*$.
In this algorithm, $r$ must divide $\text{lcm}(p-1, q-1)$, so if~$2^{m+\varsigma}$ divides~$r$ then $2^{m+\varsigma}$ must divide either $p-1$ or $q-1$, and furthermore~$2^{m}$ is of the same size as $p+q$ or greater, so the factors $p, q$ may then be deduced classically.

\subsection{Probability of observing a \texorpdfstring{$B_{\eta}$-$B_{\Delta}$}{B\_eta-B\_Delta}-good pair}
\label{section:probability-good-pair}
By~\refeq{probability-partial-sum} in Lem.~\ref{lemma:probability-distribution-truncated}, the probability of observing a $B_{\eta}$-$B_{\Delta}$-good pair is at least
\begin{align}
  &\sum_{\eta \, = \, -B_{\eta}}^{B_{\eta}} \:
  \sum_{\alpha_r' \, = \, -2^{m+\varsigma-\kappa_r-1}}^{2^{m+\varsigma-\kappa_r-1} - 1}
  2^{\kappa_r}
  F_{\eta}
  \left(
    \frac{2\pi}{2^{m+\varsigma}} \, 2^{\kappa_r} \alpha'_r
  \right)
  \sum_{\Delta \, \in \, \offsets(B_{\Delta})}
  \hspace{-2mm}
  h
  \left(
    \frac{2 \pi} {2^\ell}
    (\Delta - \delta_\eta)
  \right) \ge \notag \\
  &\quad\quad\quad
  \sum_{\eta \, = \, -B_{\eta}}^{B_{\eta}} \:
  \sum_{\alpha_r' \, = \, -2^{m+\varsigma-\kappa_r-1}}^{2^{m+\varsigma-\kappa_r-1} - 1}
  2^{\kappa_r}
  f_{\eta}
  \left(
    \frac{2\pi}{2^{m+\varsigma}} \, 2^{\kappa_r} \alpha'_r
  \right)
  \sum_{\Delta \, \in \, \offsets(B_{\Delta})}
  \hspace{-2mm}
  h
  \left(
    \frac{2 \pi} {2^\ell}
    (\Delta - \delta_\eta)
  \right)
  \label{eq:sum-Pn}
\end{align}
where we sum over all~$\alpha_r$ with multiplicity~$2^{\kappa_r}$ by introducing~$\alpha_r'$ (see Sect.~\ref{section:distribution-alpha-r}), and where~$\delta_\eta$ is a function of $\alpha_r = \{ rj \}_{2^{m+\varsigma}}$ and hence of~$\alpha'_r$ (see Sect.~\ref{section:notion-good-pair}), provided that $\kappa_r \le \ell$.\footnote{The~$2^{\kappa_r}$ frequencies~$j$ that yield a given argument~$\alpha_r$ differ by multiples of~$2^{m+\varsigma-\kappa_r}$; see Sect.~\ref{section:distribution-alpha-r}.

The corresponding values of the expression within the braces in~\refeq{delta-eta} hence differ by multiples of $2^{\ell-\kappa_r} d$, which is an integer when $\kappa_r \le \ell$, in which case~$\delta_\eta$ does not depend on which of the~$2^{\kappa_r}$ frequencies is selected.
Note that $\kappa_r = 0$ when~$r$ is odd, as is the case when~$r$ has no small prime factors; see Sect.~\ref{section:assumptions-reductions}.
Note furthermore that if $\ell = 0$, as in Sect.~\ref{section:generalizations-order}, the requirement that $\kappa_r \le \ell$ may be dropped, since $h(\phi) = 1$ for all~$\phi$ when $\ell = 0$.
The sum over~$\Delta$ in~\refeq{sum-Pn} is then equal to one irrespective of~$\delta_\eta$.}
Furthermore, as~$h(\phi_\eta)$ is $2 \pi$-periodic, $h(\phi_\eta) = h(2 \pi \{ \Delta - \delta_\eta \}_{2^\ell} / 2^\ell) = h(2 \pi (\Delta - \delta_\eta) / 2^\ell)$.

\subsection{Asymptotic soundness}
\label{section:asymptotic-soundness}
By Thm.~\ref{theorem:sum-to-one} below, the expression~\refeq{sum-Pn} for the probability of observing a $B_{\eta}$-$B_{\Delta}$-good pair tends to one asymptotically, when summed over all~$\Delta$, and~$B_{\eta}$ and/or~$\varsigma$ tend to infinity:
\begin{theorem}
  \label{theorem:sum-to-one}
  It holds that
  \begin{align*}
  \lim_{\substack{B_{\eta} \rightarrow \infty \\ \text{ and/or } \\ \varsigma \rightarrow \infty}}
  \sum_{\eta \, = \, -B_{\eta}}^{B_{\eta}} \:
  \sum_{\alpha_r' \, = \, -2^{m+\varsigma-\kappa_r-1}}^{2^{m+\varsigma-\kappa_r-1} - 1}
  2^{\kappa_r}
  f_{\eta}
  \left(
    \frac{2\pi}{2^{m+\varsigma}} \, 2^{\kappa_r} \alpha'_r
  \right)
  \sum_{\Delta \, \in \, \offsets(2^{\ell-1})}
  \hspace{-2mm}
  h
  \left(
    \frac{2 \pi} {2^\ell}
    (\Delta - \delta_\eta)
  \right)
  = 1.
  \end{align*}
\end{theorem}
\begin{proof}
  The theorem follows by combining Lem.~\ref{lemma:h-sum-to-one} and Lem.~\ref{lemma:fn-sum-to-one} in App.~\ref{appendix:supporting-lemmas-claims-proofs}.
\end{proof}

\subsection{Lower-bounded probability of observing a \texorpdfstring{$B_{\eta}$-$B_{\Delta}$}{B\_eta-B\_Delta}-good pair}
\label{section:lower-bounding-sum-Pn}
By Thm.~\ref{theorem:lower-bound} below, the expression~\refeq{sum-Pn} for the probability of observing a $B_{\eta}$-$B_{\Delta}$-good pair quickly tends to one as~$B_{\Delta}$, and~$B_{\eta}$ and/or~$\varsigma$, increase:

\begin{theorem}
  \label{theorem:lower-bound}
  Suppose that $\kappa_r \le \ell$ (see Sect.~\ref{section:probability-good-pair}), as is the case when~$r$ is odd.
  The probability of observing a $B_{\eta}$-$B_{\Delta}$-good pair is then lower bounded by
  \begin{align}
  &\sum_{\eta \, = \, -B_{\eta}}^{B_{\eta}} \:
  \sum_{\alpha_r' \, = \, -2^{m+\varsigma-\kappa_r-1}}^{2^{m+\varsigma-\kappa_r-1} - 1}
  2^{\kappa_r}
  f_{\eta}
  \left(
    \frac{2\pi}{2^{m+\varsigma}} \, 2^{\kappa_r} \alpha'_r
  \right)
  \sum_{\Delta \, \in \, \offsets(B_{\Delta})}
  \hspace{-2mm}
  h
  \left(
    \frac{2 \pi} {2^\ell}
    (\Delta - \delta_\eta)
  \right)
  \ge \rho = \notag \\
  &\quad\quad
  \underbrace{\max\left(
    0,
    1
    -
    \frac{2}{\pi^2}
    \frac{r}{2^{m}}
    \frac{1}{2^{\varsigma} (B_{\eta} + 1/2)}
    \left(
      1
      +
      \epsilon_{B_{\eta}}
    \right)
  \right)}_{= \, \rho_1} \cdot \notag \\
  &\quad\quad\quad\quad
  \underbrace{\max\left(
    0,
    1
    -
    \frac{1}{\pi^2}
    \left(
      \frac{1}{B_{\Delta} + 1/2}
      (
        1
        +
        \epsilon_{B_{\Delta}}(1/2)
      )
      +
      \frac{1}{B_{\Delta} + 3/2}
      (
        1
        +
        \epsilon_{B_{\Delta}}(3/2)
      )
    \right)
  \right)}_{= \, \rho_2}, \label{eq:theorem:lower-bound}
  \end{align}
  for~$\varsigma$ a non-negative integer, for $B_{\eta} \in [0, \ceil{r/2}) \cap \mathbb Z$ and $B_{\Delta} \in [0, 2^{\ell-1}] \cap \mathbb Z$, and for
  \begin{align*}
    \epsilon_{B_{\Delta}}(x) = \epsilon(B_{\Delta} + x)
    \quad \text{ and } \quad
    \epsilon_{B_{\eta}} = \epsilon(2^{m+\varsigma-\kappa_r} (B_{\eta} + 1/2))
    \quad \text{ where } \quad
    \epsilon(x) = \frac{1}{2x} + \frac{1}{6x^2}.
  \end{align*}
\end{theorem}
\begin{proof}
  The theorem follows by combining Lem.~\ref{lemma:h-tails} and Lem.~\ref{lemma:fn-tails} in App.~\ref{appendix:supporting-lemmas-claims-proofs}.
\end{proof}

Note that if $B_{\Delta} = \floor{2^{\ell-1}}$, then $\offsets(B_{\Delta}) = \offsets(2^{\ell-1})$, so the sum over~$\Delta$ in~\refeq{theorem:lower-bound} is equal to one by Lem.~\ref{lemma:h-sum-to-one}, and the second factor in the lower bound~\refeq{theorem:lower-bound} may then be replaced by one.

The lower bound~\refeq{theorem:lower-bound} in Thm.~\ref{theorem:lower-bound} on the expression for the probability of observing~a $B_{\eta}$-$B_{\Delta}$-good pair is tabulated in Tbls.~\ref{table:probability-strict-varsigma-0}--\ref{table:probability-strict-Bn-0} in App.~\ref{appendix:tables-lower-bound}, for $\varsigma = 0$ as a function of~$B_{\eta}$ and~$B_{\Delta}$, and for $B_{\eta} = 0$ as a function of~$\varsigma$ and~$B_{\Delta}$, respectively.
As may be seen in Tbls.~\ref{table:probability-strict-varsigma-0}--\ref{table:probability-strict-Bn-0}, it is necessary to select~$B_{\Delta}$, and~$B_{\eta}$ and/or~$\varsigma$, greater than zero for the lower bound to exceed~$99\%$, and it is easily feasible to select them so that the lower bound exceeds $1 - 10^{-4}$.

It follows from Thm.~\ref{theorem:lower-bound} that asymptotically the sum over~$h$ is $1 - O(B_{\Delta}^{-1})$, whereas the sum over~$f_\eta$ is $1 - O(2^{-\varsigma})$ and $1 - O(B_{\eta}^{-1})$ in~$\varsigma$ and~$B_{\eta}$, respectively.
By parameterizing~$B_{\Delta}$, and~$B_{\eta}$ and/or~$\varsigma$, to grow slowly as functions of~$m$, it follows that we may make the probability tend to one asymptotically while retaining the polynomial runtime of the algorithm:

\begin{theoremcorollary}
  \label{corollary:lower-bound-asymptotics}
  For~$m$ the bit length of~$r$, $\kappa_r \le \ell$, $B_{\eta} = \omega_m(1) < \ceil{r/2}$, and $B_{\Delta} = \omega_m(1) < 2^{\ell - 1}$, the probability of observing a $B_{\eta}$-$B_{\Delta}$-good pair, as lower bounded by
  \begin{align*}
  &\sum_{\eta \, = \, -B_{\eta}}^{B_{\eta}} \:
  \sum_{\alpha_r' \, = \, -2^{m+\varsigma-\kappa_r-1}}^{2^{m+\varsigma-\kappa_r-1} - 1}
  2^{\kappa_r}
  f_{\eta}
  \left(
    \frac{2\pi}{2^{m+\varsigma}} \, 2^{\kappa_r} \alpha'_r
  \right)
  \sum_{\Delta \, \in \, \offsets(B_{\Delta})}
  \hspace{-2mm}
  h
  \left(
    \frac{2 \pi} {2^\ell}
    (\Delta - \delta_\eta)
  \right),
  \end{align*}
  tends to one in the limit as $m \rightarrow \infty$.
\end{theoremcorollary}
\begin{proof}
  The corollary follows by taking the limit of the lower bound in Thm.~\ref{theorem:lower-bound}.
\end{proof}

Note that the restriction that $B_{\eta} < \ceil{r/2}$ in Thm.~\ref{theorem:lower-bound} and Cor.~\ref{corollary:lower-bound-asymptotics} may be relaxed since they use~$f_\eta$ rather than~$F_\eta$, and the terms $f_{\eta + tr}(\theta_r)$ for $t \in \mathbb Z$ sum to~$F_\eta(\theta_r)$ by~\refeq{equivalent-F}, so that no peak is overcounted, but we impose said restriction above and in what follows for simplicity.

\subsection{Expected probability of observing a \texorpdfstring{$B_{\eta}$-$B_{\Delta}$}{B\_eta-B\_Delta}-good pair}
Although the rigorous lower bound in Thm.~\ref{theorem:lower-bound} tends to one asymptotically, it does not provide a good estimate for the expected probability of obtaining a $B_{\eta}$-$B_{\Delta}$-good pair for small values of~$B_{\Delta}$, and~$B_{\eta}$ and/or~$\varsigma$.
In this section, we heuristically derive such an estimate:

Specifically, substituting~\refeq{phi-eta-extended} into~\refeq{probability-partial-sum}, heuristically replacing~$\delta_{\eta}$ with a stochastic variable~$\delta$ uniformly distributed on $\left[ -\frac{1}{2}, \frac{1}{2} \right)$, and computing the expectation value, yields
\begin{align}
  \sum_{\eta \, = \, -B_{\eta}}^{B_{\eta}}
  F_{\eta}(\theta_r)
  \int_{-1/2}^{1/2}
  h\left( \frac{2\pi}{2^\ell} (\Delta - \delta) \right)
  \,\mathrm{d}\delta.
  \label{eq:P-expectation-delta}
\end{align}

Integrating~\refeq{P-expectation-delta} over all $\theta_r = 2\pi \alpha_r / 2^{m+\varsigma}$ with multiplicity (see Sect.~\ref{section:distribution-alpha-r}), and then summing over all integers~$\Delta$ such that $|\, \Delta \,| \le B_{\Delta}$ where $B_{\Delta} < 2^{\ell-1}$, yields
\begin{align}
  &\sum_{\eta \, = \, -B_{\eta}}^{B_{\eta}}
  2^{\kappa_r}
  \int_{-2^{m+\varsigma-\kappa_r-1}}^{2^{m+\varsigma-\kappa_r-1}}
  F_{\eta}\left( \frac{2\pi}{2^{m+\varsigma}} \, 2^{\kappa_r} \alpha'_r \right)
  \,\mathrm{d}\alpha'_r \,
  \sum_{\Delta \, = \, -B_{\Delta}}^{B_{\Delta}} \, \int_{-1/2}^{1/2}
  h\left( \frac{2\pi}{2^\ell} (\Delta - \delta) \right)
  \,\mathrm{d}\delta
  \notag \\
  =
  &\sum_{\eta \, = \, -B_{\eta}}^{B_{\eta}} \,
  2^{\kappa_r}
  \int_{-2^{m+\varsigma-\kappa_r-1}}^{2^{m+\varsigma-\kappa_r-1}}
  F_{\eta}\left( \frac{2\pi}{2^{m+\varsigma}} \, 2^{\kappa_r} \alpha'_r \right)
  \,\mathrm{d}\alpha'_r
  \int_{-B_{\Delta}-1/2}^{B_{\Delta}+1/2}
  h\left( \frac{2\pi}{2^\ell} v \right)
  \,\mathrm{d}v.
  \label{eq:Pn-int}
\end{align}

The expected probability of observing a $B_{\eta}$-$B_{\Delta}$-good pair, as heuristically given by~\refeq{Pn-int}, is tabulated in Tbls.~\ref{table:probability-varsigma-0}--\ref{table:probability-Bn-0} in App.~\ref{appendix:tables-expectation}, for $\varsigma = 0$ as a function of~$B_{\eta}$ and~$B_{\Delta}$, and for $B_{\eta} = 0$ as a function of~$\varsigma$ and~$B_{\Delta}$, respectively.
As may be seen in Tbls.~\ref{table:probability-varsigma-0}--\ref{table:probability-Bn-0}, it is necessary to select~$B_{\Delta}$, and~$B_{\eta}$ and/or~$\varsigma$, greater than zero for the probability to exceed~$99\%$.

The choice of~$m$ and~$s$ is of little consequence, for as long as~$m$ is sufficiently large, and for as long as~$s$ is not very large in relation to~$m$.
The size of~$r$ in relation to~$2^m$ is important, however:
In Tbl.~\ref{table:probability-Bn-0}, the order $r = 2^{m} - 1$ is maximal.
If~$r$ is instead close to the lower end of the interval, say that $r = 2^{m-1} + 1$, then essentially all rows in the table are shifted upward one step (as halving~$r$ essentially has the same effect as incrementing~$\varsigma$ by one) yielding higher probability estimates.

In practice, the order~$r$ is often a random prime on $(2^{m-1}, 2^m) \cap \mathbb Z$ in cryptographic schemes that are based on the computational intractability of the DLP.
Notable exceptions are the NIST P-192, P-224, P-256, P-384 and P-521 elliptic curve groups~\cite[Sect.~3.2.1]{nistsp800-186} that have orders very close to~$2^m$.
This is a side effect of the base field moduli having been selected to enable fast arithmetic.
Some other elliptic curve groups also have this property.

\subsubsection{On the relation to Mosca's lower bound}
\label{section:comparison-mosca}
Mosca gives a lower bound of $((r-1)/r) \, (8/\pi^2)^2$ on the success probability in his PhD thesis~\cite[Cor.~19 on p.~58]{mosca-thesis}.
Specifically, Mosca's bound is for Shor's algorithm for computing discrete logarithms, as modified in Sect.~\ref{section:quantum-algorithm} to enable qubit recycling.
Mosca assumes~$r$ to be prime~\cite[p.~57]{mosca-thesis}, and the factor~$(r-1)/r$ accounts, in our terminology, for the probability of~$z = \round{rj / 2^{m+\varsigma}}$ being invertible modulo~$r$, as is required for the classical post-processing to recover~$d$ given~$r$ (see Sect.~\ref{section:post-processing-shor}).
Our lower bound in Thm.~\ref{theorem:lower-bound} does not account for invertibility (which we instead account for in Cor.~\ref{corollary:invertible-correct-eta} in Sect.~\ref{section:post-processing-shor-invertibility}), so our lower bound should be compared to $(8/\pi^2)^2 \approx 0.6570$, which is also the limit to which Mosca's bound tends as $r \rightarrow \infty$.

Mosca does not parameterize the algorithm in~$\varsigma$ and~$\ell$.
Furthermore, Mosca does not consider making tradeoffs, or searching in the classical post-processing.
For $r \in (2^{m-1}, 2^m) \cap \mathbb Z$, we may however fix $B_{\eta} = 0$, $\varsigma = 2$ and $\ell = m + \varsigma$ to make our algorithm equivalent to the algorithm analyzed by Mosca.
For this choice of parameters, Mosca's bound is consistent with our bound in Thm.~\ref{theorem:lower-bound}, which yields $> 0.7685$ for $B_{\Delta} = 1$ and any $m \ge 2$; see Cor.~\ref{corollary:lower-bound-comparison-mosca} below.

For comparison, our heuristically expected success probability, as given by~\refeq{Pn-int}, is
\begin{align*}
  2^{\kappa_r}
  &\int_{-2^{m+\varsigma-\kappa_r-1}}^{2^{m+\varsigma-\kappa_r-1}}
  F_0\left( \frac{2\pi}{2^{m+\varsigma}} \, 2^{\kappa_r} \alpha'_r \right)
  \,\mathrm{d}\alpha'_r \,
  \int_{-2}^{2}
  h\left( \frac{2\pi}{2^\ell} v \right)
  \,\mathrm{d}v \approx 0.9024
\end{align*}
for maximal $r = 2^m - 1$, where we have evaluated the expression numerically for $m = 128$ with Mathematica.
The same result is obtained for greater~$m$.

Note that it is required that $t = \round{r (\Delta - \delta_{\eta}) / 2^{m+\varsigma}} = 0$ and $\eta = 0$ to solve without searching as in Shor's original post-processing algorithm; see Sect.~\ref{section:post-processing-shor}.
In the limit as $m \rightarrow \infty$, it holds that $t = 0$ when $v = \Delta - \delta_{\eta} \in [-2, 2)$, giving rise to the above integration limits in~$v$, and to the choice of~$B_{\Delta} = 1$.
We consider only the contribution from~$F_0$ since $\eta = 0$.

\begin{restatable}{theoremcorollary}{corollarycomparisonmosca}
  \label{corollary:lower-bound-comparison-mosca}
  For $m \ge 2$ the bit length of $r$, $\varsigma = 2$, $\ell = m + \varsigma$, $B_{\eta} = 0$ and $B_{\Delta} = 1$,
  the probability of observing a $B_{\eta}$-$B_{\Delta}$-good pair as lower bounded by Thm.~\ref{theorem:lower-bound} is greater than $0.7685$.
\end{restatable}
\begin{proof}
  See App.~\ref{appendix:comparison-mosca} for the proof of this corollary.
\end{proof}

\section{Simulating the quantum algorithm}
\label{section:simulating-quantum-algorithm}
We now have the necessary framework in place to simulate the quantum algorithm in Sect.~\ref{section:quantum-algorithm} classically for a known problem instance, as given by~$d$ and~$r$, in the sense that we may sample the probability distribution induced by the algorithm given~$d$ and~$r$, and $m$, $\varsigma$ and $\ell$.

In what follows, we consider two simulators:
The first simulator, which is described in Sect.~\ref{section:simulating-quantum-algorithm-with-failures} below, is analogous to the simulators in our earlier works~\cite{ekera-pp, ekera-general, ekera-success-order-finding, ekera-success-short-dlp}.
It is based on the idea of pre-computing histograms for the distribution in~$\theta_r$ induced by~$F$ (or~$f$) for $\eta \in [-B_\eta, \, B_\eta] \cap \mathbb Z$.
It first samples~$\eta$ and $\theta_r = \frac{2 \pi}{2^{m+\varsigma}} \{ rj \}_{2^{m+\varsigma}}$, and hence~$j$, from said histograms.
It then samples~$k$ given~$j$ and~$\eta$ from the distribution induced by~$h$ by exploring $k = {(k_{\eta, 0} + \Delta) \text{ mod } 2^\ell}$ for $\Delta \in [-B_\Delta, B_\Delta] \cap \mathbb Z$ where $k_{\eta, 0}$ is a function of~$j$ and~$\eta$.

Said simulator has the advantage of being quite fast, but it may fail to produce a sample.
In practice, such failures occur when the pair~$(j, k)$ is not $B_\eta$-$B_\Delta$-good, so when using the simulator to, e.g., evaluate the efficiency of the classical post-processing, we may err on the side of caution by treating such sampling failures as post-processing failures.

The second simulator, which is described in Sect.~\ref{section:simulating-quantum-algorithm-without-failures} below, is based on sampling Fejér kernels.
It has the advantage of never failing to produce a sample, and it is exact under certain caveats,\footnote{Up to the precision of the arithmetic and the random number generation.} but it has the disadvantage of being slower than the first simulator.
Also, contrary to the first simulator, this simulator is not described in our earlier works~\cite{ekera-pp, ekera-general, ekera-success-order-finding, ekera-success-short-dlp}, but it may be generalized to cover said works as explained in Sect.~\ref{section:simulating-quantum-algorithm-without-failures-generalizations} (see also Sect.~\ref{section:generalizations}).

Finally, note that both simulators require~$d$ and~$r$ to be known:
They may hence not be used to simulate problem instances given in the form of two group elements~$g$ and $x = [d] \, g$.

\subsection{Sampling the probability distribution when admitting failures}
\label{section:simulating-quantum-algorithm-with-failures}
For all $\eta \in \mathbb Z$ such that~$|\, \eta \,| \le B_{\eta}$, we first construct a histogram in~$\alpha_r$ by piecewise numerical integration of~$F_{\eta}(\theta_r)$ (or of~$f_{\eta}(\theta_r)$; see Lem.~\ref{lemma:probability-distribution-f}) over small intervals in~$\alpha_r$ (where we recall that $\theta_r = 2 \pi \alpha_r / 2^{m+\varsigma}$ as defined in Sect.~\ref{section:arguments-angles}, and that~$\alpha_r$ is distributed as explained in Sect.~\ref{section:distribution-alpha-r}).

To sample~$\eta$ and~$j$, we first sample~$\eta$ and an argument~$\alpha_r$ from the resulting histograms, and then sample~$j$ from the set of all integers~$j$ that yield~$\alpha_r$, as given by
\begin{align}
  j =
  \left( \frac{\alpha_r}{2^{\kappa_r}} \left( \frac{r}{2^{\kappa_r}} \right)^{-1}
  +
  2^{m+\varsigma-\kappa_r} t_r \right) \text{ $\mathrm{mod}$ } 2^{m+\varsigma}
  \label{eq:solve-for-j}
\end{align}
as~$t_r$ runs through all integers on $[0, 2^{\kappa_r}) \cap \mathbb Z$.
This sampling procedure is analogous to that in~\cite[Lem.~A.4 in App.~A.3]{ekera-general}.
Note that all arguments~$\alpha_r$ are multiples of~$2^{\kappa_r}$ (see Sect.~\ref{section:distribution-alpha-r}), and that this fact must be accounted for when sampling~$\alpha_r$ from the histograms.

Given~$\eta$ and~$j$, we let~$k_{\eta, 0}$ be the value of $k \in [0, 2^\ell) \cap \mathbb Z$ that minimizes~$|\, \phi_{\eta} \,|$, where we recall that the expression for~$k_{\eta, 0}$ is given in~\refeq{k0}, and that
\begin{align*}
  \phi_{\eta}
  &=
  \left\{
    \theta_d - \frac{d}{r} \left( \theta_r - 2 \pi \eta \right)
  \right\}_{2\pi}
  =
  \frac{2\pi}{2^{m+\varsigma}}
  \left\{
    \alpha_d - \frac{d}{r} \left( \alpha_r - 2^{m+\varsigma} \eta \right)
  \right\}_{2^{m+\varsigma}} \\
  &=
  \frac{2\pi}{2^{m+\varsigma}}
  \left\{
    dj + 2^{m+\varsigma-\ell} k - \frac{d}{r} \left( \{ rj \}_{2^{m+\varsigma}} - 2^{m+\varsigma} \eta \right)
  \right\}_{2^{m+\varsigma}}.
\end{align*}

For all integers~$\Delta$ such that~$|\, \Delta \,| \le B_{\Delta}$ where $B_{\Delta} < 2^{\ell-1}$, we then compute the probability
\begin{align*}
  h
  \left(
  \frac{2\pi}{2^{m+\varsigma}}
  \left\{
    dj + 2^{m+\varsigma-\ell} k_{\eta, \Delta} - \frac{d}{r} \left( \{ rj \}_{2^{m+\varsigma}} - 2^{m+\varsigma} \eta \right)
  \right\}_{2^{m+\varsigma}}
  \right)
\end{align*}
of observing $k = k_{\eta, \Delta} = (k_{\eta, 0} + \Delta) \text{ mod } 2^\ell$ to construct a histogram in~$\Delta$.

We sample~$\Delta$, and hence $k = k_{\eta, \Delta}$, from the resulting histogram, and return~$(j, k)$ as the output from the simulator.
If either~$\eta$ and~$\alpha_r$, or~$\Delta$, fall outside the respective ranges of the histograms when sampled, we instead return a sampling error.
By increasing~$B_{\Delta}$, and~$B_{\eta}$ and/or~$\varsigma$, we may suppress the probability of such errors occurring in practice.

Note that it is permissible to first sample~$\eta$ and~$j$ from the distribution induced by~$F_{\eta}$ (or~$f_{\eta}$) as above, and to then sample~$k$ given~$\eta$ and~$j$ via the distribution induced by~$h$:
This is equivalent to first computing and observing~$j$, and to then computing and observing~$k$ given~$j$.

\subsubsection{Implementation remarks}
We have implemented the simulator described in Sect.~\ref{section:simulating-quantum-algorithm-with-failures} for the purpose of evaluating the efficiency of the two classical post-processing algorithms described in Sect.~\ref{section:post-processing}.
For one such implementation, see the Quaspy library for Python~\cite{quaspy}.

\subsubsection{On the relation to our earlier works}
The simulator described in Sect.~\ref{section:simulating-quantum-algorithm-with-failures} is similar to the simulators previously developed for Shor's and Seifert's order-finding algorithms~\cite[App.~A]{ekera-general}, for Ekerå's algorithm~\cite{ekera-general} for computing general discrete logarithms in groups of unknown order, and for Ekerå--Håstad's algorithm~\cite{ekera-hastad, ekera-pp} for computing short discrete logarithms.

\subsection{Sampling the probability distribution exactly without failures}
\label{section:simulating-quantum-algorithm-without-failures}
By Lem.~\ref{lemma:probability-distribution}, the probability of observing~$(j, k)$ summed over all $e \in [0, r) \cap \mathbb Z$ is
\begin{align*}
  \refeq{probability-sum}
  &=
  \frac{1}{r}
  \sum_{\mu \, = \, 0}^{r-1}
  |\, A_\mu(j) \, B_\mu(k) \,|^2,
\end{align*}
where $A_\mu(j)$ and $B_\mu(k)$ are as defined in~\refeq{Amu} and~\refeq{Bmu}, respectively, and where it holds that
\begin{align*}
  \sum_{j \, = \, 0}^{2^{m+\varsigma}-1} |\, A_\mu(j) \,|^2
  =
  \sum_{k \, = \, 0}^{2^{\ell}-1} |\, B_\mu(k) \,|^2
  =
  1,
\end{align*}
by Cl.~\ref{claim:exp-double-sum-to-one} in App.~\ref{appendix:supporting-lemmas-claims-proofs}, so~\refeq{probability-sum} is a uniform mixture of $r$~product distributions.
It follows that $(j, k)$ may be sampled according to~\refeq{probability-sum} by first sampling~$\mu$ uniformly at random from $[0, r) \cap \mathbb Z$, and by then independently sampling~$j$ and~$k$ according to~$|\, A_\mu(j) \,|^2$ and~$|\, B_\mu(k) \,|^2$, respectively.

Both $|\, A_\mu(j) \,|^2$ and $|\, B_\mu(k) \,|^2$ are Fejér kernels, which may be seen as follows:
For~$M$ a positive integer and ${0 \neq \delta_u \in [-\frac{1}{2}, \frac{1}{2})}$, let~$u$ be in any range of~$M$ consecutive integers, and let
\begin{align*}
  D_{M, \, \delta_u} (u)
  &=
  \frac{1}{M^2}
  \left|\, \sum_{a \, = \, 0}^{M-1} \e^{2 \pi \imag (u - \delta_u) a / M} \,\right|^2
  =
  \frac{1}{M^2}
  \frac{\sin^2(\pi (u - \delta_u))}{\sin^2(\pi (u - \delta_u) / M)}
\end{align*}
be the Fejér kernel, shifted by $\delta_u$.
It is $M$-periodic over $\mathbb Z$, and it sums to one over the prescribed range in~$u$ by Cl.~\ref{claim:exp-double-sum-to-one} in App.~\ref{appendix:supporting-lemmas-claims-proofs}.
If $\delta_u = 0$, then for completeness
\begin{align*}
  D_{M, \, 0} (u)
  &=
  \left\{
    \begin{array}{ll}
      1 & \text{if } u \equiv 0 \:\: (\text{mod } M), \\
      0 & \text{otherwise}.
    \end{array}
  \right.
\end{align*}

To see that $|\, A_\mu(j) \,|^2$ is a Fejér kernel, verify that
\begin{align*}
  |\, A_\mu(j) \,|^2
  &=
  \frac{1}{2^{2(m + \varsigma)}}
  \left|\,
  \sum_{a \, = \, 0}^{2^{m + \varsigma} - 1}
  \exp
  \left[
    2 \pi \imag
    \left(
      \frac{j}{2^{m + \varsigma}}
      -
      \frac{\mu}{r}
    \right)
    a
  \right]
  \,\right|^2 \\
  &=
  \frac{1}{2^{2(m + \varsigma)}}
  \frac{\sin^2(\pi 2^{m+\varsigma} (j/2^{m+\varsigma} - \mu/r))}{\sin^2(\pi (j/2^{m+\varsigma} - \mu/r))} \\
  &=
  \frac{1}{2^{2(m + \varsigma)}}
  \frac{\sin^2(\pi (j - 2^{m+\varsigma} \mu/r))}{\sin^2(\pi (j - 2^{m+\varsigma} \mu/r) / 2^{m+\varsigma})}
  =
  D_{2^{m+\varsigma}, \, \{ 2^{m + \varsigma} \mu/r \}_{1}}
  \left(
    j - \round{2^{m + \varsigma} \mu/r}
  \right),
\end{align*}
when $(j / 2^{m+\varsigma} - \mu/r) = (j - 2^{m+\varsigma} \mu/r) / 2^{m+\varsigma} = (j - \round{2^{m+\varsigma} \mu/r} - \{2^{m+\varsigma} \mu/r\}_1) / 2^{m+\varsigma} \not\in \mathbb Z$, which holds whenever $\{2^{m+\varsigma} \mu/r\}_1 \neq 0$.
Otherwise, if $\{2^{m+\varsigma} \mu/r\}_1 = 0$, then for completeness
\begin{align*}
  |\, A_\mu(j) \,|^2
  &=
  D_{2^{m+\varsigma}, \, 0}
  \left(
    j - \round{2^{m + \varsigma} \mu/r}
  \right)
  =
  \left\{
    \begin{array}{ll}
      1 & \text{if } j \equiv \round{2^{m+\varsigma} \mu/r} \:\: (\text{mod } 2^{m+\varsigma}), \\
      0 & \text{otherwise}.
    \end{array}
  \right.
\end{align*}

Analogously, to see that $|\, B_\mu(k) \,|^2$ is a Fejér kernel, verify that
\begin{align*}
  |\, B_\mu(k) \,|^2
  &=
  \frac{1}{2^{2 \ell}}
  \left|\,
  \sum_{b \, = \, 0}^{2^{\ell} - 1}
  \exp
  \left[
    2 \pi \imag
    \left(
      \frac{k}{2^{\ell}}
      +
      \frac{d \mu}{r}
    \right)
    b
  \right]
  \,\right|^2 \\
  &=
  \frac{1}{2^{2 \ell}}
  \frac{\sin^2(\pi 2^\ell (k/2^\ell + d\mu/r))}{\sin^2(\pi (k/2^\ell + d\mu/r))} \\
  &=
  \frac{1}{2^{2 \ell}}
  \frac{\sin^2(\pi (k + 2^\ell d\mu/r))}{\sin^2(\pi (k + 2^\ell d\mu/r) / 2^\ell)}
  =
  D_{2^\ell, \, \{ -2^\ell d\mu/r \}_{1}}
  \left(
    k - \round{-2^\ell d\mu/r}
  \right),
\end{align*}
when $(k/2^\ell + d\mu/r) = (k + 2^\ell d\mu/r)/2^\ell = (k - \round{-2^\ell d\mu/r} - \{ -2^\ell d\mu/r \}_1)/2^\ell \not\in \mathbb Z$, which holds whenever $\{ -2^\ell d\mu/r \}_1 \neq 0$.
Otherwise, if $\{ -2^\ell d\mu/r \}_1 = 0$, then for completeness
\begin{align*}
  |\, B_\mu(k) \,|^2
  &=
  D_{2^\ell, \, 0}
  \left(
    k - \round{-2^\ell d\mu/r}
  \right)
  =
  \left\{
    \begin{array}{ll}
      1 & \text{if } k \equiv \round{-2^{\ell} d\mu/r} \:\: (\text{mod } 2^{\ell}), \\
      0 & \text{otherwise}.
    \end{array}
  \right.
\end{align*}

Above, we have used that the sums are geometric, and that $v = \round{v} + \{v\}_1$ for any real~$v$ by our rounding convention; see Sect.~\ref{section:notation}.
(It does not always hold that $\round{-v} = -\round{v}$, however, as explained in Sect.~\ref{section:notation}, which is why we do not cancel the minus signs above.)

It follows that~$j$ may be sampled from $|\, A_\mu(j) \,|^2$ by sampling~$u$ from
$D_{2^{m+\varsigma}, \, \{ 2^{m + \varsigma} \mu/r \}_{1}} \left( u \right)$
and computing $j = (u + \round{2^{m + \varsigma} \mu/r}) \text{ mod } 2^{m + \varsigma}$, and analogously that~$k$ may be sampled from $|\, B_\mu(k) \,|^2$ by sampling~$u$ from
$D_{2^\ell, \, \{ -2^\ell d\mu/r \}_{1}} \left( u \right)$
and computing $k = (u + \round{-2^\ell d\mu/r}) \text{ mod } 2^\ell$.

The pair~$(j, k)$ thus sampled is associated with the peak index ${\eta = \{ \mu - z \}_r}$ by the definition of~$\eta$ in Sect.~\ref{section:arguments-angles}, where $z = \round{rj / 2^{m+\varsigma}}$.
Furthermore, since $\{ rj \}_{2^{m+\varsigma}} = rj - 2^{m+\varsigma} z$,
\begin{align*}
  k_{\eta, 0} = \refeq{k0} &= \round{\frac{1}{2^{m + \varsigma - \ell}} \left(-dj + \frac{d}{r} \left( \{ rj \}_{2^{m+\varsigma}} - 2^{m + \varsigma} \eta \right) \right)} \text{ mod } 2^\ell \\
              &= \round{-2^\ell d (\eta + z) / r} \text{ mod } 2^\ell = \round{-2^\ell d \mu / r} \text{ mod } 2^\ell,
\end{align*}
and $k = k_{\eta, \Delta} = (k_{\eta, 0} + \Delta) \text{ mod } 2^\ell$ where $\Delta \in [-2^{\ell-1}, 2^{\ell-1}) \cap \mathbb Z$ (see Sect.~\ref{section:notion-good-pair}), which implies that $\Delta = \{ u \}_{2^\ell}$ for~$u$ the value sampled from $D_{2^\ell, \, \{ -2^\ell d\mu/r \}_{1}} \left( u \right)$ when sampling~$k$.

\subsubsection{Sampling the Fejér kernel exactly}
For $u - \delta_u \in [-\frac{M}{2}, \frac{M}{2})$, where $0 \neq \delta_u \in [-\frac{1}{2}, \frac{1}{2})$, the Fejér kernel is
\begin{align*}
  D_{M, \delta_u}(u)
  =
  \underbrace{\frac{\sin^2(\pi(u - \delta_u))}{(\pi(u - \delta_u))^2}}_{S_{\delta_u}(u)}
  \underbrace{\frac{(\pi (u - \delta_u) / M)^2}{\sin^2({\pi(u-\delta_u)/M})}}_{T_{M, \delta_u}(u)},
\end{align*}
where $T_{M, \delta_u}(u) \in [1, \frac{\pi^2}{4}]$ since $\frac{2}{\pi} \, |\, x \,| \le |\, \sin x \,| \le |\, x \,|$ for $x = \pi (u - \delta_u) / M \in [-\frac{\pi}{2}, \frac{\pi}{2})$.

Furthermore, it holds that $\sum_{u \, = \, -\infty}^{\infty} S_{\delta_u}(u) = 1$ by Cl.~\ref{claim:h-tails-sums-to-one} in App.~\ref{appendix:supporting-lemmas-claims-proofs}, so $S_{\delta_u}(u)$ is a probability distribution over~$\mathbb Z$.
It follows that~$u$ may be sampled from $D_{M, \delta_u}(u)$ using rejection sampling, by first sampling~$u$ from $S_{\delta_u}(u)$ and rejecting if $u - \delta_u \not\in [-\frac{M}{2}, \frac{M}{2})$, and by then accepting with probability $\frac{4}{\pi^2} T_{M, \delta_u}(u)$ and otherwise rejecting and starting over.

By Cl.~\ref{claim:bound-psi} in App.~\ref{appendix:supporting-lemmas-claims-proofs}, the fact that $\sin^2(\pi(u - \delta_u)) = \sin^2(\pi \delta_u)$, and the fact that $\sum_{u \, = \, -\infty}^{\infty} S_{\delta_u}(u) = 1$, the cumulative distribution function~$C_{\delta_u}$ of~$S_{\delta_u}$ is given by
\begin{align*}
  C_{\delta_u}(u')
  =
  \sum_{u \, = \, -\infty}^{u'}
  S_{\delta_u}(u)
  =
  \left\{
    \begin{array}{ll}
      \sin^2(\pi \delta_u) / \pi^2 \cdot \psi'(\delta_u - u') & \text{if } u' < 0, \\
      1 - \sin^2(\pi \delta_u) / \pi^2 \cdot \psi'(u' + 1 - \delta_u) & \text{otherwise},
    \end{array}
  \right.
\end{align*}
allowing~$u$ to be sampled from~$S_{\delta_u}$, conditioned on $u - \delta_u \in [-\frac{M}{2}, \frac{M}{2})$, using bisection in $O(\log_2 M)$ evaluations of the trigamma function $\psi'$ with $O(\log_2 M + \log_2 r)$ bits of precision in each evaluation;\footnote{As~$\delta_u$ is a non-zero multiple of~$1/r$ in Alg.~\ref{alg:simulating-quantum-algorithm-without-failures}, it holds that $|\, \delta_u \,| \ge 1/r$, and hence that $|\, \sin(\pi \delta_u) \,| \ge 2 \, |\, \delta_u \,| \ge 2/r$, since $|\, \sin x \,| \ge \frac{2}{\pi} \, |\, x \,|$ for $x = \pi \delta_u \in [-\frac{\pi}{2}, \frac{\pi}{2})$.
Furthermore, it holds that $|\, u - \delta_u \,| \le \frac{M}{2}$ for all~$u$ such that $u - \delta_u \in [-\frac{M}{2}, \frac{M}{2})$, so ${S_{\delta_u}(u) = \sin^2(\pi \delta_u) / (\pi (u - \delta_u))^2 \ge 16 / (\pi^2 r^2 M^2)}$ for all such~$u$.
With $2 \log_2 M + 2 \log_2 r + \lambda$ bits of precision, the errors in the computed values of~$C_{\delta_u}$ hence perturb each such probability by a relative error of at most~$O(2^{-\lambda})$, where we may, e.g., let $\lambda = 64$.} see Alg.~\ref{alg:simulating-quantum-algorithm-without-failures} in Sect.~\ref{section:simulating-quantum-algorithm-without-failures-pseudocode} below for the complete sampling procedure.
Note that the above definition of $C_{\delta_u}$ ensures that~$\psi'$ is only evaluated for positive arguments.

\subsubsection{The complete sampling procedure in pseudocode}
\label{section:simulating-quantum-algorithm-without-failures-pseudocode}
The complete sampling procedure is described in pseudocode in Alg.~\ref{alg:simulating-quantum-algorithm-without-failures} below:
\begin{breakablealgorithm}
  \caption{Samples~$(j, k)$ (or~$j$ only if $\ell = 0$) from the probability distribution induced by the quantum algorithm in Sect.~\ref{section:quantum-algorithm}, which induces the state~\refeq{psi} and reads out~$(j, k)$ (or~$j$ only).
  \\
  {\bfseries \emph{Inputs:}} $m$, $\varsigma$, $\ell$, $d$, $r$
  $\:$
  {\bfseries \emph{Returns:}} $(j, k)$ and optionally $\eta$ and $\Delta$ (or $j$ and optionally $\eta$ if $\ell = 0$)}
  \label{alg:simulating-quantum-algorithm-without-failures}
  \begin{pseudocode}
    \item Let \textsc{SampleFejérKernel}($M$, $\delta_u$) be the function:
    \begin{pseudocode}
      \item If $\delta_u = 0$:
      \begin{pseudocode}
        \item Return $0$.
      \end{pseudocode}

      \item Perform repeated rejection sampling: \label{alg:sample-j-k-step:restart}
      \begin{pseudocode}
        \item  Let $u_0 \leftarrow \ceil{\delta_u - M/2}$ and $u_1 \leftarrow u_0 + M - 1$.

        \emph{Note: It holds that $u - \delta_u \in [-M/2, M/2)$ if and only if $u \in [u_0, u_1] \cap \mathbb Z$.}

        \item Sample~$\rho$ uniformly at random from $[0, 1)$.

        \item If $\rho < C_{\delta_u}(u_0 - 1)$ or $\rho \ge C_{\delta_u}(u_1)$:

        \emph{Note: $u$ such that $C_{\delta_u}(u - 1) \le \rho < C_{\delta_u}(u)$ is outside $[u_0, u_1] \cap \mathbb Z$.}

        \begin{pseudocode}
          \item Go to step~\ref{alg:sample-j-k-step:restart}.

          \emph{Note: Reject and start over.}
        \end{pseudocode}

        \item While $u_0 < u_1$:

        \emph{Note: It holds that $C_{\delta_u}(u_0 - 1) \le \rho < C_{\delta_u}(u_1)$ throughout this loop.}

        \begin{pseudocode}
          \item Let $u' \leftarrow \floor{(u_0 + u_1) / 2}$.
          \item If $\rho < C_{\delta_u}(u')$, let $u_1 \leftarrow u'$, otherwise let $u_0 \leftarrow u' + 1$.
        \end{pseudocode}

        \item Sample~$\rho$ uniformly at random from $[0, 1)$.

        \item If $\rho < 4 / \pi^2 \cdot T_{M, \delta_u}(u_0)$:
        \begin{pseudocode}
          \item Return~$u_0$.
        \end{pseudocode}

        \item Go to step~\ref{alg:sample-j-k-step:restart}.

        \emph{Note: Reject and start over.}
      \end{pseudocode}
    \end{pseudocode}

    \item Sample~$\mu$ uniformly at random from $[0, r) \cap \mathbb Z$. \label{alg:sample-j-k-step:sample-mu}

    \item Let $\delta_u \leftarrow \{ 2^{m+\varsigma} \mu/r \}_1$ and $M \leftarrow 2^{m+\varsigma}$.

    Let $u \leftarrow \textsc{SampleFejérKernel}(M, \, \delta_u)$ to sample~$u$ from $D_{M, \delta_u}(u)$.

    Let $j \leftarrow (u + \round{2^{m+\varsigma} \mu/r}) \text{ mod } 2^{m+\varsigma}$ and $\eta \leftarrow \{ \mu - z \}_r$ where $z = \round{rj / 2^{m+\varsigma}}$.

    \item If $\ell = 0$:
    \begin{pseudocode}
      \item Return~$j$ and optionally~$\eta$.
    \end{pseudocode}

    \item Let $\delta_u \leftarrow \{ -2^{\ell} d\mu/r \}_1$ and $M \leftarrow 2^{\ell}$.

    Let $u \leftarrow \textsc{SampleFejérKernel}(M, \, \delta_u)$ to sample~$u$ from $D_{M, \delta_u}(u)$.

    Let $k \leftarrow (u + \round{-2^{\ell} d \mu/r}) \text{ mod } 2^{\ell}$ and $\Delta \leftarrow \{ u \}_{2^\ell}$.

    \item Return~$(j, k)$ and optionally~$\eta$ and~$\Delta$.
  \end{pseudocode}
\end{breakablealgorithm}

\subsubsection{On generalizations to our earlier works}
\label{section:simulating-quantum-algorithm-without-failures-generalizations}
The procedure in Alg.~\ref{alg:simulating-quantum-algorithm-without-failures} may be used to sample the distribution induced not only by the modified version of Shor's algorithm for the DLP that we primarily consider in this paper, which computes the logarithm~$d$ given the order~$r$, but also the variation in~\cite{ekera-general} which computes both~$d$ and~$r$, by letting~$m$ be an upper bound on the bit length of~$r$ and $\varsigma = \ell \sim m / s$.

Furthermore, said procedure may be used to sample the distribution induced by Ekerå--Håstad's algorithm~\cite{ekera-hastad, ekera-pp, ekera-success-short-dlp} for the short DLP, which computes short~$d$ when~$r$ is unknown, by letting~$m$ be an upper bound on the bit length of~$d$ and $\varsigma = \ell \sim m /s$.

Finally, said procedure may be used to sample the distribution induced by Shor's order-finding algorithm~\cite{shor1994, shor1997} as described and analyzed in~\cite{ekera-success-order-finding} (including Seifert's variation~\cite{seifert} thereof, as described and analyzed in~\cite[App.~A]{ekera-general}), which computes~$r$, by letting~$m$ be an upper bound on the bit length of~$r$, $\varsigma \sim m / s$ and $\ell = 0$ (in which case Alg.~\ref{alg:simulating-quantum-algorithm-without-failures} returns only~$j$).

Note that there is no need (and it makes no sense) to compute and return~$\eta$ and/or~$\Delta$ for some of the above quantum algorithms.
Returning~$\eta$ and/or~$\Delta$ is therefore optional in Alg.~\ref{alg:simulating-quantum-algorithm-without-failures}.

\section{Classical post-processing}
\label{section:post-processing}
When not making tradeoffs (i.e.,\ when~$s = 1$), we may individually post-process pairs~$(j, k)$ yielded by the quantum algorithm in the manner originally proposed by Shor~\cite{shor1994}:

Shor's original post-processing algorithm recovers~$d$ given~$r$ from any pair~$(j, k)$ for which the offset $\Delta = \{k - k_{\eta, 0}\}_{2^\ell}$ is close to zero when $\eta = 0$, provided that~$z$ is invertible modulo~$r$.
Specifically, said algorithm\footnote{When modified to account for the uniform initialization of the two control registers, and to account for the registers being of length $m+\varsigma$ and $\ell$~qubits, respectively, but not to account for searching.} requires $t = \round{r(\Delta - \delta_0) / 2^{\ell}} = 0$ to be successful.
In Sect.~\ref{section:post-processing-shor} below, we describe a slightly modified version of Shor's post-processing algorithm.
The modified algorithm performs two searches, over~$t$ and~$\eta$ respectively, to recover~$d$ from any $B_{\eta}$-$B_{\Delta}$-good pair~$(j, k)$, provided that $z + \eta$ is invertible modulo~$r$; see Sect.~\ref{section:post-processing-shor-invertibility}.

To handle tradeoff factors $s \ge 1$, we adapt the lattice-based post-processing algorithms from our earlier works~\cite{ekera-modifying, ekera-hastad, ekera-pp, ekera-general}, so as to recover~$d$ given~$r$ from a set of~$n$ pairs $\{ (j_1, k_1), \, \ldots, \, (j_n, k_n) \}$ yielded by~$n$ independent runs of the quantum algorithm.
This lattice-based post-processing algorithm is described in Sect.~\ref{section:post-processing-lattice}.

\subsection{Solving individual runs by modifying Shor's post-processing}
\label{section:post-processing-shor}
Suppose that the quantum algorithm in Sect.~\ref{section:quantum-algorithm} yields a $B_{\eta}$-$B_{\Delta}$-good pair~$(j, k)$:

For the arguments~$\alpha_d$, $\alpha_r$ and angle~$\phi_{\eta}$ associated with this pair, it then holds that
\begin{align*}
  \frac{2^{m+\varsigma}}{2 \pi} \phi_{\eta} = \left\{ \alpha_d - \frac{d}{r} \left( \alpha_r - 2^{m+\varsigma} \eta \right) \right\}_{2^{m+\varsigma}} &= 2^{m+\varsigma-\ell} (\Delta - \delta_{\eta}) \\
  \Rightarrow
  \left\{ \{ dj + 2^{m+\varsigma-\ell} k \}_{2^{m+\varsigma}} - \frac{d}{r} \left( \{ rj \}_{2^{m+\varsigma}} - 2^{m+\varsigma} \eta \right) \right\}_{2^{m+\varsigma}} &= 2^{m+\varsigma-\ell} (\Delta - \delta_{\eta}) \\
  \Rightarrow
  dj + 2^{m+\varsigma-\ell} k - \frac{d}{r} \left( \{ rj \}_{2^{m+\varsigma}} - 2^{m+\varsigma} \eta \right) + 2^{m+\varsigma} \nu &= 2^{m+\varsigma-\ell} (\Delta - \delta_{\eta}) \\
  \Rightarrow
  rdj + 2^{m+\varsigma-\ell} rk - \{ rj \}_{2^{m+\varsigma}} d + 2^{m+\varsigma} \eta d + 2^{m+\varsigma} \nu r &= 2^{m+\varsigma-\ell} r (\Delta - \delta_{\eta}) \\
  \Rightarrow
  d (rj - \{ rj \}_{2^{m+\varsigma}}) + 2^{m+\varsigma-\ell} rk + 2^{m+\varsigma} \eta d + 2^{m+\varsigma} \nu r &= 2^{m+\varsigma-\ell} r (\Delta - \delta_{\eta}) \\
  \Rightarrow
  d \underbrace{\frac{rj - \{ rj \}_{2^{m+\varsigma}}}{2^{m+\varsigma}}}_{\text{known } z \, = \, \round{\frac{rj}{2^{m+\varsigma}}} \, \in \, \mathbb Z} + \frac{rk}{2^{\ell}} + \eta d + \nu r &= \frac{r (\Delta - \delta_{\eta})}{2^{\ell}} \\
  \Rightarrow
  d (z + \eta) + \round{\frac{rk}{2^{\ell}}} + \nu r &= \underbrace{\round{\frac{r (\Delta - \delta_{\eta})}{2^{\ell}}}}_{\text{unknown small } t} \\
  \Rightarrow
  d (z + \eta) + \round{\frac{rk}{2^{\ell}}} &\equiv t \quad (\text{mod } r)
\end{align*}
for some~$\nu \in \mathbb Z$, which implies that we may solve for~$d$ if $z + \eta$ is invertible modulo~$r$.
Specifically
\begin{align}
  d &\equiv
  \left( t - \round{\frac{rk}{2^{\ell}}} \right) (z + \eta)^{-1}
  \quad
  (\text{mod } r)
  \label{eq:solve-d}
\end{align}
where $|\, t \,| \le \round{r (B_{\Delta} + 1/2) / 2^{\ell}}$ since $|\, \Delta - \delta_\eta \,| \le B_{\Delta} + 1/2$, so we need to search at most $2 \cdot \round{r (B_{\Delta} + 1/2) / 2^{\ell}} + 1$ values of~$t$ for each of the at most $2 B_{\eta} + 1$ values of~$\eta$ to recover~$d$ from~$(j, k)$ given~$r$.
Suppose that $\ell = m - \ellint$ for some fixed~$\ellint \in \mathbb Z$ such that~$|\, \ellint \,|$ is small.
Then we need to search at most
\begin{align*}
  2 \cdot \underbrace{\round{r (B_{\Delta} + 1/2) / 2^{m-\ellint}}}_{= \, B_t} + 1
  \le
  2 \cdot \round{2^{\ellint} (B_{\Delta} + 1/2)} + 1
\end{align*}
values of~$t$ for each of the $2 B_{\eta} + 1$ values of~$\eta$, which is feasible if~$B_{\eta}$ and~$B_{\Delta}$ are sufficiently small.
A minimum requirement for the search to be efficient is that $B_{\eta}, B_{\Delta} \in O(\text{poly}(m))$.

If~$\ellint$ is positive, work in the quantum algorithm is traded for work in the classical post-processing, and vice versa.
A reasonable choice is to fix $\ellint = 0$, or to fix some positive~$\ellint$.

\subsubsection{On searching efficiently}
As explained in~\cite{ekera-success-short-dlp}, meet-in-the-middle and random-walk techniques may be leveraged to speed up the limited searches in the classical post-processing.

In the case of the above searches over~$\eta$ and~$t$, this may for instance be accomplished as follows:
For the correct~$\eta \in [-B_\eta, B_\eta] \cap \mathbb Z$, use that
\begin{align*}
  [d] \, g
  =
  \left[ \left( t - \round{\frac{rk}{2^\ell}} \right) \big( (z + \eta)^{-1} \text{ mod } r \big) \right]
  g
  =
  x
  \quad
  \Rightarrow
  \quad
  [\, t \,] \, g
  &=
  \left[ \round{\frac{rk}{2^\ell}} \right] \, g
  \odot
  [z + \eta] \, x
  =
  x'
\end{align*}
where $t \in [-B_t, B_t] \cap \mathbb Z$ to see that we need to solve the short DLP $\log_g x'$ in $\langle g \rangle$ of known order~$r$ to find~$t$ and hence~$d$.
This may, e.g., be accomplished in $O(\sqrt{B_t})$ group operations by using Shanks' algorithm~\cite{shanks} (which is deterministic), or Pollard's $\lambda$-algorithm~\cite{pollard-rho-lambda, oorschot-wiener} (which is probabilistic, but features a much smaller space footprint than Shanks' algorithm).\footnote{Pollard's $\lambda$-algorithm~\cite{pollard-rho-lambda} only requires $O(1)$ group elements to be stored (when not parallelized as described by van Oorschot and Wiener~\cite{oorschot-wiener}; such implementations require a bit more space), whereas Shanks' algorithm~\cite{shanks} requires $O(\sqrt{B_t})$ exponents to be stored in a lookup table indexed by group elements.}

To exemplify, suppose that we target a $\ge 99\%$ success probability.
According to Tbl.~\ref{table:probability-strict-Bn-0}, an option is then to pick $\varsigma = 7$, $B_\eta = 0$ and $B_\Delta = 100$, which yields a complexity of roughly
\begin{align}
  2 \cdot (2 B_\eta + 1)
  \cdot
  \sqrt{2 \cdot B_t + 1}
  \le
  2 \cdot \sqrt{2 \cdot \round{2^{\ellint} (100 + 1/2)} + 1}
  \label{eq:sqrt-complexity}
\end{align}
group operations in the post-processing.\footnote{For Pollard's $\lambda$-algorithm, the constant~$2$ typically needs to be increased slightly to reach the~$99\%$ success probability targeted, but here we are only interested in a rough ballpark complexity estimate.}
Suppose that we pick $\ellint = 50 + \varsigma = 57$.
Then, by~\refeq{sqrt-complexity}, the post-processing complexity is roughly $\approx 2^{33}$ group operations, which is clearly feasible in practice, and $m + \varsigma + \ell = 2m - 50$ group operations are then performed quantumly.
It is feasible to further increase $\ellint$, or to target a greater success probability, or both, up to some limit that depends on the computational resources available.\footnote{As a point of reference, in 2020, Pons~\cite{pons} reported having solved a 114-bit short DLP in the \texttt{secp256k1} 256-bit curve in 13~days using 256 GPUs of the model Tesla~V100.
Based on this, it should be feasible to pick $\varsigma = 7$, $B_\eta = 0$, $B_\Delta = 100$ and $\ellint = 100 + \varsigma = 107$ for said curve when targeting a $99\%$ single-run success probability.
Then, $m + \varsigma + \ell = 2m - 100 = 412$ group operations are performed quantumly in the run (which yields a relative reduction of $\approx 20\%$ compared to performing $2m$~group operations quantumly).}

It should be noted in this context that when implementing Pollard's $\lambda$-algorithm in parallel as described by van Oorschot and Wiener~\cite{oorschot-wiener}, a database of distinguished points is constructed.
The short discrete logarithm ${t = \log_g x'}$ is found when a collision between two distinguished points, produced by a wild walk and a tame walk, respectively, is detected, where tame walks depend on~$g$ and wild walks on~$g$ and~$x'$.
Distinguished points produced by tame walks may be pre-computed.
If multiple ${t_i = \log_g x'_i}$ are to be computed with respect to the same generator~$g$ and interval for~$t_i$, then the same pre-computed database may be used for all~$t_i$, enabling the pre-computation cost to be amortized across all~$t_i$.\footnote{Furthermore, once~$t_i$ has been computed, distinguished points produced by wild walks involving~$x'_i$ may be cheaply converted into points produced by tame walks and added to the database.
This implies that it becomes progressively cheaper to compute logarithms, assuming there is space to store the database.} This is advantageous in practice.

Yet another option is of course to solve the short DLP $\log_g x'$ quantumly, using Ekerå--Håstad's algorithm~\cite{ekera-hastad, ekera-pp, ekera-success-short-dlp}, effectively resulting in a two-stage quantum algorithm for the DLP.
In practice, it is however likely better to make two runs of the quantum algorithm for the DLP, and to post-process these runs jointly using lattice-based techniques; see Sect.~\ref{section:post-processing-lattice}.

\subsubsection{On the probability of \texorpdfstring{$z + \eta$}{z + eta} being invertible modulo~\texorpdfstring{$r$}{r}}
\label{section:post-processing-shor-invertibility}
The classical post-processing requires $z + \eta$ to be invertible modulo~$r$ for the correct~$\eta$.
As the probability of this requirement being met is not accounted for in the lower bound in Thm.~\ref{theorem:lower-bound}, we account for it in the following corollary:

\begin{theoremcorollary}
  Let~$\rho$ be the lower bound in Thm.~\ref{theorem:lower-bound} on the probability of observing a $B_{\eta}$-$B_{\Delta}$-good pair in a run of the quantum algorithm in Sect.~\ref{section:quantum-algorithm}.
  Then, the probability of recovering~$d$ given~$r$ from the pair~$(j, k)$ observed in such a run, with the classical post-processing in Sect.~\ref{section:post-processing-shor}, which searches all $\eta, t \in \mathbb Z$ such that $|\, \eta \,| \le B_{\eta}$ and $|\, t \,| \le B_t = \round{r (B_{\Delta} + 1/2) / 2^{\ell}}$, is at least
  \begin{align*}
    \rho - (r - \phi(r))/r
  \end{align*}
  for~$\phi$ Euler's totient function.
  In particular, it is at least $\rho - 1/r$ if~$r$ is prime, and at least $\rho - \log_2(r) / (B_r \log_2(B_r))$ if~$r$ has no prime factor less than~$B_r$ for $B_r \ge 2$ a bound.
  \label{corollary:invertible-correct-eta}
\end{theoremcorollary}
\begin{proof}
  First recall that by Lem.~\ref{lemma:probability-distribution}, the probability of observing~$(j, k)$ and~$y = [e] \, g$, summed over all $e \in [0, r) \cap \mathbb Z$, is
  \begin{align*}
    \frac{1}{r}
    \sum_{\mu \, = \, 0}^{r-1}
    |\, A_\mu(j) \, B_\mu(k) \,|^2
    =
    \sum_{\eta \, = \, -\floor{r/2}}^{\ceil{r/2} - 1}
    F_{\eta}(\theta_r)
    \,
    h(\phi_{\eta}),
  \end{align*}
  where the term for~$\eta$ on the right-hand side is equal to the term for $\mu = (z + \eta) \text{ mod } r$ on the left-hand side.
  Hence $\gcd(z + \eta, r) = \gcd(\mu, r)$.

  Furthermore, as explained in Sect.~\ref{section:simulating-quantum-algorithm-without-failures}, for each $\mu \in [0, r) \cap \mathbb Z$, it holds that
  \begin{align*}
    \sum_{j \, = \, 0}^{2^{m+\varsigma}-1}
    |\, A_\mu(j) \,|^2
    =
    \sum_{k \, = \, 0}^{2^\ell - 1}
    |\, B_\mu(k) \,|^2
    =
    1.
  \end{align*}

  Hence, when summed over all~$(j, k)$, the probability of sampling a given $\mu \in [0, r) \cap \mathbb Z$ (in the sense of the pair $(j, k)$ observed arising from the probability mass stemming from the term for $\mu = (z + \eta) \text{ mod } r$; see also step~\ref{alg:sample-j-k-step:sample-mu} in Alg.~\ref{alg:simulating-quantum-algorithm-without-failures}) is~$1/r$.
  It follows that the probability of sampling~$\mu$ such that $\gcd(\mu, r) \neq 1$, or, equivalently, such that $\gcd(z + \eta, r) \neq 1$, is~$(r - \phi(r))/r$.

  Let~$\mathrm{B}$ denote this event, and let~$\mathrm{A}$ denote the event of~$(j, k)$ being a $B_\eta$-$B_\Delta$-good pair with respect to the~$\eta$ associated with~$\mu$, i.e., of $|\, \eta \,| \le B_\eta$ and $k = k_{\eta, \Delta}$ for some $\Delta \in \offsets(B_{\Delta})$, where $\eta = \{ \mu - z \}_r$.
  By the above correspondence between the terms, and the derivation of~\refeq{sum-Pn} in Sect.~\ref{section:probability-good-pair}, $\Pr(\mathrm{A})$ is at least~\refeq{sum-Pn}, and hence at least~$\rho$ by Thm.~\ref{theorem:lower-bound}.
  The events~$\mathrm{A}$ and~$\mathrm{B}$ are not independent, however, since both depend on~$\mu$, so we use that
  \begin{align*}
    \Pr(\mathrm{A} \cap \overline{\mathrm{B}})
    =
    \Pr(\mathrm{A}) - \Pr(\mathrm{A} \cap \mathrm{B})
    \ge
    \Pr(\mathrm{A}) - \Pr(\mathrm{B})
    \ge
    \rho
    -
    (
      r - \phi(r)
    ) / r.
  \end{align*}

  If the event $\mathrm{A} \cap \overline{\mathrm{B}}$ occurs, then~$(j, k)$ is a $B_\eta$-$B_\Delta$-good pair with respect to~$\eta$ such that $\gcd(z + \eta, r) = 1$.
  The post-processing in Sect.~\ref{section:post-processing-shor} then recovers~$d$ given~$r$ via~\refeq{solve-d}, as $|\, \eta \,| \le B_{\eta}$, and as $|\, t \,| \le B_t$ for $t = \round{r (\Delta - \delta_{\eta}) / 2^{\ell}}$ (see Sect.~\ref{section:post-processing-shor}), so the searches find~$\eta$ and~$t$.
  As whether~$d$ is recovered depends only on~$(j, k)$, the probability of recovering~$d$ is at least $\Pr(\mathrm{A} \cap \overline{\mathrm{B}})$.

  Lastly, $(r - \phi(r))/r = 1/r$ if~$r$ is prime, and $(r - \phi(r)) / r \le \log_2(r) / (B_r \log_2(B_r))$ by Cl.~\ref{claim:r-minus-phi-r-bound} in App.~\ref{appendix:invertibility-classical-post-processing} if~$r$ has no prime factor less than~$B_r$, and so the corollary follows.
\end{proof}

Recall that~$r$ may be assumed to have no small prime factors without loss of generality (see Sect.~\ref{section:assumptions-reductions}), since we can efficiently find all factors less than~$B_r$ when $B_r = O(\text{poly}(m))$ and perform Pohlig--Hellman decomposition.
For this reason, $r$ is typically prime in cryptographic schemes that are based on the computational intractability of the DLP.

Note that the resulting post-processing algorithm is identical to that originally described by Shor~\cite{shor1994}, except for the searches over~$t$, $\eta$ and the partial factorization of~$r$.

\subsubsection{On the asymptotic success probability}
\label{section:post-processing-shor-asymptotic-limit}
As $r \rightarrow \infty$, the probability of recovering~$d$ given~$r$ in a single run of the quantum algorithm tends to one, if we first perform Pohlig--Hellman decomposition for all factors of~$r$ less than $B_r = \Omega(m)$, pick $\ell \sim m$, and perform two searches in the post-processing:
\begin{theorem}
  The probability of recovering~$d$ given~$r$ in a single run of the quantum algorithm in Sect.~\ref{section:quantum-algorithm}, with~$m$ the bit length of~$r$, $\varsigma$ a non-negative integer, and $\ell = m - \ellint$ for some fixed $\ellint \in \mathbb Z$ such that~$|\, \ellint \,|$ is small, tends to one in the limit as $r \rightarrow \infty$, when
  \begin{romanlist}
    \item $r$ has no prime factors $< B_r = \Omega(m)$, and
    \item using the post-processing in Sect.~\ref{section:post-processing-shor} and searching all $\eta, t \in \mathbb Z$ such that $|\, \eta \,| \le B_{\eta}$ and $|\, t \,| \le \round{2^{\ellint} (B_{\Delta} + 1/2)}$, where $B_{\eta} = \omega_m(1) < \ceil{r/2}$ and $B_{\Delta} = \omega_m(1) < 2^{\ell - 1}$.
  \end{romanlist}
  \label{theorem:solve-in-single-run}
\end{theorem}
\begin{proof}
  By Cor.~\ref{corollary:invertible-correct-eta}, the probability of recovering~$d$ given~$r$ is at least $\rho - \log_2(r) / (B_r \log_2(B_r))$, for~$\rho$ the lower bound in Thm.~\ref{theorem:lower-bound}, as~$r$ has no prime factor less than~$B_r$.
  As $B_r = \Omega(m)$ exceeds two asymptotically, $r$ is odd asymptotically, and $\kappa_r = 0 \le \ell$.
  Hence, Thm.~\ref{theorem:lower-bound} applies.
  As in the proof of Cor.~\ref{corollary:lower-bound-asymptotics}, $\rho$ tends to one in the limit in the theorem, whereas $\log_2(r) / (B_r \log_2(B_r)) < m / (B_r \log_2(B_r))$ tends to zero when $B_r = \Omega(m)$, and so the theorem follows.
\end{proof}

\subsubsection{On simulating the complete algorithm}
We have implemented the post-processing algorithm and tested it using the simulator in Sect.~\ref{section:simulating-quantum-algorithm-with-failures}.
For one such implementation, see the Quaspy library for Python~\cite{quaspy}.

Unsurprisingly, running the post-processing algorithm with respect to simulated quantum algorithm outputs recovers~$d$, provided that no sampling errors occur, and that sufficiently large intervals in~$t$ and~$\eta$ are searched as functions of~$B_{\Delta}$ and~$\ellint$, and of~$B_{\eta}$ and~$\varsigma$.

As~$B_{\Delta}$, and~$B_{\eta}$ and/or~$\varsigma$, increase, the probability of sampling errors occurring tends to zero; see the lower bound in Thm.~\ref{theorem:lower-bound}, the asymptotics in Cor.~\ref{corollary:lower-bound-asymptotics}, and the tables in App.~\ref{appendix:tabulated-results}.

\subsubsection{On the success probability of Shor's original algorithm}
Recall that it is required that $t = \round{r (\Delta - \delta_{\eta}) / 2^{\ell}} = 0$ and $\eta = 0$ to solve without searching as in Shor's original post-processing algorithm; see Sect.~\ref{section:post-processing-shor}.

For $\varsigma = 0$ and $\ell = m$, as in Shor's original algorithm\footnote{Modified to induce a uniform superposition over~$[0, 2^m) \cap \mathbb Z$ instead of~$[0, r) \cap \mathbb Z$ for~$m$ the bit length of~$r$.}, and maximal $r = 2^{m} - 1$, we have that $t = 0$ when $\Delta - \delta_0 \in [-\frac{1}{2}, \frac{1}{2})$ in the limit as $m \rightarrow \infty$.
This implies that Shor's original classical post-processing algorithm succeeds if~$(j, k)$ is a $B_{\eta}$-$B_{\Delta}$-good pair for $B_{\eta} = B_{\Delta} = 0$, and~$z$ is invertible modulo~$r$.
Such a $B_{\eta}$-$B_{\Delta}$-good pair is yielded with heuristically expected probability approximately~$0.5986$ by Tbl.~\ref{table:probability-Bn-0}.

At the other extreme, for $r = 2^{m-1} + 1$ (where we add one to avoid making~$r$ a perfect power of two), we have that $t = 0$ when $v = \Delta - \delta_0 \in [-1, 1)$, yielding the expected probability
\begin{align*}
  2^{\kappa_r}
  &\int_{-2^{m+\varsigma-\kappa_r-1}}^{2^{m+\varsigma-\kappa_r-1}}
  F_0\left( \frac{2\pi}{2^{m+\varsigma}} \, 2^{\kappa_r} \alpha'_r \right)
  \,\mathrm{d}\alpha'_r \,
  \int_{-1}^{1}
  h\left( \frac{2\pi}{2^\ell} v \right)
  \,\mathrm{d}v
  \approx
  0.8151
\end{align*}
by the heuristic~\refeq{Pn-int}, where we have evaluated the expression numerically for $m = 128$ with Mathematica.\footnote{In the initial pre-print of this paper, for $r = 2^{m-1} + 1$, we accidentally took $v = \Delta - \delta_0 \in [-\frac{1}{2}, \frac{1}{2})$ as opposed to $v \in [-1, 1)$, leading us to estimate the success probability to approximately~$70\%$ as opposed to~$82\%$.}
The same result is obtained for greater~$m$.

By the same argument as in the proof of Cor.~\ref{corollary:invertible-correct-eta}, requiring~$z$ to be invertible modulo~$r$ reduces the success probability by at most $(r - \phi(r))/r$ compared to the above probabilities.
As follows from Cor.~\ref{corollary:invertible-correct-eta}, and in particular from Cl.~\ref{claim:r-minus-phi-r-bound} in App.~\ref{appendix:invertibility-classical-post-processing}, this reduction is negligible if~$r$ has no small prime factors, which may be assumed without loss of generality; see Sect.~\ref{section:assumptions-reductions}.

We conclude from the above that Shor's original algorithm, modified to allow the semi-classical QFT to be used with control qubit recycling, achieves an expected success probability of approximately~$60$--$82\%$ according to our heuristic.

\subsection{Solving using lattice-based post-processing}
\label{section:post-processing-lattice}
Suppose that $s \ge 1$, $\ell \sim m/s$, and that $n$ independent runs of the quantum algorithm are performed, yielding pairs $(j_i, k_i)$ for $i \in [1, n] \cap \mathbb Z$ with associated arguments
\begin{align*}
  \alpha_{r,i} = \{ rj_i \}_{2^{m+\varsigma}}
  \quad \text{ and } \quad
  \alpha_{d,i} = \{ dj_i + 2^{m+\varsigma-\ell} k_i \}_{2^{m+\varsigma}}.
\end{align*}

Furthermore, let~$\eta_i$ for $i \in [1, n] \cap \mathbb Z$ be the unknown peak index associated with the pair~$(j_i, k_i)$, i.e., the index~$\eta$ of the term on the right-hand side of~\refeq{probability-sum-lemma-equated} in Lem.~\ref{lemma:probability-distribution} that gives rise to~$(j_i, k_i)$, and let
\begin{align}
  \phi_{\eta_i, i}
  &=
  \frac{2\pi}{2^{m+\varsigma}}
  \left\{
    \alpha_{d,i} - \frac{d}{r} \left( \alpha_{r,i} - 2^{m+\varsigma} \eta_i \right)
  \right\}_{2^{m+\varsigma}}.
\end{align}

As previously explained, if~$\varsigma$ is selected sufficiently large, then $|\, \eta_i \,| \le B_\eta = 0$ with high probability for all $i \in [1, n] \cap \mathbb Z$.
(In fact, the probability of $\eta_i \neq 0$ is exponentially suppressed in~$\varsigma$, as shown in Sect.~\ref{section:lower-bounding-sum-Pn}, so it is fairly easy to select $\varsigma$ sufficiently large.)
In practice, when $n$ is large, it is necessary to select~$\varsigma$ so that $\eta_1 = \cdots = \eta_n = 0$ with high probability, allowing $B_\eta = 0$ to be used:
The post-processing depends on $\eta_1, \, \ldots, \, \eta_n$ being known, as explained below, and the cost of exhaustively running it for up to $(2 B_\eta + 1)^n$ possible combinations of $\eta_1, \, \ldots, \, \eta_n$ grows exponentially in $n$ if $B_\eta > 0$.

Let~$\mathcal L^\tau$ be the lattice given by integer combinations of the rows $\mathbf{a}_0, \, \ldots, \, \mathbf{a}_n$ of
\begin{align}
  \label{eq:A}
  \mathbf{A}
  =
  \left(
  \begin{array}{c}
    \mathbf{a}_0 \\
    \mathbf{a}_1 \\
    \vdots \\
    \mathbf{a}_{n}
  \end{array}
  \right)
  =
  \left(
  \begin{array}{cccc}
    j_{1} + (2^{m+\varsigma} \eta_1 - \alpha_{r,1} ) / r & \cdots &
    j_{n} + (2^{m+\varsigma} \eta_n - \alpha_{r,n}) / r & 2^{\varsigma-\ell+\tau} \\
    2^{m+\varsigma} & \cdots & 0 & 0 \\
    & \ddots & & \vdots \\
    0 & \cdots & 2^{m+\varsigma} & 0
  \end{array}
  \right)
\end{align}
for~$2^{\varsigma-\ell+\tau}$ a scaling factor, with $\tau$~an integer parameter elaborated on below, and enumerate the vectors in~$\mathcal L^\tau$ closest to
\begin{align*}
  \mathbf{v} = (-2^{m+\varsigma-\ell} k_1, \, \ldots, \, -2^{m+\varsigma-\ell} k_n, \, 0) \in \mathbb Z^{n+1}
\end{align*}
with the aim of finding
\begin{align*}
  \mathbf{u} = d \cdot \mathbf{a}_0 + \sum_{i \, = \, 1}^{n} \nu_i \cdot \mathbf{a}_i \in \mathcal L^\tau
  \quad \text{ for some } \quad
  \nu_1, \, \ldots, \, \nu_n \in \mathbb Z.
\end{align*}

The vector~$\mathbf{u}$ yields~$d$, as its last component is $2^{\varsigma-\ell+\tau} d$.
Furthermore, $\mathbf{u}$ is within distance
\begin{align*}
  R
  &=
  \left|\, \mathbf{u} - \mathbf{v} \,\right|
  =
  \left|\, (\gamma_1, \, \ldots, \, \gamma_n, \, 2^{\varsigma-\ell+\tau} d) \,\right|
\end{align*}
of~$\mathbf{v}$, where, for some appropriately selected $\nu_i \in \mathbb Z$ for $i \in [1, n] \cap \mathbb Z$, each
\begin{align*}
  \gamma_i
  &=
  d j_i + 2^{m+\varsigma-\ell} k_i - \frac{d}{r} \left( \alpha_{r,i} - 2^{m+\varsigma} \eta_i \right) + 2^{m+\varsigma} \nu_i \\
  &=
  \left\{
    d j_i + 2^{m+\varsigma-\ell} k_i - \frac{d}{r} \left( \alpha_{r,i} - 2^{m+\varsigma} \eta_i \right)
  \right\}_{2^{m+\varsigma}}
\end{align*}
which implies that
\begin{align*}
  |\, \gamma_i \,| = |\, 2^{m+\varsigma-\ell} (\Delta_i - \delta_{\eta_i, i}) \,|
\end{align*}
for~$\Delta_i$ the unknown offset such that $k_i = k_{\eta_i, \Delta_i, i}$, where~$k_{\eta, \Delta, i}$ and~$\delta_{\eta, i}$ denote~$k_{\eta, \Delta}$ and~$\delta_\eta$ in Sect.~\ref{section:notion-good-pair} with~$j$ replaced by~$j_i$.
Hence ${|\, \gamma_i \,| \le 2^{m+\varsigma-\ell} (B_{\Delta} + 1/2)}$ when $|\, \Delta_i \,| \le B_{\Delta}$.
Furthermore $2^{\varsigma-\ell+\tau} d \in [0, 2^{m+\varsigma-\ell+\tau})$.
This implies
\begin{align*}
  R
  =
  \left|\, \mathbf{u} - \mathbf{v} \,\right|
  &=
  \sqrt{
    2^{2(m+\varsigma-\ell)} \sum_{i \, = \, 1}^n (\Delta_i - \delta_{\eta_i, i})^2
    +
    2^{2(\varsigma-\ell+\tau)} d^2
  } \\
  &<
  2^{m+\varsigma-\ell}
  \sqrt{
    n (B_\Delta + 1/2)^2
    +
    2^{2\tau}
  }
\end{align*}
so~$\mathbf u$, and hence~$d$, may be found by enumerating all vectors in~$\mathcal L^\tau$ within radius~$R$ of~$\mathbf v$.

The volume of an $(n + 1)$-dimensional ball of radius~$R$ is
\begin{align*}
  V_{n+1}(R) &= \frac{\pi^{(n + 1) / 2}}{\Gamma \left( \frac{n + 1}{2} + 1 \right)} R^{n + 1},
\end{align*}
whereas the volume of the fundamental parallelepiped in $\mathcal L^\tau$, which contains a single lattice vector, is ${\det \mathcal L^\tau = 2^{n (m + \varsigma) + \varsigma-\ell+\tau}}$.
It follows that, heuristically, approximately $V_{n + 1}(R) / \det \mathcal L^\tau$ vectors in~$\mathcal L^\tau$ have to be enumerated to find~$d$.
Hence, by picking parameters so that with probability $\ge q$, we have ${V_{n + 1}(R) / \det \mathcal L^\tau < 2}$, we heuristically expect to find~$\mathbf u$ and hence~$d$ with probability $\ge q$ immediately by finding the closest vector to~$\mathbf v$ in~$\mathcal L^\tau$.
In practice, an approximation to the closest vector may be found by first using the Lenstra--Lenstra--Lovász (LLL)~\cite{lll} algorithm to reduce the lattice basis, and then using Babai's nearest plane algorithm~\cite{babai}.

As for the integer parameter~$\tau$, it and $B_\Delta$ may, e.g., be selected so that $B_\Delta = 2^\tau - 1$:
The last component of $\mathbf u - \mathbf v$ is then of roughly the same size as the first~$n$ components in the worst case, and $R = |\, \mathbf u - \mathbf v \,| < 2^{m+\varsigma-\ell+\tau} \sqrt{n+1}$.
Another option is to select~$\tau$ based on the expected sizes of $|\, \Delta_i - \delta_{\eta_i, i} \,|$ in average-case analyses, or to simply select~$\tau = 0$.

\subsubsection{Handling artificially short vectors}
\label{section:post-processing-lattice-artificially-short-vectors}
A complication for $u = \gcd(z_1 + \eta_1, \, \ldots, \, z_n + \eta_n, \, r)$, where
\begin{align*}
  z_i
  =
  \frac{r j_i - \alpha_{r,i}}{2^{m+\varsigma}}
  =
  \frac{r j_i - \{rj_i\}_{2^{m+\varsigma}}}{2^{m+\varsigma}}
  =
  \round{\frac{rj_i}{2^{m+\varsigma}}} \in \mathbb Z,
\end{align*}
is that the artificially short non-zero vector $\mathbf s = (0, \, \ldots, \, 0, \, 2^{\varsigma-\ell+\tau} r / u) \in \mathcal L^\tau$.
It follows that for~$d'$ the candidate for~$d$ as given by the last component of the vector in $\mathcal L^\tau$ closest to $\mathbf v$, we expect $d' = \{ d \}_{r/u}$, so it is necessary to test if $d'' = (d' + i \cdot r / u) \text{ mod } r$ is equal to~$d$ where~$i$ runs over $[0, u) \cap \mathbb Z$.
In practice, said tests may be performed by checking if $[d''] \, g = x = [d] \, g$.

To see that $\mathbf s \in \mathcal L^\tau$, note that $r/u \in \mathbb Z$, and that $r/u \cdot \mathbf a_0$ has
\begin{align*}
  r / u \cdot (j_i + (2^{m + \varsigma} \eta_i - \alpha_{r,i}) / r)
  =
  (r j_i + 2^{m + \varsigma} \eta_i - \alpha_{r,i}) / u
  =
  2^{m+\varsigma} (z_i + \eta_i) / u
\end{align*}
in the first~$n$ components, and $2^{\varsigma-\ell+\tau} r/u$ in the last component.
The first~$n$ components are hence divisible by $2^{m+\varsigma}$ since~$u$ divides $z_i + \eta_i$, allowing them to be canceled by taking suitable multiples of $\mathbf a_1, \, \ldots, \, \mathbf a_n$.
As $(z_i + \eta_i) \text{ mod } r$ may be modeled as uniform on $[0, r) \cap \mathbb Z$, the probability of a prime~$p$ that divides~$r$ also dividing $z_1 + \eta_1, \, \ldots, \, z_n + \eta_n$ may be modeled as~$p^{-n}$.
Testing more than one candidate~$d''$ for~$d$ is hence in practice only necessary when~$r$ has small prime factors\footnote{Note that~$r$ is typically prime in cryptographic applications; see Sect.~\ref{section:assumptions-reductions}.} and~$n$ is small (due to the exponential suppression of $p^{-n}$ in $n$).

\subsubsection{Heuristically estimating the expected number of runs required}
For given $d$ and $r$, $s \ge 1$ and $\ell \sim m/s$, and a target success probability~$q$, the number of runs required may be heuristically estimated as described, e.g., in~\cite[Sect.~5.5.10]{ekera-thesis}:
For $n = s + 1, \, s + 2, \, \ldots \,$, first select~$\varsigma$ via Tbl.~\ref{table:varsigma-given-n-qeta} with $B_q \ge q$, and then estimate a bound~$\tilde v$ that the volume quotient in~$\mathcal L^0$ is below with probability $\ge q$.
Return the least~$n$ for which $\tilde v < 2$.

Then verify the estimate by running the classical post-processing on~$M_S$ sets of~$n$ pairs~$(j, k)$ generated by the simulator in Sect.~\ref{section:simulating-quantum-algorithm-with-failures}, with $B_\eta = 0$ and sufficiently large $B_\Delta$.
If the sampling fails for one or more pairs in a set, treat this sampling failure as a failure of the classical post-processing to solve the set for~$d$.
If the verification succeeds for at least~$\round{M_S q}$ sets, decrement~$n$ and try again until the least~$n$ for which it succeeds for at least~$\round{M_S q}$ sets is identified.
Otherwise, increment~$n$ and try again until the least~$n$ for which it succeeds for at least~$\round{M_S q}$ sets is identified.
Return the least~$n$ thus identified.

In practice, the initial estimate of the least~$n$ for which $\tilde v < 2$ can be found more quickly by interpolating and jumping ahead, as noted in~\cite[Sect.~5.5.10]{ekera-thesis}.
The estimation and verification procedures are both implemented in the Qunundrum suite of MPI programs~\cite{qunundrum}.

\subsubsubsection{Estimating a bound on the volume quotient}
To estimate an upper bound~$\tilde v$ on the volume quotient such that
\begin{align*}
  \Pr
  \left(
    \frac{V_{n+1}(R)}{\det \mathcal L^0} \le \tilde v
  \right) \ge q,
\end{align*}
sample~$M_E$ sets of~$n$ pairs~$(j, k)$ by using the simulator for the quantum algorithm in Sect.~\ref{section:simulating-quantum-algorithm-with-failures} (which is fast, but which can fail to produce a valid sample), with $B_\eta = 0$ and sufficiently large $B_\Delta$, compute the volume quotient for each set (where~$R$ is yielded by the set), sort the results in ascending order, and return the quotient at index $\round{(M_E - 1) q}$.
If the sampling fails, set the quotient to~$\infty$.
If the quotient at index $\round{(M_E - 1) q}$ is~$\infty$, no estimate is produced.
This procedure is analogous to that described in~\cite[Sect.~5.5.10.1]{ekera-thesis}.

\subsubsubsection{Selecting~\texorpdfstring{$\varsigma$}{varsigma} so that \texorpdfstring{$\eta_1 = \cdots = \eta_n = 0$}{eta\_1 = ... = eta\_n = 0}}
To select~$\varsigma$ so that $\eta_1 = \cdots = \eta_n = 0$ with probability ${\ge B_q}$ for $\eta_i$ the peak index associated with the pair~$(j_i, k_i)$, and for~$B_q$ a given bound, it suffices, by~\refeq{sum-Pn} and the analysis in Sect.~\ref{section:probability-good-pair} (with~$h$ in~\refeq{sum-Pn} summed over all $\Delta \in \offsets(2^{\ell-1}) = [-2^{\ell-1}, 2^{\ell-1}) \cap \mathbb Z$, making the sum equal one by Lem.~\ref{lemma:h-sum-to-one} in App.~\ref{appendix:supporting-lemmas-claims-proofs}), to pick~$\varsigma$ so that
\begin{align}
  \sum_{\eta \, = \, -B_{\eta}}^{B_{\eta}} \:
  \sum_{\alpha_r' \, = \, -2^{m+\varsigma-\kappa_r-1}}^{2^{m+\varsigma-\kappa_r-1} - 1}
  2^{\kappa_r}
  f_{\eta}
  \left(
    \frac{2\pi}{2^{m+\varsigma}} \, 2^{\kappa_r} \alpha'_r
  \right)
  &>
  B_q^{1/n}
  \label{eq:requirement-q-varsigma}
\end{align}
which, by Lem.~\ref{lemma:fn-tails} in App.~\ref{appendix:supporting-lemmas-claims-proofs}, holds if
\begin{align}
  1
  -
  \frac{4}{\pi^2}
  \frac{1}{2^{\varsigma}}
  \left(
    1
    +
    \frac{1}{2^{m+\varsigma}} + \frac{2}{3 \cdot 2^{2(m+\varsigma)}}
  \right)
  &>
  B_q^{1/n},
  \label{eq:requirement-q-varsigma-simplified}
\end{align}
where we have used that~$m$ is the bit length of~$r$ when computing general discrete logarithms, which implies $r < 2^m$, and that $B_\eta = 0$.
Furthermore, we have assumed that~$r$ is odd so that $\kappa_r = 0$ when invoking the lower bound in Lem.~\ref{lemma:fn-tails}.
Tbl.~\ref{table:varsigma-given-n-qeta} gives, for various~$B_q$, the intervals in~$n$ for which~$\varsigma$ is the least integer that satisfies~\refeq{requirement-q-varsigma-simplified}.

\begin{table}[h!]
  \begin{center}
    {\renewcommand{\arraystretch}{1.1}
    \begin{tabular}{c|c|c|c|c|c|c|c}
      \hline
      & \multicolumn{7}{c}{$\varsigma$} \\
      $B_q$ & $6$ & $7$ & $8$ & $9$ & $10$ & $11$ & $12$ \\
      \thickhline
      $0.990$ & $1$ & $2$--$3$ & $4$--$6$ & $7$--$12$ & $13$--$25$ & $26$--$50$ & $51$--$101$ \\
      $0.991$ & $1$ &      $2$ & $3$--$5$ & $6$--$11$ & $12$--$22$ & $23$--$45$ &  $46$--$91$ \\
      $0.992$ & $1$ &      $2$ & $3$--$5$ & $6$--$10$ & $11$--$20$ & $21$--$40$ &  $41$--$81$ \\
      $0.993$ & $1$ &      $2$ & $3$--$4$ &  $5$--$8$ &  $9$--$17$ & $18$--$35$ &  $36$--$70$ \\
      $0.994$ &     &      $1$ & $2$--$3$ &  $4$--$7$ &  $8$--$15$ & $16$--$30$ &  $31$--$60$ \\
      $0.995$ &     &      $1$ & $2$--$3$ &  $4$--$6$ &  $7$--$12$ & $13$--$25$ &  $26$--$50$ \\
      \hline
    \end{tabular}}
  \end{center}
  \caption{The intervals in~$n$ for which~$\varsigma$ is the least integer that satisfies~\refeq{requirement-q-varsigma-simplified}, for various~$B_q$.
  The analysis on which this table is based assumes that $m \in [128, \, 8192] \cap \mathbb Z$ and that~$r$ is odd.
  (This table is identical to~\cite[Tbl.~5.6]{ekera-thesis} but for minor changes in notation.)}
  \label{table:varsigma-given-n-qeta}
\end{table}

\subsubsubsection{Tabulated estimates}
To exemplify their use, Tbl.~\ref{table:runs-estimated-simulated} yields~$n$ as estimated by the above procedures as a function of~$m \in \{128, \, 256, \, \ldots, \, 8192 \}$ and~$s \in \{1, \, \ldots, \, 8, \, 10, \, 20, \, \ldots, \, 50 \}$, with $\ell = \ceil{m/s}$, $q = 0.99$ and~$\varsigma$ selected as a function of~$n$ according to the row in Tbl.~\ref{table:varsigma-given-n-qeta} for which $B_q = 0.991$, with~$d$ and~$r$ selected as functions of~$m$ as in~\cite[Sect.~7.3]{ekera-general}, and with $M_E = 10^6$ and $M_S = 10^3$.

Both when computing and verifying the estimates, the lattice~$\mathcal L^0$ with ${\eta_1 = \cdots = \eta_n = 0}$ was used.
When verifying the estimates, the lattice basis was first reduced with the LLL~\cite{lll} algorithm, after which Babai's nearest plane algorithm~\cite{babai} was applied, followed by the small search required to handle the artificially short lattice vector~$\mathbf s$ as described in Sect.~\ref{section:post-processing-lattice-artificially-short-vectors}.

As may be seen in Tbl.~\ref{table:runs-estimated-simulated}, the initial estimates correspond well to the number of runs actually required when verifying the estimates by running the classical post-processing with respect to simulated quantum algorithm outputs.
Note that slightly fewer runs suffice if one is willing to enumerate a limited number of vectors in~$\mathcal L^0$.
In particular, a single run typically suffices for $s = 1$ when performing a limited enumeration of vectors in~$\mathcal L^0$, or when solving by performing one or two limited searches as described in Sect.~\ref{section:post-processing-shor}.
Asymptotically, as may be seen in Tbl.~\ref{table:runs-estimated-simulated} (and as is shown analytically in~\cite[Sect.~5.5.8]{ekera-thesis} under the Gaussian heuristic), the number of runs~$n$ required tends to~$s + 1$ as $m \rightarrow \infty$ for fixed integral~$s$.

\begin{table}[h!]
  \begin{center}
  {\renewcommand{\arraystretch}{1.1}
  \begin{tabular}{cr|r|r|r|r|r|r|r}
    \hline
    && \multicolumn{7}{c}{\bfseries logarithm and order length~$m$} \\
    && \multicolumn{1}{r|}{128} & \multicolumn{1}{r|}{256} & \multicolumn{1}{r|}{512} & \multicolumn{1}{r|}{1024} & \multicolumn{1}{r|}{2048} & \multicolumn{1}{r|}{4096} & \multicolumn{1}{r}{8192} \\
    \thickhline
    \multirow{13}{*}{\rotatebox{90}{\bfseries tradeoff factor~$s$}}
    %              128      256        512       1024       2048       4096      8192
    &   $1$ &        2 &      2 &        2 &        2 &        2 &        2 &       2 \\
    &   $2$ &        3 &      3 &        3 &        3 &        3 &        3 &       3 \\
    &   $3$ &        4 &      4 &        4 &        4 &        4 &        4 &       4 \\
    &   $4$ &        6 &      5 &        5 &        5 &        5 &        5 &       5 \\
    &   $5$ &        7 &      6 &        6 &        6 &        6 &        6 &       6 \\
    &   $6$ &   9 / 10 &  7 / 8 &        7 &        7 &        7 &        7 &       7 \\
    &   $7$ &  11 / 12 &      9 &        8 &        8 &        8 &        8 &       8 \\
    &   $8$ &       -- &     11 &   9 / 10 &        9 &        9 &        9 &       9 \\
    &  $10$ &       -- &     15 &       12 &       11 &       11 &       11 &      11 \\
    &  $20$ &       -- &     -- &  31 / 30 &       24 &       22 &       21 &      21 \\
    &  $30$ &       -- &     -- &  -- / 56 &       40 &  35 / 34 &  33 / 32 & 32 / 31 \\
    &  $40$ &       -- &     -- &       -- &  -- / 60 &  49 / 48 &       44 &      42 \\
    &  $50$ &       -- &     -- &       -- &       -- &  -- / 64 &  57 / 56 & 54 / 53 \\
    \hline
  \end{tabular}}
  \end{center}
  \caption{The heuristically estimated number of runs~$n$ required to solve for~$d$ with success probability $\ge q = 0.99$ in~$\mathcal L^0$ without enumerating the lattice.
  For~$A$ the initial and~$B$ the verified estimate, we print~$A$ if $A = B$, otherwise we print $B / A$.
  The dashes indicate no estimate.
  (This table is identical to~\cite[Tbl.~5.4]{ekera-thesis}.)
  }
  \label{table:runs-estimated-simulated}
\end{table}

It should be noted that in practice, for elliptic curve groups, it is typically the case that $m \in \{ 224, \, 253, \, 256, \, 384, \, 446, \, 512, \, 521 \}$, whereas $m \in \{ 224, \, 256, \, 304, \, 352, \, 400 \}$ for Schnorr groups (see~\cite[Sects.~7.2.1.1 and~7.2.3]{ekera-thesis} for further details).
The choice of~$m$ in Tbl.~\ref{table:runs-estimated-simulated} was made to illustrate the asymptotic behavior as~$m$ grows large, not to correspond to practical use cases.
Furthermore, it should be noted that in practice it is better not to restrict~$s$ to be an integer, and to tabulate the least~$\ell$ for which $n = 1, \, 2, \, \ldots$ runs suffice instead of tabulating the least number of runs~$n$ that suffices for $\ell = \ceil{m/s}$ for $s = 1, \, 2, \, \ldots$ as in Tbl.~\ref{table:runs-estimated-simulated}.

\subsubsection{On searching efficiently}
As explained in~\cite{ekera-success-short-dlp}, meet-in-the-middle and random-walk techniques may be leveraged to speed up the enumeration of the lattice.
In a nutshell, the idea is to re-write the enumeration problem as a multidimensional short DLP that may then be solved efficiently classically.
As the idea presented in~\cite{ekera-success-short-dlp} generalizes directly to the setting we consider in this work, and as this fact is already explicitly stated in~\cite{ekera-success-short-dlp}, we do not further elaborate on said idea here.

\subsubsection{On the relation to our earlier works}
Compared to~\cite{ekera-general}, where $n \sim s$ runs of a quantum algorithm that performs $\sim (1 + 2/s)m$ group operations quantumly are required to solve for~$d$ and~$r$, the algorithm in this work performs only $\sim (1 + 1/s)m$ group operations quantumly when solving for~$d$ given~$r$ in $n \sim s$ runs.

This is a slight improvement, entirely explained by the fact that~$r$ is required to be known in this algorithm when computing~$d$, whereas the algorithm in~\cite{ekera-general} computes both~$d$ and~$r$.

\subsubsection{On the relation to Kaliski's work}
Kaliski~\cite{kaliski-magic-box} has modified Shor's quantum algorithm for computing general discrete logarithms to produce an oracle for the most significant half-bit of the logarithm:
For~$d$ the logarithm and~$r$ the group order, the oracle estimates whether $d/r \ge 1/2$ with a non-negligible advantage.

The oracle is used by Kaliski to instantiate the Blum--Micali~\cite{blum-micali} reduction, as generalized by Kaliski~\cite[Ch.~5]{kaliski-phd}, acknowledging ideas from~Goldreich.
This enables general discrete logarithms to be computed in groups of known prime order by repeatedly querying the oracle.

Specifically, to compute~$d$ given $x = [d] \, g$ in a group of $m$-bit order~$r$, the reduction queries the oracle $2m \, c(m)$ times in total:
For each $i \in [0, m) \cap \mathbb Z$, it makes~$2 c(m)$ queries for elements of the form $[2^i] \, x \odot [t] \, g$, where~$t$ is selected uniformly at random from~$[0, r) \cap \mathbb Z$ in each query.
This allows~$d$ to be classically reconstructed with a success probability that depends on~$c(m)$.
In~\cite[Figs.~5.2--5.3]{kaliski-phd} Kaliski lets $c(m) = \ceil{4 \sqrt{m} \, \epsilon^{-1}} \ceil{\epsilon^{-1}}$ for $\epsilon \in (0, 1/2)$ the average advantage of the oracle.
By~\cite[Thm.~5.6]{kaliski-phd} the success probability is then at least~$1/2$.

As for the oracle, we consider an optimized version of Kaliski's quantum algorithm~\cite[Sect.~6.3]{kaliski-magic-box}, in which the first stage computes and reads out~$j$, after which the second stage performs a single controlled group operation by the element $[z^{-1} \: (\text{mod } r)] \, x$ for $z = \round{rj / 2^{m}}$.
A phase-shift gate and a Hadamard gate are then applied to the control qubit, after which it is read out.
The result is correlated with $d/r$, producing the required half-bit oracle.

In summary, $m+1$ group operations are evaluated quantumly in each of the $2m \, c(m) = 2m \ceil{4 \sqrt{m} \, \epsilon^{-1}} \ceil{\epsilon^{-1}}$ runs, where $\epsilon = \pi^{-1}$ in the ideal case~\cite[Sect.~4]{kaliski-magic-box}.
In practice, it is usually possible to solve in a much smaller, but still substantial, number of runs.\footnote{By, e.g., only querying the oracle once in each bit position $i \in [0, m) \cap \mathbb Z$ in each iteration, while storing the results so that all queries can be used when estimating the cross-correlation in each bit position in each iteration.
The same query results may be used for both cross-correlation estimates, and for all initial guesses.}

By comparison, our algorithm for computing general discrete logarithms in groups of known order is parameterized under a tradeoff factor~$s$ that controls the extent to which we make tradeoffs.
This is in analogy with our earlier works~\cite{ekera-hastad, ekera-pp, ekera-general} on computing short and general discrete logarithms in groups of unknown order, and with Seifert's work on order finding~\cite{seifert}.

Specifically, our quantum algorithm evaluates $\sim (1 + 1/s)m$ group operations quantumly in each run, and requires~$s$ or slightly more than~$s$ runs, as each run yields $\sim m/s$ bits of information on~$d$.
Expressed in this language of tradeoffs, Kaliski may hence be said to make \emph{maximal} tradeoffs for~$s = m$, whereas we make \emph{variable} tradeoffs for~$s$ a parameter.

Besides~$s$, our algorithm takes two additional parameters, in the form of~$\varsigma$ and the bound on the search space when enumerating the lattice.
For small to moderate~$s$, and for appropriate choices of~$\varsigma$ and the bound on the search space, we expect to find good overall tradeoffs between the requirements on the quantum computer, the requirements on the classical computer performing the post-processing, and the number of runs required.

Note that as we post-process the outputs from all~$n$ runs simultaneously using lattice-based techniques, there is a limit to how large we can grow~$s$:
The dimension of the lattice~$\mathcal L^\tau$ is $n + 1$ where $n \ge s$.
Hence, the fact that we must be able to compute a sufficiently good reduced basis for~$\mathcal L^\tau$ restricts~$s$.
Kaliski's algorithm does not suffer from this restriction.

If maximal tradeoffs are sought, Kaliski's algorithm provides the necessary means to achieve such tradeoffs.
In practice, one would however arguably seek to make a good overall tradeoff, rather than a maximal tradeoff, in which case the algorithm proposed in this work is useful.
It may be possible to extend Kaliski's approach to cover a range of different tradeoffs.

\section{Generalizing the analysis}
\label{section:generalizations}
As stated in Sect.~\ref{section:quantum-algorithm}, the integer parameters~$\ell$ and~$\varsigma$ roughly correspond to the number of bits of information that are computed on~$d$ and~$r$, respectively, in each run of the quantum algorithm.

The main focus in this paper is to compute~$d$ given $r$, as in Shor's original algorithm for the general DLP, so we let~$m$ be the bit length of~$r$.
Furthermore, we let~$\varsigma$ be small (i.e., sufficiently large to suppress subsidiary peaks in the probability distribution induced by the quantum algorithm, when required by the classical post-processing, but nowhere near large enough to yield~$r$), and we let $\ell = m - \ellint$ for some~$\ellint$ when solving for~$d$ given~$r$ in a single run (see Sect.~\ref{section:post-processing-shor}), or $\ell \sim m/s$ for~$s$ a tradeoff factor when solving in $\sim s$ runs (see Sect.~\ref{section:post-processing-lattice}).

As previously explained in Sect.~\ref{section:quantum-algorithm-generalizations}, it is however possible to relax these requirements, and to more freely select~$m$, $\varsigma$ and~$\ell$, without voiding the analysis in this paper:
Specifically, the exact analysis in Sect.~\ref{section:analysis-probability-distribution} generalizes to several of our earlier works~\cite{ekera-hastad, ekera-pp, ekera-general, ekera-success-order-finding, ekera-success-short-dlp}, providing a unified lens on all of these algorithms.
In what follows below, we exemplify why this is useful.

\subsection{Computing general~\texorpdfstring{$d$}{d} and optionally~\texorpdfstring{$r$}{r} when~\texorpdfstring{$r$}{r} is unknown}
\label{section:generalizations-general}
If~$r$ is unknown, we may let~$m$ be an upper bound on the bit length of~$r$, and let $\varsigma = \ell = m - \ellint$ to compute both~$d$ and~$r$ in a single run, at the expense of evaluating $3m-2\ellint$ group operations in this single run (in the vocabulary of Sect.~\ref{section:earlier-works}) as opposed to $2m-\ellint$ operations when~$r$ is known.
Furthermore, we may make tradeoffs, and let $\varsigma = \ell \sim m / s$ for $s$~a tradeoff factor to compute both~$d$ and~$r$ in $\sim s$ runs, at the expense of evaluating $\sim m + 2m/s$ group operations per run as opposed to $\sim m + \epsilon + m/s$ operations for some small~$\epsilon$ when~$r$ is known.\footnote{Note also that when~$r$ is known, $m$ is the bit length of~$r$, whereas~$m$ is an upper bound on the bit length of~$r$ when~$r$ is unknown, so the relative cost difference is greater than what this coarse comparison indicates.}

The variation of Shor's algorithm that results from selecting the parameters $m$, $\varsigma$ and~$\ell$ in this manner is exactly the variation described and analyzed in~\cite{ekera-general}.
The exact analysis in Sect.~\ref{section:analysis-probability-distribution} and simulator in Sect.~\ref{section:simulating-quantum-algorithm-without-failures} of this paper immediately generalize to said variation.

The analysis in~\cite{ekera-general} is based upon several approximation steps, where the error introduced in each approximation step is upper bounded.
By comparison, the analysis in this paper is exact.
It is valid also for parameter regimes (in~$m$, $\varsigma$ and~$\ell$) that are beyond the reach of the analysis in~\cite{ekera-general} due to the error bounds on the approximation errors growing too large.

Furthermore, it should be noted that the simulator in~\cite{ekera-general} for said variation can fail to sample, whereas the exact simulator in Sect.~\ref{section:simulating-quantum-algorithm-without-failures} of this paper never fails to sample.

\subsection{Computing short~\texorpdfstring{$d$}{d} when~\texorpdfstring{$r$}{r} is unknown}
\label{section:generalizations-short}
If~$r$ is unknown, and~$d$ is short (see Sect.~\ref{section:preliminaries}), we may let~$m$ be an upper bound on the bit length of~$d$, and let $\varsigma = \ell = m - \ellint$ to compute~$d$ in a single run without knowing or computing~$r$.
Furthermore, we may make tradeoffs, and let $\varsigma = \ell \sim m / s$ for~$s$ a tradeoff factor to compute~$d$ in~$\sim s$ runs without knowing or computing~$r$.
The variation of Shor's algorithm that results from selecting the parameters $m$, $\varsigma$ and~$\ell$ in this manner is exactly the algorithm of Ekerå and Håstad described and analyzed in~\cite{ekera-hastad, ekera-pp, ekera-success-short-dlp}.
The exact analysis in Sect.~\ref{section:analysis-probability-distribution} and simulator in Sect.~\ref{section:simulating-quantum-algorithm-without-failures} of this paper immediately generalize to said variation.

The analyses in~\cite{ekera-hastad, ekera-pp, ekera-success-short-dlp} are exact, but they rely on~$d \sim \sqrt{r}$ or smaller when $s = 1$.
Specifically, they require that $r \ge 2^{m + \varsigma} + (2^\ell - 1)d$, which implies $\log_2(d) \sim \log_2(r) / (1 + 1/s)$ or smaller.
By comparison, the exact analysis in this work also covers the regime where $d \sim \sqrt{r}$ or greater:
It covers all~$d$ and~$r$, including the case where $r < 2^{m + \varsigma} + (2^\ell - 1)d$.

Furthermore, it should be noted that the simulators in~\cite{ekera-pp, ekera-success-short-dlp} for said variation can fail to sample, whereas the exact simulator in Sect.~\ref{section:simulating-quantum-algorithm-without-failures} of this paper never fails to sample.

\subsection{Computing~\texorpdfstring{$r$}{r} only when~\texorpdfstring{$r$}{r} is unknown}
\label{section:generalizations-order}
We may let~$m$ be an upper bound on the bit length of~$r$, and let $\varsigma = m - \ellint$ and $\ell = 0$ to compute only~$r$ in a single run.
Furthermore, we may make tradeoffs, and let $\varsigma \sim m / s$ for $s$~a tradeoff factor and $\ell = 0$ to compute only~$r$ in $\sim s$ runs.

The variations of Shor's algorithm that result from selecting the parameters $m$, $\varsigma$ and~$\ell$ in this manner are exactly the variations described and analyzed in, e.g.,~\cite{shor1994, shor1997, seifert, ekera-general, ekera-success-order-finding}.
The exact analysis in Sect.~\ref{section:analysis-probability-distribution} and simulator in Sect.~\ref{section:simulating-quantum-algorithm-without-failures} of this paper immediately generalize to said variations.
The exact analysis in~\cite{ekera-success-order-finding} of said variations is essentially equivalent to the analysis in this paper.
That said, it should be noted that the simulator in~\cite{ekera-success-order-finding} for said variations can fail to sample, whereas the exact simulator in Sect.~\ref{section:simulating-quantum-algorithm-without-failures} of this paper never fails to sample.

\section{Summary and conclusion}
\label{section:summary-conclusion}
We have heuristically shown that Shor's quantum algorithm for the DLP achieves an expected success probability of approximately~$60$--$82\%$ in a single run when modified to enable efficient implementation with the semi-classical QFT with control qubit recycling.

By slightly increasing the number of group operations that are evaluated quantumly and performing a single limited search in the post-processing, or by performing two limited searches in the post-processing, we have rigorously proved that the modified algorithm can be made to achieve a success probability that, e.g., exceeds~$1 - 10^{-4}$ in a single run.
Specifically, we have given an exact analysis of the probability distribution induced by the algorithm and used it to prove a lower bound on the success probability, as a function of the group order~$r$, the size of the search space in the post-processing, and the number of additional group operations evaluated quantumly.
In the limit as~$r \rightarrow \infty$, we have shown that the lower bound on the success probability can be made to tend to one.

In analogy with our earlier work~\cite{ekera-success-short-dlp}, we have explained how the limited searches in the post-processing can be performed efficiently using meet-in-the-middle or random-walk techniques, and we have shown how slightly better tradeoffs can be achieved, compared to our earlier work~\cite{ekera-general}, when~$r$ is known.
We have generalized our exact analysis to cover several quantum algorithms analyzed in our earlier works~\cite{ekera-hastad, ekera-pp, ekera-general, ekera-success-order-finding, ekera-thesis, ekera-success-short-dlp}, including for parameter regimes not covered by our earlier analyses, yielding a unified lens on all of these algorithms.
Furthermore, we have generalized our exact analysis to the setting where information on multiple discrete logarithms is computed in each run of the algorithm so as to amortize costs; see App.~\ref{appendix:multiple-logarithms}.
Last but not least, we have given a novel exact simulator, which is based on our generalized exact analysis, and which hence works for all quantum algorithms covered by said analysis.

\section*{Acknowledgments}
I am grateful to Johan Håstad for valuable comments.
I thank Craig Gidney for asking questions about the success probabilities of Shor's algorithms, Ryan Babbush for asking questions about the algorithm of van Dam~\cite{dam} and of van Dam and Seroussi~\cite{dam-seroussi}, and Burt Kaliski Jr.\ for discussions regarding tradeoffs.
I thank Lennart Brynielsson, Andreas Minne and other colleagues who have proofread early versions of this manuscript for useful suggestions and review comments.
I thank Burt Kaliski Jr.\ for spotting a minor issue in Sect.~\ref{section:generalizations-short}.

Funding and support were provided by the Swedish NCSA that is a part of the Swedish Armed Forces.
Some work on this manuscript was conducted when the author was a visiting researcher at the Simons Institute at UC Berkeley, Berkeley, USA, within the framework of the Summer Cluster on Quantum Computing~2026.
Certain computations were enabled by resources at PDC at KTH Royal Institute of Technology.
Access to these resources was provided by the National Academic Infrastructure for Supercomputing in Sweden (NAISS), and by its predecessor the Swedish National Infrastructure for Computing (SNIC), funded by the Swedish Research Council, e.g., through grant agreements no.~2022-06725 and~2018-05973.

An earlier pre-print version of this manuscript is a part of the author's PhD thesis~\cite{ekera-thesis}.

\paragraph{Use of artificial intelligence~(AI)}
In earlier versions of this manuscript, the lower bound on the success probability was heuristically derived.
AI in the form of Claude Opus~5.5 and Fable~5.2 has been used to help to derive the same bound in a rigorous manner, and to tighten it.
AI has furthermore, e.g., been used to help develop the efficient exact sampling procedure in Sect.~\ref{section:simulating-quantum-algorithm-without-failures} (which is based on sampling Fejér kernels) and Cor.~\ref{corollary:invertible-correct-eta}, as well as the generalization to computing information on multiple discrete logarithms in App.~\ref{appendix:multiple-logarithms}.
Finally, AI has been used to edit this manuscript, and to review it for consistency, correctness and rigor.
Needless to say, the author assumes responsibility for the correctness of this manuscript.

\paragraph{Status of this manuscript}
This manuscript exists in pre-print form in five earlier versions, the first of which dates back to 2019.
None of the earlier versions of this manuscript have been submitted for publication, however, since they either did not capture the full probability distribution or were based on heuristics.
This is the first non-heuristic version of the manuscript to capture the full probability distribution.

\clearpage

\appendix
\counterwithin{table}{section}

\section{Computing multiple discrete logarithms}
\label{appendix:multiple-logarithms}
In this appendix, we extend the quantum algorithm in Sect.~\ref{section:quantum-algorithm} to the setting where information on multiple discrete logarithms $d_i = \log_g x_i$ for $i \in [1, c] \cap \mathbb Z$ is computed in each run, with respect to the same generator~$g$, by first computing~$j$ and then computing~$k_i$ given~$j$ for each~$d_i$.

We show that each pair~$(j, k_i)$ thus yielded is distributed exactly as the pair~$(j, k)$ yielded for $d = d_i$ by the quantum algorithm in Sect.~\ref{section:quantum-algorithm}; i.e., the analysis in this paper applies to each pair~$(j, k_i)$, as does the classical post-processing in Sect.~\ref{section:post-processing}, and the simulator in Sect.~\ref{section:simulating-quantum-algorithm-without-failures}.

In essence, this allows the cost of exponentiating~$g$ quantumly to be amortized over all~$c$ logarithms:
In each run, $g$ is exponentiated only once, so $(c - 1)(m + \varsigma)$ group operations are saved compared to performing~$c$ runs of the quantum algorithm in Sect.~\ref{section:quantum-algorithm}; see App.~\ref{appendix:multiple-logarithms-algorithm}.

\paragraph{Earlier works}
It should be stated that extensions of the kind that we consider in this appendix have been proposed earlier in the literature, e.g., by Litinski~\cite{litinski}, and by Hhan--Yamakawa--Yun~\cite{hyy24} (see also Hhan~\cite{hhan25} for follow-up work with a matching lower bound).

Litinski~\cite{litinski} considers the case where the control registers are uniform up to powers of two, and the QFTs are of corresponding power-of-two sizes (to enable the use of the semi-classical QFT with control qubit recycling and windowing), but gives no analysis of the resulting algorithm.
Hhan--Yamakawa--Yun~\cite{hyy24} consider only the comparatively simple case where the control registers are uniform up to~$r$, and the QFTs are of size~$r$.
In this appendix, we give an analysis for the case where the control registers are uniform up to powers of two, and the QFTs are of corresponding power-of-two sizes.

It should furthermore be noted that in work that predates~\cite{litinski, hyy24}, van Dam~\cite{dam} and van Dam and Seroussi~\cite{dam-seroussi} consider amortizing the cost of exponentiating~$g$ over multiple discrete logarithms.
The algo\-rithm proposed in~\cite{dam, dam-seroussi} is essentially Shor's algorithm, in the extended form considered in~\cite{hyy24}, but with the inverse of~$j$ modulo~$r$ applied quantumly, by means of an additional exponentiation, enabling the logarithm $-j^{-1} k_i \text{ mod } r$ (in our notation) to be directly read out from the quantum state.
In practice, it appears desirable to instead apply said inverse classically, as in~\cite{hyy24} and in Shor's original classical post-processing~\cite{shor1994, shor1997}.

\subsection{The extended quantum algorithm}
\label{appendix:multiple-logarithms-algorithm}
Given~$g$ and $x_i = [d_i] \, g$ where $d_i \in [0, r) \cap \mathbb Z$ for $i \in [1, c] \cap \mathbb Z$, for~$c$ a positive integer, the extended quantum algorithm first initializes one control register of length $m + \varsigma$ qubits, and~$c$ control registers of length $\ell \le m + \varsigma$ qubits each, to uniform superpositions of all $a \in [0, 2^{m+\varsigma}) \cap \mathbb Z$ and $b_1, \, \ldots, \, b_c \in [0, 2^{\ell}) \cap \mathbb Z$, respectively.
It then computes $[a] \, g \odot [-b_1] \, x_1 \odot \cdots \odot [-b_c] \, x_c$ to a work register in what amounts to a generalized exponentiation.
Applying QFTs of size~$2^{m+\varsigma}$ and~$2^{\ell}$ to the first and to the remaining $c$~control registers, respectively, then maps
\begin{align}
  &\frac{1}{\sqrt{2^{m+\varsigma+c\ell}}} \hspace{-1.5mm}
  \sum_{a \, = \, 0}^{2^{m+\varsigma}-1}
  \sum_{b_1, \, \ldots, \, b_c \, = \, 0}^{2^{\ell}-1}
  \hspace{-4mm} \ket{a, \, b_1, \, \ldots, \, b_c, \, [a] \, g \odot [-b_1] \, x_1 \odot \ldots \odot [-b_c] \, x_c}
  \quad \xrightarrow{\text{QFT}} \notag \\
  &\quad\quad \frac{1}{2^{m+\varsigma+c\ell}} \hspace{-1.5mm}
  \sum_{a, \, j \, = \, 0}^{2^{m+\varsigma}-1}
  \sum_{\substack{b_1, \, \ldots, \, b_c \, = \, 0 \\ k_1, \, \ldots, \, k_c \, = \, 0}}^{2^{\ell}-1}
  \hspace{-4mm} \e^{ \frac{2 \pi \imag}{2^{m+\varsigma}} \, \left( a j + 2^{m+\varsigma-\ell} \sum_{i = 1}^c b_i k_i \right)}
  \ket{j, \, k_1, \, \ldots, \, k_c, \, \left[a - \Sigma_{i = 1}^c b_i d_i \right] \, g}.
  \label{eq:psi-multiple}
\end{align}

For $c = 1$, the extended quantum algorithm reduces to the algorithm in Sect.~\ref{section:quantum-algorithm}.
In total, $m + \varsigma + c \ell$ group operations are evaluated quantumly in a single run of the extended algorithm, compared to $c (m + \varsigma + \ell)$ group operations in~$c$ runs of the quantum algorithm in Sect.~\ref{section:quantum-algorithm}, so $(c - 1)(m + \varsigma)$ group operations are saved by amortizing the cost of exponentiating~$g$.\footnote{Note that if the extended quantum algorithm is implemented using the space-saving technique of Chevignard, Fouque and Schrottenloher~\cite{cfs, cfs-ecc}, the control registers cannot be recycled; all $m + \varsigma + c \ell$ control qubits must then be kept for the duration of the computation, causing the space usage to grow with~$c$.}

As a further generalization, it is worth noting that the~$c$ control registers to which $x_1, \, \ldots, \, x_c$ are exponentiated may be of different lengths $\ell_1, \, \ldots, \, \ell_c$, in which case the analysis in the next section still holds with~$\ell$ replaced by $\ell_1, \, \ldots, \, \ell_c$ for each respective register.

\subsection{Analysis of the probability distribution}
\label{appendix:multiple-logarithms-distribution}
When the state~\refeq{psi-multiple} is observed, it collapses to $(j, k_1, \ldots, k_c)$ and $y = [e] \, g$ for $e \in [0, r) \cap \mathbb Z$ with probability
\begin{align}
  \frac{1}{2^{2(m + \varsigma + c \ell)}}
  \left|\,
  \sum_{(a, \, b_1, \, \ldots, \, b_c)}
  \exp{\left[ \frac{2 \pi \imag}{2^{m + \varsigma}} \left(aj + 2^{m + \varsigma - \ell} \sum_{i \, = \, 1}^c b_i k_i \right) \right]}
  \,\right|^2
  \label{eq:probability-multiple}
\end{align}
where $(a, \, b_1, \, \ldots, \, b_c)$ runs over all $a \in [0, 2^{m + \varsigma}) \cap \mathbb Z$ and $b_1, \ldots, b_c \in [0, 2^{\ell}) \cap \mathbb Z$ such that ${e \equiv a - \sum_{i \, = \, 1}^c b_i d_i \:\: (\text{mod } r)}$.
To simplify~\refeq{probability-multiple}, we follow Sect.~\ref{section:analysis-probability-distribution-simplifying}, and use the fact that
\begin{align*}
  \frac{1}{r}
  \sum_{\mu \, = \, 0}^{r-1}
  \exp
  \left[
    \frac{2 \pi \imag}{r} \left( e - \left( a - \sum_{i \, = \, 1}^c b_i d_i \right) \right) \, \mu
  \right]
  =
  \left\{
    \begin{array}{cl}
      1 & \text{if } e \equiv a - \sum_{i \, = \, 1}^{c} b_i d_i \:\: (\text{mod } r), \\
      0 & \text{otherwise,}
    \end{array}
  \right.
\end{align*}
yielding the equivalence
\begin{align}
  \refeq{probability-multiple}
  &=
  \frac{1}{2^{2(m+\varsigma+c\ell)} r^2}
  \left|\,
  \sum_{\mu \, = \, 0}^{r-1}
  \sum_{a \, = \, 0}^{2^{m + \varsigma} - 1}
  \sum_{b_1, \, \ldots, \, b_c \, = \, 0}^{2^{\ell} - 1}
  \exp
  \left[
    \frac{2 \pi \imag}{r} \left( e - \left( a - \sum_{i \, = \, 1}^c b_i d_i \right) \right) \, \mu
  \right] \cdot \right. \notag \\
  &\phantom{=} \quad\quad\quad\quad\quad\quad\quad\quad\quad\:
  \left. \exp{\left[ \frac{2 \pi \imag}{2^{m + \varsigma}} \left( aj + 2^{m + \varsigma - \ell} \sum_{i \, = \, 1}^c b_i k_i \right) \right]} \,\right|^2 \notag \\
  &=
  \frac{1}{2^{2(m+\varsigma+c\ell)} r^2}
  \left|\,
  \sum_{\mu \, = \, 0}^{r-1}
  \exp
  \left[
    \frac{2 \pi \imag}{r} e \mu
  \right]
  \sum_{a \, = \, 0}^{2^{m + \varsigma} - 1}
  \hspace{-2mm}
  \exp
  \left[
    2 \pi \imag
    \left(
      \frac{j}{2^{m + \varsigma}}
      -
      \frac{\mu}{r}
    \right)
    a
  \right]
  \right. \notag \cdot \\
  &\phantom{=} \quad\quad\quad\quad\quad\quad\quad\quad\quad\:
  \left.
  \prod_{i \, = \, 1}^{c}
  \,
  \sum_{b_i \, = \, 0}^{2^{\ell} - 1}
  \exp
  \left[
    2 \pi \imag
    \left(
      \frac{k_i}{2^{\ell}}
      +
      \frac{d_i \mu}{r}
    \right)
    b_i
  \right]
  \,\right|^2 \notag \\
  &=
  \frac{1}{r^2}
  \left|\,
  \sum_{\mu \, = \, 0}^{r-1}
  \exp
  \left[
    \frac{2 \pi \imag}{r} e \mu
  \right]
  A_\mu(j)
  \prod_{i \, = \, 1}^{c}
  B_{\mu, i}(k_i)
  \,\right|^2
  \label{eq:probability2-multiple}
\end{align}
where $A_\mu(j)$ is as in~\refeq{Amu} in Sect.~\ref{section:analysis-probability-distribution-simplifying}, and
\begin{align}
  B_{\mu, i}(k_i)
  =
  \frac{1}{2^{\ell}}
  \sum_{b_i \, = \, 0}^{2^{\ell} - 1}
  \exp
  \left[
    2 \pi \imag
    \left(
      \frac{k_i}{2^{\ell}}
      +
      \frac{d_i \mu}{r}
    \right)
    b_i
  \right]
  \label{eq:Bmui}
\end{align}
in analogy with the definition of~$B_{\mu}(k)$ in~\refeq{Bmu} in Sect.~\ref{section:analysis-probability-distribution-simplifying}.

Summing~\refeq{probability-multiple}, or equivalently~\refeq{probability2-multiple}, over all $e \in [0, r) \cap \mathbb Z$ hence yields
\begin{align}
  &\frac{1}{2^{2(m + \varsigma + c \ell)}}
  \sum_{e \, = \, 0}^{r-1}
  \left|\,
  \sum_{(a, \, b_1, \, \ldots, \, b_c)}
  \exp{\left[ \frac{2 \pi \imag}{2^{m + \varsigma}} \left(aj + 2^{m + \varsigma - \ell} \sum_{i \, = \, 1}^c b_i k_i \right) \right]}
  \,\right|^2 \notag \\
  &\quad\:\: = \frac{1}{r^2}
  \sum_{e \, = \, 0}^{r-1}
  \left|\,
  \sum_{\mu \, = \, 0}^{r-1}
  \exp
  \left[
    \frac{2 \pi \imag}{r} e \mu
  \right]
  A_\mu(j)
  \prod_{i \, = \, 1}^{c}
  B_{\mu, i}(k_i)
  \,\right|^2 \notag \\
  &\quad\:\: =
  \frac{1}{r^2}
  \sum_{\mu, \, \mu' \, = \, 0}^{r-1}
  A_\mu(j) \, \overline{A_{\mu'}(j)}
  \left(\:
    \prod_{i \, = \, 1}^{c}
    B_{\mu, i}(k_i) \, \overline{B_{\mu', i}(k_i)}
  \right)
  \sum_{e \, = \, 0}^{r-1}
  \exp
  \left[
    \frac{2 \pi \imag}{r} (\mu - \mu') e
  \right]
  \notag \\
  &\quad\:\: =
  \frac{1}{r}
  \sum_{\mu \, = \, 0}^{r - 1}
  |\, A_{\mu}(j) \,|^2
  \prod_{i \, = \, 1}^{c}
  |\, B_{\mu, i}(k_i) \,|^2
  \label{eq:probability-sum-multiple}
\end{align}
where we have used that
\begin{align*}
  \sum_{e \, = \, 0}^{r-1}
  \exp
  \left[
    \frac{2 \pi \imag}{r} (\mu - \mu') e
  \right]
  &=
  \left\{
    \begin{array}{cl}
      r & \text{if } \mu \equiv \mu' \:\: (\text{mod } r), \\
      0 & \text{otherwise,}
    \end{array}
  \right.
\end{align*}
as in the derivation of~\refeq{probability-sum} in Sect.~\ref{section:analysis-probability-distribution-simplifying}.

Finally, in analogy with the definition of~$\phi_\eta$ in~\refeq{phi-eta} in Sect.~\ref{section:arguments-angles}, let
\begin{align}
  \phi_{\eta, i}
  =
  \left\{
  \frac{2 \pi}{2^{m+\varsigma}}
  \left(
    \alpha_{d,i}
    -
    \frac{d_i}{r} \left( \alpha_r - 2^{m+\varsigma} \eta \right)
  \right)
  \right\}_{2\pi}
  =
  \left\{
    \theta_{d, i}
    -
    \frac{d_i}{r} \left( \theta_r - 2 \pi \eta \right)
  \right\}_{2\pi}
  \label{eq:phi-etai}
\end{align}
where~$\alpha_r$ and~$\theta_r$ are as in~\refeq{arguments-angles} in Sect.~\ref{section:arguments-angles}, and where
\begin{align*}
  \alpha_{d, i} = \{ d_i j + 2^{m+\varsigma-\ell} k_i \}_{2^{m+\varsigma}},
  \:\quad\:
  \theta_{d, i} = \frac{2 \pi \alpha_{d, i}}{2^{m+\varsigma}},
\end{align*}
in analogy with the definition of~$\alpha_d$ and~$\theta_d$, respectively, in~\refeq{arguments-angles} in Sect.~\ref{section:arguments-angles}.

By the derivation in Sect.~\ref{section:arguments-angles}, it then holds for $\eta = \{ \mu - z \}_r$ that $|\, A_\mu(j) \,|^2 / r = F_{\eta}(\theta_r)$, and that $|\, B_{\mu, i}(k_i) \,|^2 = h(\phi_{\eta, i})$ for all $i \in [1, c] \cap \mathbb Z$, as said derivation of $|\, B_\mu(k) \,|^2 = h(\phi_\eta)$ holds for any~$d$ and~$k$.
As~$\mu$ runs over $[0, r) \cap \mathbb Z$ in~\refeq{probability-sum-multiple}, the peak index $\eta = \{ \mu - z \}_r$ runs over $[-\lfloor r/2 \rfloor, \lceil r/2 \rceil) \cap \mathbb Z$, for $z = \round{rj / 2^{m+\varsigma}}$.
Hence, it follows that
\begin{align}
  \refeq{probability-sum-multiple}
  &=
  \sum_{\eta \, = \, -\floor{r/2}}^{\ceil{r/2} - 1}
  F_{\eta}(\theta_r)
  \prod_{i \, = \, 1}^{c}
  h(\phi_{\eta, i}).
  \label{eq:probability-sum2-multiple}
\end{align}

The above derivation is summarized and formalized in the below lemma:

\begin{applemma}
\label{lemma:multiple-joint-distribution}
The probability of observing $(j, \, k_1, \, \ldots, \, k_c)$ and $y = [e] \, g$, summed over all $e \in [0, r) \cap \mathbb Z$, in a run of the quantum algorithm in App.~\ref{appendix:multiple-logarithms-algorithm}, is
\begin{align}
  \frac{1}{r}
  \sum_{\mu \, = \, 0}^{r - 1}
  |\, A_{\mu}(j) \,|^2
  \prod_{i \, = \, 1}^{c}
  |\, B_{\mu, i}(k_i) \,|^2
  =
  \sum_{\eta \, = \, -\floor{r/2}}^{\ceil{r/2} - 1}
  F_{\eta}(\theta_r)
  \prod_{i \, = \, 1}^{c}
  h(\phi_{\eta, i}),
  \label{eq:probability-sum-multiple-lemma-equated}
\end{align}
where the term for~$\eta$ on the right-hand side is equal to the term for $\mu = (z + \eta) \text{ mod } r$ on the left-hand side, for $z = \round{rj / 2^{m+\varsigma}}$, for~$A_\mu(j)$ and~$B_{\mu,i}(k_i)$ as in~\refeq{Amu} and~\refeq{Bmui}, respectively, and for~$F_{\eta}(\theta_r)$ and~$h(\phi_{\eta, i})$ as in~\refeq{probability-sum3}, where~$\theta_r$ and $\phi_{\eta, i}$ are as in~\refeq{arguments-angles} and~\refeq{phi-etai}, respectively.
More specifically, it holds for $\eta = \{ \mu - z \}_r$ that $|\, A_\mu(j) \,|^2 / r = F_{\eta}(\theta_r)$ and that $|\, B_{\mu, i}(k_i) \,|^2 = h(\phi_{\eta, i})$ for all $i \in [1, c] \cap \mathbb Z$.
\end{applemma}
\begin{proof}
  As shown in App.~\ref{appendix:multiple-logarithms-distribution}, it holds that $\refeq{probability-multiple} = \refeq{probability2-multiple}$ is the probability of observing the tuple $(j, \, k_1, \, \ldots, \, k_c)$ and $y = [e] \, g$ for $e \in [0, r) \cap \mathbb Z$.
  Summing~\refeq{probability2-multiple} over all $e \in [0, r) \cap \mathbb Z$ yields $\refeq{probability-sum-multiple} = \refeq{probability-sum2-multiple}$.
  The expressions equated in~\refeq{probability-sum-multiple-lemma-equated} are~\refeq{probability-sum-multiple} and~\refeq{probability-sum2-multiple}, and so that part follows.
  The termwise equalities follow from the expressions for~$A_\mu(j)$ and~$B_{\mu, i}(k_i)$, which are {$r$-periodic} in $\mu - z$, and from the definitions of~$F_{\eta}(\theta_r)$ and~$h(\phi_{\eta, i})$, and so the lemma follows.
\end{proof}

It furthermore follows from the above derivation that each pair~$(j, k_i)$ for $i \in [1, c] \cap \mathbb Z$ is distributed exactly as the pair~$(j, k)$ yielded for $d = d_i$ by the quantum algorithm in Sect.~\ref{section:quantum-algorithm}:

\begin{applemma}
\label{lemma:multiple-marginal-distribution}
For each $i \in [1, c] \cap \mathbb Z$, the pair $(j, k_i)$ yielded by the extended quantum algorithm in App.~\ref{appendix:multiple-logarithms-algorithm} is distributed exactly as the pair~$(j, k)$ yielded for $d = d_i$ by the quantum algorithm in Sect.~\ref{section:quantum-algorithm}, i.e., according to~\refeq{probability-sum-lemma-equated} in Lem.~\ref{lemma:probability-distribution} with~$d$ and~$k$ replaced by~$d_i$ and~$k_i$, respectively.
\end{applemma}
\begin{proof}
  By Cl.~\ref{claim:exp-double-sum-to-one}, it holds that $\sum_{k \, = \, 0}^{2^{\ell}-1} |\, B_{\mu, i'}(k) \,|^2 = 1$ for all $\mu \in [0, r) \cap \mathbb Z$ and all $i' \in [1, c] \cap \mathbb Z$; see also Sect.~\ref{section:simulating-quantum-algorithm-without-failures}.
  Summing the left-hand side of~\refeq{probability-sum-multiple-lemma-equated} in Lem.~\ref{lemma:multiple-joint-distribution} over all $k_{i'} \in [0, 2^{\ell}) \cap \mathbb Z$ for all $i' \in [1, c] \cap \mathbb Z$ such that $i' \neq i$ therefore yields
  \begin{align*}
    \frac{1}{r}
    \sum_{\mu \, = \, 0}^{r - 1}
    |\, A_{\mu}(j) \, B_{\mu, i}(k_i) \,|^2
    &=
    \sum_{\eta \, = \, -\floor{r/2}}^{\ceil{r/2} - 1}
    F_{\eta}(\theta_r)
    \,
    h(\phi_{\eta, i}),
  \end{align*}
  which is~\refeq{probability-sum-lemma-equated} with~$d$ and~$k$ replaced by~$d_i$ and~$k_i$, respectively, and so the lemma follows.
\end{proof}

It follows from Lem.~\ref{lemma:multiple-joint-distribution} that $(j, \, k_1, \, \ldots, \, k_c)$ may be sampled by first sampling~$\mu$ uniformly at random from $[0, r) \cap \mathbb Z$, and by then independently sampling~$j$ according to~$|\, A_{\mu}(j) \,|^2$, and~$k_i$ according to~$|\, B_{\mu, i}(k_i) \,|^2$ for each $i \in [1, c] \cap \mathbb Z$, in analogy with Sect.~\ref{section:simulating-quantum-algorithm-without-failures}.
Note that a single~$\mu$ is sampled, and that it is shared by all problem instances.
Intuitively, $j / 2^{m+\varsigma}$ is then typically close to~$\mu / r$, and $k_i / 2^{\ell}$ to~$-d_i \mu / r$, modulo one, for all $i \in [1, c] \cap \mathbb Z$.

Hence, although each pair~$(j, k_i)$ is distributed as in Sect.~\ref{section:analysis-probability-distribution} for $d = d_i$ by Lem.~\ref{lemma:multiple-marginal-distribution}, the pairs are not independent:
They share~$j$, and hence also $z = \round{rj / 2^{m+\varsigma}}$, and they share~$\mu$, and hence also the peak index $\eta = \{ \mu - z \}_r$.
Given~$\mu$, or equivalently given~$j$ and~$\eta$, the frequencies $k_1, \ldots, k_c$ are independent, but they are in general not independent given only~$j$.
The pairs $(j, k_1), \, \ldots, \, (j, k_c)$ may therefore not be treated as if they were yielded by~$c$ independent runs of the quantum algorithm in Sect.~\ref{section:quantum-algorithm}.
That said, the classical post-processing in Sect.~\ref{section:post-processing} applies unchanged when the pairs associated with each logarithm~$d_i$ are passed to it separately, both for a single run with $\ell = m - \ellint$ and for~$n$ runs with $\ell \sim m/s$.
Each~$d_i$ is then recovered with the same probability as a single logarithm; only the joint success probability differs.

\subsection{Simulating the extended quantum algorithm}
\label{appendix:multiple-logarithms-simulation}
The exact simulator in Alg.~\ref{alg:simulating-quantum-algorithm-without-failures} generalizes directly to the extended quantum algorithm in App.~\ref{appendix:multiple-logarithms-algorithm}, see Alg.~\ref{alg:simulating-extended-quantum-algorithm} below, as follows from the analysis in App.~\ref{appendix:multiple-logarithms-distribution}.

\begin{breakablealgorithm}
  \caption{Samples~$(j, \, k_1), \, \ldots, \, (j, \, k_c)$ from the probability distribution induced by the quantum algorithm in App.~\ref{appendix:multiple-logarithms-algorithm}, which induces the state~\refeq{psi-multiple} and reads out~$j, \, k_1, \, \ldots, \, k_c$.
  \\
  {\bfseries \emph{Inputs:}} $m$, $\varsigma$, $\ell$, $c$, $d_1, \ldots, d_c$, $r$
  $\:$
  {\bfseries \emph{Returns:}} $(j, \, k_1), \, \ldots, \, (j, \, k_c)$ and optionally $\eta, \, \Delta_1, \, \ldots, \, \Delta_c$}
  \label{alg:simulating-extended-quantum-algorithm}
  \begin{pseudocode}
    \item Let \textsc{SampleFejérKernel}($M$, $\delta_u$) be the function defined in Alg.~\ref{alg:simulating-quantum-algorithm-without-failures}.

    \item Sample~$\mu$ uniformly at random from $[0, r) \cap \mathbb Z$.

    \emph{Note: A single~$\mu$ is sampled, and it is shared by all~$c$ problem instances.}

    \item Let $\delta_u \leftarrow \{ 2^{m+\varsigma} \mu/r \}_1$ and $M \leftarrow 2^{m+\varsigma}$.

    Let $u \leftarrow \textsc{SampleFejérKernel}(M, \, \delta_u)$ to sample~$u$ from $D_{M, \delta_u}(u)$.

    Let $j \leftarrow (u + \round{2^{m+\varsigma} \mu/r}) \text{ mod } 2^{m+\varsigma}$ and $\eta \leftarrow \{ \mu - z \}_r$ where $z = \round{rj / 2^{m+\varsigma}}$.

    \item For $i = 1, \, \ldots, \, c$:
    \begin{pseudocode}
      \item Let $\delta_u \leftarrow \{ -2^{\ell} d_i \mu/r \}_1$ and $M \leftarrow 2^{\ell}$.

      Let $u \leftarrow \textsc{SampleFejérKernel}(M, \, \delta_u)$ to sample~$u$ from $D_{M, \delta_u}(u)$.

      Let $k_i \leftarrow (u + \round{-2^{\ell} d_i \mu/r}) \text{ mod } 2^{\ell}$ and $\Delta_i \leftarrow \{ u \}_{2^\ell}$.
    \end{pseudocode}

    \item Return~$(j, \, k_1), \, \ldots, \, (j, \, k_c)$ and optionally~$\eta$ and~$\Delta_1, \, \ldots, \, \Delta_c$.
  \end{pseudocode}
\end{breakablealgorithm}

Analogously, the histogram-based simulator in Sect.~\ref{section:simulating-quantum-algorithm-with-failures} may be generalized to cover the extended quantum algorithm by first sampling~$\eta$ and~$j$, and by then, given~$j$ and~$\eta$, sampling~$\Delta_i$, and hence $k_i = k_{\eta, \Delta_i}$, for each $i \in [1, c] \cap \mathbb Z$, as described in Sect.~\ref{section:simulating-quantum-algorithm-with-failures} with~$d$ replaced by~$d_i$.

\subsection{Lower-bounding the success probability}
It should be noted that for~$c$ related pairs $(j, k_1), \, \ldots, \, (j, k_c)$, the probability that all~$c$ pairs are $B_\eta$-$B_\Delta$-good with respect to the same~$\eta$, with $d = d_i$ for the pair~$(j, k_i)$, is at least $\rho_1 \rho_2^c$, where $\rho = \rho_1 \rho_2$ is the lower bound in Thm.~\ref{theorem:lower-bound} for a single pair~$(j, k)$.
Similarly, the probability of recovering all $c$ logarithms $d_1, \, \ldots, \, d_c$ with the classical post-processing in Sect.~\ref{section:post-processing-shor} is at least $\rho_1 \rho_2^c - (r - \phi(r))/r$, compared to $\rho - (r - \phi(r))/r$ for a single logarithm in Cor.~\ref{corollary:invertible-correct-eta}.

This is because the pairs share~$j$ and~$\eta$:
The factor~$\rho_1$, which lower bounds the probability that $|\, \eta \,| \le B_\eta$, and the probability of $z + \eta$ not being invertible modulo~$r$, need only be accounted for once, whereas~$\rho_2$ is accounted for once for each pair, as $k_1, \, \ldots, \, k_c$ are independent given~$j$ and~$\eta$; see App.~\ref{appendix:multiple-logarithms-distribution}.
The argument is formalized in the below lemma:

\begin{applemma}
\label{lemma:multiple-lower-bound}
Let $\rho = \rho_1 \rho_2$ be the lower bound~\refeq{theorem:lower-bound} in Thm.~\ref{theorem:lower-bound}, for~$B_\eta$, $B_\Delta$ and~$\varsigma$ as in Thm.~\ref{theorem:lower-bound}.
Then, for $(j, \, k_1, \, \ldots, \, k_c)$ observed in a run of the extended quantum algorithm in App.~\ref{appendix:multiple-logarithms-algorithm}, the probability that all~$c$ pairs $(j, k_1), \, \ldots, \, (j, k_c)$ are $B_\eta$-$B_\Delta$-good with respect to the same~$\eta$, with $d = d_i$ for the pair~$(j, k_i)$, is at least~$\rho_1 \rho_2^c$.
Furthermore, the probability of recovering all~$c$ logarithms $d_1, \, \ldots, \, d_c$ given~$r$, with the classical post-processing in Sect.~\ref{section:post-processing-shor} applied to~$(j, k_i)$ for each $i \in [1, c] \cap \mathbb Z$, which searches~$\eta$ and~$t$ as in Cor.~\ref{corollary:invertible-correct-eta}, is at least $\rho_1 \rho_2^c - (r - \phi(r))/r$.
\end{applemma}
\begin{proof}
  The extended quantum algorithm may be simulated exactly and $(j, \, k_1, \, \ldots, \, k_c)$ sampled by first sampling~$\mu$ uniformly at random from $[0, r) \cap \mathbb Z$, and by then independently sampling~$j$ and $k_1, \, \ldots, \, k_c$ given~$\mu$, as explained in App.~\ref{appendix:multiple-logarithms-distribution}.
  Let $\eta = \{ \mu - z \}_r$ for $z = \round{rj / 2^{m+\varsigma}}$, and let~$\mathrm{A}$ denote the event that $|\, \eta \,| \le B_\eta$, and that $k_i = k_{\eta, \Delta_i}$ for some $\Delta_i \in \offsets(B_\Delta)$ for all $i \in [1, c] \cap \mathbb Z$, where~$k_{\eta, \Delta_i}$ is as in Sect.~\ref{section:notion-good-pair} with~$d$ replaced by~$d_i$.
  If the event~$\mathrm{A}$ occurs, then $(j, k_i)$ is a $B_\eta$-$B_\Delta$-good pair with respect to~$\eta$ for $d = d_i$ for all $i \in [1, c] \cap \mathbb Z$.

  By Lem.~\ref{lemma:multiple-joint-distribution}, the probability of sampling~$j$ and $\mu = (z + \eta) \text{ mod } r$ is $|\, A_\mu(j) \,|^2 / r = F_\eta(\theta_r)$, where distinct $\eta \in [-B_\eta, B_\eta] \cap \mathbb Z$ yield distinct~$\mu$ as $B_\eta < \ceil{r/2}$.
  Given~$j$ and~$\eta$, the frequencies $k_1, \, \ldots, \, k_c$ are furthermore independent, and $k_i = k_{\eta, \Delta}$ is sampled with probability $|\, B_{\mu, i}(k_{\eta, \Delta}) \,|^2 = h(\phi_{\eta, i}) = h(2 \pi (\Delta - \delta_{\eta, i}) / 2^{\ell})$ by Lem.~\ref{lemma:multiple-joint-distribution}, by~\refeq{phi-eta-extended} with~$d$ replaced by~$d_i$, and by the $2 \pi$-periodicity of~$h$, for~$\delta_{\eta, i}$ the value of~$\delta_\eta$ in~\refeq{delta-eta} with~$d$ replaced by~$d_i$.
  As the frequencies~$k_{\eta, \Delta}$ are distinct for distinct $\Delta \in \offsets(B_\Delta)$, it follows that
  \begin{align*}
    \Pr(\mathrm{A})
    &=
    \sum_{j \, = \, 0}^{2^{m+\varsigma} - 1}
    \sum_{\eta \, = \, -B_{\eta}}^{B_{\eta}}
    F_{\eta}(\theta_r)
    \prod_{i \, = \, 1}^{c}
    \sum_{\Delta \, \in \, \offsets(B_{\Delta})}
    \hspace{-2mm}
    h
    \left(
      \frac{2 \pi}{2^{\ell}}
      (\Delta - \delta_{\eta, i})
    \right) \\
    &\ge
    \rho_2^c
    \sum_{j \, = \, 0}^{2^{m+\varsigma} - 1}
    \sum_{\eta \, = \, -B_{\eta}}^{B_{\eta}}
    f_{\eta}(\theta_r)
    \ge
    \rho_1 \rho_2^c,
  \end{align*}
  where~$\theta_r$ and~$\delta_{\eta, i}$ are functions of~$j$.
  In the first inequality above, we have used that each sum over~$\Delta$ is at least~$\rho_2$ by Lem.~\ref{lemma:h-tails}, and that $F_\eta(\theta_r) \ge f_\eta(\theta_r)$; see the proof of Lem.~\ref{lemma:probability-distribution-truncated}.
  In the second inequality, we have used that the sum over~$j$ and~$\eta$ is at least~$\rho_1$ by Cl.~\ref{claim:alpha-r} and Lem.~\ref{lemma:fn-tails}.

  Let~$\mathrm{B}$ denote the event that $\gcd(\mu, r) \neq 1$, or, equivalently, that $\gcd(z + \eta, r) \neq 1$.
  As~$\mu$ is uniformly distributed on $[0, r) \cap \mathbb Z$, it holds that $\Pr(\mathrm{B}) = (r - \phi(r))/r$.
  If $\mathrm{A} \cap \overline{\mathrm{B}}$ occurs, then the post-processing recovers~$d_i$ from~$(j, k_i)$ given~$r$ for each $i \in [1, c] \cap \mathbb Z$, by the argument in the proof of Cor.~\ref{corollary:invertible-correct-eta} with~$d$ and~$k$ replaced by~$d_i$ and~$k_i$, respectively.
  As whether all of $d_1, \, \ldots, \, d_c$ are recovered depends only on $(j, \, k_1, \, \ldots, \, k_c)$, the probability of recovering all of $d_1, \, \ldots, \, d_c$ is at least $\Pr(\mathrm{A} \cap \overline{\mathrm{B}}) \ge \Pr(\mathrm{A}) - \Pr(\mathrm{B}) \ge \rho_1 \rho_2^c - (r - \phi(r))/r$, and so the lemma follows.
\end{proof}

\clearpage

\section{Tabulated lower bounds and expectation values}
\label{appendix:tabulated-results}
This appendix contains tabulated lower bounds and expectation values.

\subsection{Lower-bounded probabilities}
\label{appendix:tables-lower-bound}
Tbls.~\ref{table:probability-strict-varsigma-0}--\ref{table:probability-strict-Bn-0} below were created by evaluating~\refeq{theorem:lower-bound} in Thm.~\ref{theorem:lower-bound} numerically.

\begin{table}[h!]
  \begin{center}
    {\renewcommand{\arraystretch}{1.1}\setlength{\tabcolsep}{5.5pt}
    \begin{tabular}{cr|ccccccccc}
      \hline
      &  & \multicolumn{9}{c}{$B_{\Delta}$} \\
      &       &       0 &     10 &    100 &    250 &    500 &   1000 &   2500 &   5000 &  10000 \\
      \thickhline
      \multirow{10}{*}{$B_{\eta}$}
      &    0 & 0.2168 & 0.5832 & 0.5935 & 0.5942 & 0.5944 & 0.5945 & 0.5946 & 0.5946 & 0.5947 \\
      &    1 & 0.3153 & 0.8481 & 0.8631 & 0.8642 & 0.8645 & 0.8647 & 0.8648 & 0.8648 & 0.8648 \\
      &    2 & 0.3350 & 0.9011 & 0.9170 & 0.9181 & 0.9185 & 0.9187 & 0.9188 & 0.9189 & 0.9189 \\
      &   10 & 0.3575 & 0.9617 & 0.9787 & 0.9799 & 0.9803 & 0.9805 & 0.9806 & 0.9806 & 0.9806 \\
      &  100 & 0.3638 & 0.9786 & 0.9959 & 0.9971 & 0.9975 & 0.9977 & 0.9979 & 0.9979 & 0.9979 \\
      &  250 & 0.3642 & 0.9798 & 0.9971 & 0.9983 & 0.9987 & 0.9989 & 0.9991 & 0.9991 & 0.9991 \\
      &  500 & 0.3644 & 0.9802 & 0.9975 & 0.9987 & 0.9991 & 0.9993 & 0.9995 & 0.9995 & 0.9995 \\
      & 1000 & 0.3644 & 0.9804 & 0.9977 & 0.9989 & 0.9993 & 0.9995 & 0.9997 & 0.9997 & 0.9997 \\
      & 2500 & 0.3645 & 0.9805 & 0.9979 & 0.9991 & 0.9995 & 0.9997 & 0.9998 & 0.9998 & 0.9998 \\
      & 5000 & 0.3645 & 0.9806 & 0.9979 & 0.9991 & 0.9995 & 0.9997 & 0.9998 & 0.9999 & 0.9999 \\
      \hline
    \end{tabular}}
  \end{center}
  \caption{The probability of observing a $B_{\eta}$-$B_{\Delta}$-good pair, as given by the lower bound in Thm.~\ref{theorem:lower-bound}, for $\varsigma = 0$ and as $m \rightarrow \infty$ in the worst case where $r = 2^m - 1$.
  This table was created by evaluating the bound using Mathematica and rounding down.}
  \label{table:probability-strict-varsigma-0}
\end{table}

\begin{table}[h!]
  \begin{center}
    {\renewcommand{\arraystretch}{1.1}\setlength{\tabcolsep}{5.5pt}
    \begin{tabular}{cr|ccccccccc}
      \hline
      & & \multicolumn{9}{c}{$B_{\Delta}$} \\
      &                  &       0 &     10 &    100 &    250 &    500 &   1000 &   2500 &   5000 &  10000 \\
      \thickhline
      \multirow{14}{*}{$\varsigma$}
      & \hphantom{1000}0 & 0.2168 & 0.5832 & 0.5935 & 0.5942 & 0.5944 & 0.5945 & 0.5946 & 0.5946 & 0.5947 \\
      &                1 & 0.2906 & 0.7819 & 0.7957 & 0.7967 & 0.7970 & 0.7971 & 0.7972 & 0.7973 & 0.7973 \\
      &                2 & 0.3276 & 0.8813 & 0.8968 & 0.8979 & 0.8983 & 0.8984 & 0.8986 & 0.8986 & 0.8986 \\
      &                3 & 0.3460 & 0.9309 & 0.9474 & 0.9485 & 0.9489 & 0.9491 & 0.9492 & 0.9493 & 0.9493 \\
      &                4 & 0.3553 & 0.9558 & 0.9727 & 0.9738 & 0.9742 & 0.9744 & 0.9745 & 0.9746 & 0.9746 \\
      &                5 & 0.3599 & 0.9682 & 0.9853 & 0.9865 & 0.9869 & 0.9871 & 0.9872 & 0.9872 & 0.9873 \\
      &                6 & 0.3622 & 0.9744 & 0.9916 & 0.9928 & 0.9932 & 0.9934 & 0.9935 & 0.9936 & 0.9936 \\
      &                7 & 0.3633 & 0.9775 & 0.9948 & 0.9960 & 0.9964 & 0.9966 & 0.9967 & 0.9967 & 0.9968 \\
      &                8 & 0.3639 & 0.9791 & 0.9964 & 0.9976 & 0.9980 & 0.9982 & 0.9983 & 0.9983 & 0.9983 \\
      &                9 & 0.3642 & 0.9798 & 0.9971 & 0.9984 & 0.9988 & 0.9990 & 0.9991 & 0.9991 & 0.9991 \\
      &               10 & 0.3644 & 0.9802 & 0.9975 & 0.9987 & 0.9991 & 0.9994 & 0.9995 & 0.9995 & 0.9995 \\
      &               11 & 0.3644 & 0.9804 & 0.9977 & 0.9989 & 0.9993 & 0.9995 & 0.9997 & 0.9997 & 0.9997 \\
      &               12 & 0.3645 & 0.9805 & 0.9978 & 0.9990 & 0.9994 & 0.9996 & 0.9998 & 0.9998 & 0.9998 \\
      &               13 & 0.3645 & 0.9806 & 0.9979 & 0.9991 & 0.9995 & 0.9997 & 0.9998 & 0.9999 & 0.9999 \\
      \hline
    \end{tabular}}
  \end{center}
  \caption{The probability of observing a $B_{\eta}$-$B_{\Delta}$-good pair, as given by the lower bound in Thm.~\ref{theorem:lower-bound}, for $B_{\eta} = 0$ and as $m \rightarrow \infty$ in the worst case where $r = 2^m - 1$.
  This table was created by evaluating the bound using Mathematica and rounding down.}
  \label{table:probability-strict-Bn-0}
\end{table}

\clearpage

\subsection{Heuristically derived expected probabilities}
\label{appendix:tables-expectation}
Tbls.~\ref{table:probability-varsigma-0}--\ref{table:probability-Bn-0} below were created by evaluating~\refeq{Pn-int} numerically.

\begin{table}[h!]
  \begin{center}
    {\renewcommand{\arraystretch}{1.1}\setlength{\tabcolsep}{5.5pt}
    \begin{tabular}{cr|ccccccccc}
      \hline
      & & \multicolumn{9}{c}{$B_{\Delta}$} \\
      &      &      0 &      1 &      2 &     10 &    100 &    250 &    500 &   1000 &   2500 \\
      \thickhline
      \multirow{8}{*}{$B_{\eta}$}
      &    0 & 0.5986 & 0.7204 & 0.7421 & 0.7662 & 0.7729 & 0.7734 & 0.7735 & 0.7736 & 0.7737 \\
      &    1 & 0.7204 & 0.8669 & 0.8931 & 0.9221 & 0.9302 & 0.9307 & 0.9309 & 0.9310 & 0.9311 \\
      &    2 & 0.7421 & 0.8931 & 0.9200 & 0.9499 & 0.9582 & 0.9588 & 0.9590 & 0.9591 & 0.9591 \\
      &   10 & 0.7662 & 0.9221 & 0.9499 & 0.9808 & 0.9893 & 0.9899 & 0.9901 & 0.9902 & 0.9903 \\
      &  100 & 0.7729 & 0.9302 & 0.9582 & 0.9893 & 0.9980 & 0.9986 & 0.9988 & 0.9989 & 0.9990 \\
      &  250 & 0.7734 & 0.9307 & 0.9588 & 0.9899 & 0.9986 & 0.9992 & 0.9994 & 0.9995 & 0.9996 \\
      &  500 & 0.7735 & 0.9309 & 0.9590 & 0.9901 & 0.9988 & 0.9994 & 0.9996 & 0.9997 & 0.9998 \\
      & 1000 & 0.7736 & 0.9310 & 0.9591 & 0.9902 & 0.9989 & 0.9995 & 0.9997 & 0.9998 & 0.9999 \\
      \hline
    \end{tabular}}
  \end{center}
  \caption{The probability of observing a $B_{\eta}$-$B_{\Delta}$-good pair, as given by the heuristically derived expectation value in~\refeq{Pn-int}, for $\varsigma = 0$, $m = 128$, $s = 1$ and $r = 2^m-1$.
  The same table is obtained for greater~$m$.
  The choice of~$s$ has no effect as long as~$s$ is not selected very large in relation to~$m$.
  Smaller $r \in [2^{m-1}, 2^m) \cap \mathbb Z$ yield larger probabilities.
  This table was created by evaluating~\refeq{Pn-int} using Mathematica and rounding to the closest decimal.
  The symmetry is explained by~$r$ being close to~$2^m$.
  For random~$r$, see Tbl.~\ref{table:probability-varsigma-0-random-r}.}
  \label{table:probability-varsigma-0}
\end{table}

\begin{table}[h!]
  \begin{center}
    {\renewcommand{\arraystretch}{1.1}\setlength{\tabcolsep}{5.5pt}
    \begin{tabular}{cc|ccccccccc}
      \hline
      &    &      \multicolumn{9}{c}{$B_{\Delta}$} \\
      &    &      0 &      1 &      2 &     10 &    100 &    250 &    500 &   1000 &   2500 \\
      \thickhline
      \multirow{12}{*}{$\varsigma$}
      &  0 & 0.5986 & 0.7204 & 0.7421 & 0.7662 & 0.7729 & 0.7734 & 0.7735 & 0.7736 & 0.7737 \\
      &  1 & 0.6985 & 0.8406 & 0.8659 & 0.8941 & 0.9019 & 0.9025 & 0.9026 & 0.9027 & 0.9028 \\
      &  2 & 0.7350 & 0.8845 & 0.9111 & 0.9408 & 0.9490 & 0.9496 & 0.9497 & 0.9498 & 0.9499 \\
      &  3 & 0.7542 & 0.9076 & 0.9349 & 0.9653 & 0.9738 & 0.9744 & 0.9746 & 0.9746 & 0.9747 \\
      &  4 & 0.7639 & 0.9193 & 0.9470 & 0.9778 & 0.9863 & 0.9869 & 0.9871 & 0.9872 & 0.9873 \\
      &  5 & 0.7688 & 0.9252 & 0.9531 & 0.9841 & 0.9927 & 0.9933 & 0.9935 & 0.9936 & 0.9936 \\
      &  6 & 0.7712 & 0.9281 & 0.9561 & 0.9872 & 0.9958 & 0.9964 & 0.9966 & 0.9967 & 0.9968 \\
      &  7 & 0.7725 & 0.9296 & 0.9576 & 0.9888 & 0.9974 & 0.9980 & 0.9982 & 0.9983 & 0.9984 \\
      &  8 & 0.7731 & 0.9304 & 0.9584 & 0.9896 & 0.9982 & 0.9988 & 0.9990 & 0.9991 & 0.9992 \\
      &  9 & 0.7734 & 0.9307 & 0.9588 & 0.9900 & 0.9986 & 0.9992 & 0.9994 & 0.9995 & 0.9996 \\
      & 10 & 0.7735 & 0.9309 & 0.9590 & 0.9901 & 0.9988 & 0.9994 & 0.9996 & 0.9997 & 0.9998 \\
      & 11 & 0.7736 & 0.9310 & 0.9591 & 0.9902 & 0.9989 & 0.9995 & 0.9997 & 0.9998 & 0.9999 \\
      \hline
    \end{tabular}}
  \end{center}
  \caption{The probability of observing a $B_{\eta}$-$B_{\Delta}$-good pair, as given by the heuristically derived expectation value in~\refeq{Pn-int}, for $B_{\eta} = 0$, $m = 128$, $s = 1$ and $r = 2^m-1$.
  The same table is obtained for greater~$m$.
  The choice of~$s$ has no effect as long as~$s$ is not selected very large in relation to~$m$.
  Smaller $r \in [2^{m-1}, 2^m) \cap \mathbb Z$ yield larger probabilities.
  This table was created by evaluating~\refeq{Pn-int} using Mathematica and rounding to the closest~decimal.}
  \label{table:probability-Bn-0}
\end{table}

\clearpage

\subsubsection{Supplementary table for a random choice of~\texorpdfstring{$r$}{r}}
The choice of $r = 2^{m} - 1$ in Tbl.~\ref{table:probability-varsigma-0} induces symmetry.
To show that this is only an artifact of the choice of~$r$, we also include Tbl.~\ref{table:probability-varsigma-0-random-r} below that features a random choice of~$r$.

\begin{table}[h!]
  \begin{center}
    {\renewcommand{\arraystretch}{1.1}\setlength{\tabcolsep}{5.5pt}
    \begin{tabular}{cr|ccccccccc}
      \hline
      & & \multicolumn{9}{c}{$B_{\Delta}$} \\
      &      &      0 &      1 &      2 &     10 &    100 &    250 &    500 &   1000 &   2500 \\
      \thickhline
      \multirow{8}{*}{$B_{\eta}$}
      &    0 & 0.6841 & 0.8232 & 0.8480 & 0.8756 & 0.8833 & 0.8838 & 0.8840 & 0.8841 & 0.8841 \\
      &    1 & 0.7356 & 0.8852 & 0.9119 & 0.9416 & 0.9498 & 0.9504 & 0.9505 & 0.9506 & 0.9507 \\
      &    2 & 0.7528 & 0.9059 & 0.9332 & 0.9635 & 0.9720 & 0.9725 & 0.9727 & 0.9728 & 0.9729 \\
      &   10 & 0.7685 & 0.9248 & 0.9527 & 0.9837 & 0.9923 & 0.9929 & 0.9931 & 0.9932 & 0.9932 \\
      &  100 & 0.7732 & 0.9304 & 0.9585 & 0.9897 & 0.9983 & 0.9989 & 0.9991 & 0.9992 & 0.9993 \\
      &  250 & 0.7735 & 0.9308 & 0.9589 & 0.9901 & 0.9987 & 0.9993 & 0.9995 & 0.9996 & 0.9997 \\
      &  500 & 0.7736 & 0.9310 & 0.9590 & 0.9902 & 0.9989 & 0.9995 & 0.9997 & 0.9998 & 0.9998 \\
      & 1000 & 0.7736 & 0.9310 & 0.9591 & 0.9903 & 0.9989 & 0.9995 & 0.9997 & 0.9998 & 0.9999 \\
      \hline
    \end{tabular}}
  \end{center}
  \caption{The probability of observing a $B_{\eta}$-$B_{\Delta}$-good pair, as given by the heuristically derived expectation value in~\refeq{Pn-int}, for $\varsigma = 0$, $m = 128$, $s = 1$ and a random choice of the order $r = 234176320093007559271185988522878687746 \in [2^{m-1}, 2^m) \cap \mathbb Z$.
  This table was created by evaluating~\refeq{Pn-int} using Mathematica and rounding to the closest decimal.}
  \label{table:probability-varsigma-0-random-r}
\end{table}

\clearpage

\section{Supporting lemmas, claims and proofs}
\label{appendix:supporting-lemmas-claims-proofs}

This appendix contains proofs of supporting lemmas and claims.
It furthermore contains the proof of Cor.~\ref{corollary:lower-bound-comparison-mosca} which is stated without a proof in the main body of the paper.

\subsection{The \texorpdfstring{$h$}{h}-function}
\begin{applemma}
  \label{lemma:h-sum-to-one}
  For any real~$\delta$, it holds that
  \begin{align*}
    \sum_{\Delta \, \in \, \offsets(2^{\ell-1})}
    \hspace{-2mm}
    h
    \left(
    \frac{2 \pi} {2^\ell}
    (\Delta - \delta)
    \right) = 1.
  \end{align*}
\end{applemma}
\begin{proof}
If $\ell = 0$, then $\offsets(2^{\ell-1}) = \{ 0 \}$ and $h(\phi) = 1$ for any real~$\phi$, so the lemma holds trivially.

Suppose therefore that $\ell \ge 1$.
It then holds that
\begin{align*}
  \sum_{\Delta \, \in \, \offsets(2^{\ell-1})}
  \hspace{-2mm}
  h
  \left(
  \frac{2 \pi} {2^\ell}
  (\Delta - \delta)
  \right)
  &=
  \frac{1}{2^{2 \ell}}
  \sum_{\Delta \, = \, -2^{\ell - 1}}^{2^{\ell - 1} - 1}
  \left|\,
  \sum_{b \, = \, 0}^{2^\ell - 1}
  \e^{2 \pi \imag (\Delta - \delta) b / 2^\ell}
  \,\right|^2 \\
  &=
  \frac{1}{2^{2 \ell}}
  \sum_{\Delta \, = \, 0}^{2^{\ell} - 1}
  \left|\,
  \sum_{b \, = \, 0}^{2^\ell - 1}
  \e^{2 \pi \imag (\Delta - 2^{\ell-1} - \delta) b / 2^\ell}
  \,\right|^2 \\
  &=
  \frac{1}{2^{2 \ell}}
  \sum_{\Delta \, = \, 0}^{2^{\ell} - 1}
  \left|\,
  \sum_{b \, = \, 0}^{2^\ell - 1}
  \e^{\imag (2 \pi \Delta / 2^\ell + \varphi) b}
  \,\right|^2
  = 1
\end{align*}
where we have used Cl.~\ref{claim:exp-double-sum-to-one} with $\varphi = -\pi (1 + \frac{\delta}{2^{\ell-1}})$, and so the lemma follows.
\end{proof}

\subsubsection{Supporting claims}
\begin{appclaim}
  \label{claim:exp-double-sum-to-one}
  For any positive integer~$N$ and any real~$\varphi$, it holds that
  \begin{align*}
    \frac{1}{N^2}
    \sum_{j \, = \, 0}^{N - 1}
    \left|\,
      \sum_{a \, = \, 0}^{N - 1}
      \e^{\imag (2 \pi j / N + \varphi) a}
    \,\right|^2
    =
    1.
  \end{align*}
\end{appclaim}
\begin{proof}
This claim is a variation of~\cite[Lem.~C.1 in App.~C.2]{ekera-general}:
Specifically, it holds that
\begin{align*}
  \left|\,
  \sum_{a \, = \, 0}^{N - 1}
  \e^{\imag (2 \pi j / N + \varphi) a}
  \,\right|^2
  &=
  \left(\:
    \sum_{a \, = \, 0}^{N - 1}
    \e^{\imag (2 \pi j / N + \varphi) a}
  \right)
  \left(\:
    \sum_{a \, = \, 0}^{N - 1}
    \e^{-\imag (2 \pi j / N + \varphi) a}
  \right) \\
  &=
  \sum_{a \,= \, -(N-1)}^{N - 1}
  (N - |\, a \,|) \,
  \e^{\imag (2 \pi j / N + \varphi) a}
  \\
  &=
  N
  +
  \sum_{a \, = \, 1}^{N - 1}
  (N - a)
  \left(
    \e^{\imag (2 \pi j / N + \varphi) a}
    +
    \e^{-\imag (2 \pi j / N + \varphi) a}
  \right)
\end{align*}
from which it follows that
\begin{align*}
  &
  \frac{1}{N^2}
  \sum_{j \, = \, 0}^{N-1}
  \left|\,
  \sum_{a \, = \, 0}^{N - 1}
  \e^{\imag (2 \pi j / N + \varphi) a}
  \,\right|^2
  = \\
  &\quad\quad\quad
  1
  +
  \frac{1}{N^2}
  \sum_{a \, = \, 1}^{N - 1}
  (N - a)
  \underbrace{\:\sum_{j \, = \, 0}^{N - 1}
  \left(
    \e^{\imag (2 \pi j / N + \varphi) a}
    +
    \e^{-\imag (2 \pi j / N + \varphi) a}
  \right)}_{= \, 0}
  = 1,
\end{align*}
and so the claim follows.
\end{proof}

\subsubsection{Bounding the tails}
\begin{applemma}
  \label{lemma:h-tails}
  For any~$\delta_\eta \in [-\frac{1}{2}, \frac{1}{2})$ and $B_{\Delta} \in [0, 2^{\ell-1}] \cap \mathbb Z$, it holds that
  \begin{align*}
    \sum_{\Delta \, \in \, \offsets(B_{\Delta})}
    \hspace{-2mm}
    h
    \left(
    \frac{2 \pi} {2^\ell}
    (\Delta - \delta_\eta)
    \right)
    &\ge
    1
    -
    \frac{1}{\pi^2}
    \left(
      \psi'(B_{\Delta} + 1/2)
      +
      \psi'(B_{\Delta} + 3/2)
    \right) \\
    &>
    1
    -
    \frac{1}{\pi^2}
    \left(
      \frac{1}{B_{\Delta} + 1/2}
      (
        1
        +
        \epsilon_{B_{\Delta}}(1/2)
      )
      +
      \frac{1}{B_{\Delta} + 3/2}
      (
        1
        +
        \epsilon_{B_{\Delta}}(3/2)
      )
    \right)
  \end{align*}
  for~$\psi'$ the trigamma function and
  \begin{align*}
    \epsilon_{B_{\Delta}}(x) = \epsilon(B_{\Delta} + x)
    \quad \text{ where } \quad
    \epsilon(x) = \frac{1}{2x} + \frac{1}{6x^2}.
  \end{align*}
\end{applemma}
\begin{proof}
  If $B_{\Delta} = 2^{\ell-1}$, then the sum over~$h$ is exactly one by Lem.~\ref{lemma:h-sum-to-one}, and so the lemma follows.
  Suppose therefore that $B_{\Delta} < 2^{\ell-1}$, in which case $\offsets(B_{\Delta}) = [-B_{\Delta}, B_{\Delta}] \cap \mathbb Z$.

  Furthermore, the lemma trivially holds for $\delta_\eta = 0$, since $h(0) = 1$ whereas $h(2 \pi \Delta / 2^\ell) = 0$ for all $\Delta \in ([-B_\Delta, B_\Delta] \cap \mathbb Z) \, \backslash \, \{ 0 \}$, i.e., the sum over~$h$ is exactly one also when $\delta_\eta = 0$.
  Suppose therefore that $\delta_\eta \neq 0$.
  For simplicity, let $u = \Delta - \delta_\eta$.
  As $u \not\in \mathbb Z$, it holds that
  \begin{align*}
    h
    \left(
      \frac{2 \pi} {2^\ell} \, u
    \right)
    &=
    \frac{1}{2^{2 \ell}}
    \frac{1 - \cos(2 \pi u)}{1 - \cos(2 \pi u / 2^\ell)}
    =
    \frac{1}{2^{2 \ell}}
    \frac{\sin^2(\pi u)}{\sin^2(\pi u / 2^\ell)}
    \ge
    \frac{\sin^2(\pi u)}{(\pi u)^2}
  \end{align*}
  since $1 - \cos(\varphi) = 2 \sin^2(\varphi / 2)$, and $\sin^2(\varphi) \le \varphi^2$ for all $\varphi \in \mathbb R$.
  It follows that
  \begin{align}
    \sum_{\Delta \, \in \, \offsets(B_\Delta)}
    \hspace{-2mm}
    h
    \left(
      \frac{2 \pi} {2^\ell}
      (\Delta - \delta_\eta)
    \right)
    &=
    \sum_{\Delta \, = \, -B_{\Delta}}^{B_{\Delta}}
    \hspace{-1mm}
    h
    \left(
      \frac{2 \pi} {2^\ell}
      (\Delta - \delta_\eta)
    \right)
    \ge
    \sum_{\Delta \, = \, -B_{\Delta}}^{B_{\Delta}}
    \frac{\sin^2(\pi (\Delta - \delta_\eta))}{(\pi (\Delta - \delta_\eta))^2} \notag \\
    &=
    1
    -
    \sum_{\Delta \, = \, -\infty}^{-B_{\Delta}-1}
    \frac{\sin^2(\pi (\Delta - \delta_\eta))}{(\pi (\Delta - \delta_\eta))^2}
    -
    \sum_{\Delta \, = \, B_{\Delta} + 1}^{\infty}
    \frac{\sin^2(\pi (\Delta - \delta_\eta))}{(\pi (\Delta - \delta_\eta))^2} \label{eq:h-tails-reverse} \\
    &\ge
    1
    -
    \sum_{\Delta \, = \, -\infty}^{-B_{\Delta}-1}
    \frac{1}{(\pi (\Delta - \delta_\eta))^2}
    -
    \sum_{\Delta \, = \, B_{\Delta} + 1}^{\infty}
    \frac{1}{(\pi (\Delta - \delta_\eta))^2}
    \label{eq:h-tails-one} \\
    &=
    1
    -
    \frac{1}{\pi^2}
    \sum_{\Delta \, = \, B_{\Delta} + 1}^{\infty}
    \left(\:
      \frac{1}{(-\Delta - \delta_\eta)^2}
      +
      \frac{1}{(\Delta - \delta_\eta)^2}
    \right) \notag \\
    &=
    1
    -
    \frac{1}{\pi^2}
    \sum_{\Delta \, = \, 0}^{\infty}
    \left(\:
      \frac{1}{(-(\Delta + B_{\Delta} + 1) - \delta_\eta)^2}
      +
      \frac{1}{( (\Delta + B_{\Delta} + 1) - \delta_\eta)^2}
    \right) \notag \\
    &=
    1
    -
    \frac{1}{\pi^2}
    \sum_{\Delta \, = \, 0}^{\infty}
    \left(\:
      \frac{1}{(\Delta + (B_{\Delta} + 1 + \delta_\eta))^2}
      +
      \frac{1}{( \Delta + (B_{\Delta} + 1 - \delta_\eta))^2}
    \right) \notag \\
    &=
    1
    -
    \frac{1}{\pi^2}
    \left(
      \psi'(B_\Delta + 1 + \delta_\eta)
      +
      \psi'(B_\Delta + 1 - \delta_\eta)
    \right) \label{eq:h-tails-psi-delta} \\
    &\ge
    1
    -
    \frac{1}{\pi^2}
    \left(
      \psi'(B_\Delta + 1/2)
      +
      \psi'(B_\Delta + 3/2)
    \right) \label{eq:h-tails-psi-delta2} \\
    &>
    1
    -
    \frac{1}{\pi^2}
    \left(
      \frac{1}{B_{\Delta} + 1/2}
      (
        1
        +
        \epsilon_{B_{\Delta}}(1/2)
      )
      +
      \frac{1}{B_{\Delta} + 3/2}
      (
        1
        +
        \epsilon_{B_{\Delta}}(3/2)
      )
    \right) \notag
  \end{align}
  where we have used Cl.~\ref{claim:h-tails-sums-to-one} in~\refeq{h-tails-reverse}, and that $\sin^2(\pi (\Delta - \delta_\eta)) \le 1$ in~\refeq{h-tails-one}.
  Furthermore, in~\refeq{h-tails-psi-delta}, we have used Cl.~\ref{claim:bound-psi} with $x = B_\Delta + 1 \pm \delta_\eta > 0$ since $B_\Delta \ge 0$ and $\delta_\eta \in [-\frac{1}{2}, \frac{1}{2})$, and, in~\refeq{h-tails-psi-delta2}, that~\refeq{h-tails-psi-delta} is minimized for $\delta_\eta = -\frac{1}{2}$ (as~\refeq{h-tails-psi-delta} is even, and decreasing in $|\, \delta_\eta \,|$ since $\psi'$ is convex).
  In the final step, we have again used Cl.~\ref{claim:bound-psi}, and so the lemma follows.
\end{proof}

\subsubsection{Supporting claims}
\begin{appclaim}
  \label{claim:h-tails-sums-to-one}
  For any real $\delta \notin \mathbb Z$, it holds that
  \begin{align*}
    \sum_{\Delta \, = \, -\infty}^{\infty}
    \frac{\sin^2(\pi (\Delta - \delta))}{(\pi (\Delta - \delta))^2}
    =
    1.
  \end{align*}
\end{appclaim}
\begin{proof}
  Consider the $2\pi$-periodic function $\xi(x) = \e^{\imag \delta x}$, where $x \in [-\pi, \pi)$, extended to $\mathbb R$.

  Its Fourier coefficients are
  \begin{align*}
    \hat \xi_\Delta(\delta)
    &=
    \frac{1}{2 \pi}
    \int_{-\pi}^{\pi}
    \e^{\imag \delta x} \, \e^{-\imag \Delta x}
    \,
    \mathrm{d}x
    =
    \frac{1}{2 \pi}
    \left[
      \frac{\imag \, \e^{-\imag (\Delta - \delta) x}}{\Delta - \delta}
    \right]_{-\pi}^{\pi}
    =
    \frac{\sin(\pi (\Delta - \delta))}{\pi (\Delta - \delta)}
  \end{align*}
  as $\Delta - \delta \neq 0$ for any $\Delta \in \mathbb Z$ since $\delta \not\in \mathbb Z$.
  By Parseval's theorem, it thus holds that
  \begin{align*}
    \sum_{\Delta \, = \, -\infty}^{\infty}
    \frac{\sin^2(\pi (\Delta - \delta))}{(\pi (\Delta - \delta))^2}
    &=
    \sum_{\Delta \, = \, -\infty}^{\infty}
    \left|\,
    \hat \xi_\Delta(\delta)
    \,\right|^2
    =
    \frac{1}{2 \pi}
    \int_{-\pi}^{\pi}
    \left|\, \xi(x) \,\right|^2
    \,
    \mathrm{d}x
    =
    \frac{1}{2 \pi}
    \int_{-\pi}^{\pi}
    \left|\, \e^{\imag \delta x} \,\right|^2
    \,
    \mathrm{d}x
    =
    1,
  \end{align*}
  and so the claim follows.
\end{proof}

\begin{appclaim}
  \label{claim:bound-psi}
  For any real $x > 0$ and~$\psi'(x)$ the trigamma function, it holds that
  \begin{align*}
    \psi'(x)
    &=
    \sum_{\Delta \, = \, 0}^{\infty} \frac{1}{(x + \Delta)^{2}}
    < \frac{1}{x} + \frac{1}{2 x^2} + \frac{1}{6 x^3}.
  \end{align*}
\end{appclaim}
\begin{proof}
  This claim is from~\cite{ekera-success-order-finding} via~\cite{nemes}:
  See~\cite[Cl.~3.2 in App.~D.2]{ekera-success-order-finding} for the proof of the upper bound, and, e.g., \cite[Eq.~6.4.10 on p.~260]{abramowitz-stegun} for the series representation of~$\psi'(x)$.
\end{proof}

\subsection{The \texorpdfstring{$f_\eta$}{f\_eta}-function}
\begin{applemma}
  \label{lemma:fn-sum-to-one}
  It holds that
  \begin{align*}
    \lim_{\substack{B_{\eta} \rightarrow \infty \\ \text{ and/or } \\ \varsigma \rightarrow \infty}}
    \sum_{\eta \, = \, -B_{\eta}}^{B_{\eta}} \:
    \sum_{\alpha_r' \, = \, -2^{m+\varsigma-\kappa_r-1}}^{2^{m+\varsigma-\kappa_r-1} - 1}
    2^{\kappa_r}
    f_{\eta}
    \left(
      \frac{2\pi}{2^{m+\varsigma}} \, 2^{\kappa_r} \alpha'_r
    \right)
    =
    1.
  \end{align*}
\end{applemma}
\begin{proof}
Suppose that $2^{\kappa_r} \alpha'_r - 2^{m+\varsigma} \eta \neq 0$.
It then holds that
\begin{align}
  f_{\eta}
  \left(
    \frac{2\pi}{2^{m+\varsigma}} \, 2^{\kappa_r} \alpha'_r
  \right)
  &=
  \frac{r}{2^{2(m+\varsigma)}}
  \frac{2(1 - \cos(2\pi (2^{\kappa_r} \alpha'_r - 2^{m+\varsigma} \eta) / r))}{(2\pi (2^{\kappa_r} \alpha'_r / 2^{m+\varsigma} - \eta))^2} \notag \\
  &=
  \frac{1}{r}
  \frac{2(1 - \cos(2\pi (2^{\kappa_r} \alpha'_r - 2^{m+\varsigma} \eta) / r))}{(2\pi (2^{\kappa_r} \alpha'_r / 2^{m+\varsigma} - \eta)(2^{m+\varsigma} / r))^2} \notag \\
  &=
  \frac{1}{r}
  \frac{2(1 - \cos(2\pi (2^{\kappa_r} \alpha'_r - 2^{m+\varsigma} \eta) / r))}{(2\pi (2^{\kappa_r} \alpha'_r - 2^{m+\varsigma} \eta) / r)^2} \notag \\
  &=
  \frac{1}{r}
  \frac{\sin^2(\pi (2^{\kappa_r} \alpha'_r - 2^{m+\varsigma} \eta) / r)}{(\pi (2^{\kappa_r} \alpha'_r - 2^{m+\varsigma} \eta) / r)^2}, \label{eq:simplify-fn}
\end{align}
and furthermore, since $\alpha'_r \in [-2^{m + \varsigma - \kappa_r - 1}, 2^{m + \varsigma - \kappa_r - 1}) \cap \mathbb Z$ and $\eta \in \mathbb Z$, and
\begin{align*}
  f_0(0) = \frac{1}{r}
  =
  \lim_{\alpha'_r \rightarrow 0}
  \frac{1}{r}
  \frac{\sin^2(2^{\kappa_r} \pi \alpha'_r / r)}{(2^{\kappa_r} \pi \alpha'_r / r)^2},
\end{align*}
we have that~\refeq{simplify-fn} also holds in the limit as $2^{\kappa_r} \alpha'_r - 2^{m+\varsigma} \eta \rightarrow 0$.
It follows that
\begin{align*}
  &\phantom{=} \:
  \lim_{\substack{B_{\eta} \rightarrow \infty \\ \text{ and/or } \\ \varsigma \rightarrow \infty}}
  \sum_{\eta \, = \, -B_{\eta}}^{B_{\eta}} \:
  \sum_{\alpha_r' \, = \, -2^{m+\varsigma-\kappa_r-1}}^{2^{m+\varsigma-\kappa_r-1} - 1}
  2^{\kappa_r}
  f_{\eta}
  \left(
    \frac{2\pi}{2^{m+\varsigma}} \, 2^{\kappa_r} \alpha'_r
  \right) \\
  &=
  \lim_{\substack{B_{\eta} \rightarrow \infty \\ \text{ and/or } \\ \varsigma \rightarrow \infty}}
  \frac{2^{\kappa_r}}{r}
  \sum_{\eta \, = \, -B_{\eta}}^{B_{\eta}} \:
  \sum_{\alpha_r' \, = \, -2^{m+\varsigma-\kappa_r-1}}^{2^{m+\varsigma-\kappa_r-1} - 1}
  \frac{\sin^2(\pi (2^{\kappa_r} \alpha'_r - 2^{m+\varsigma} \eta) / r)}{(\pi (2^{\kappa_r} \alpha'_r - 2^{m+\varsigma} \eta) / r)^2} \\
  &=
  \lim_{\substack{B_{\eta} \rightarrow \infty \\ \text{ and/or } \\ \varsigma \rightarrow \infty}}
  \frac{1}{r'}
  \sum_{u \, = \, -2^{m+\varsigma-\kappa_r} (B_{\eta} + 1/2)}^{2^{m+\varsigma-\kappa_r} (B_{\eta} + 1/2) - 1}
  \frac{\sin^2(\pi u / r')}{(\pi u / r')^2}
  =
  \frac{1}{r'}
  \sum_{u \, = \, -\infty}^{\infty}
  \frac{\sin^2(\pi u / r')}{(\pi u / r')^2}
  =
  1,
\end{align*}
where we have introduced $r' = r / 2^{\kappa_r}$, and used Cl.~\ref{claim:sin2u-u2-sums-to-one}, and so the lemma follows.
\end{proof}

\subsubsection{Supporting claims}
\begin{appclaim}
  \label{claim:sin2u-u2-sums-to-one}
  For any positive integer~$r'$, it holds that
  \begin{align*}
    \sum_{u \, = \, -\infty}^{\infty}
    \frac{\sin^2(\pi u / r')}{(\pi u / r')^2}
    =
    r'.
  \end{align*}
\end{appclaim}
\begin{proof}
  By Parseval's theorem, it holds that
  \begin{align*}
    \sum_{u \, = \, -\infty}^{\infty}
    |\, \hat \xi_u \,|^2
    =
    \frac{1}{2 \pi}
    \int_{-\pi}^{\pi}
    \left|\, \xi(x) \,\right|^2
    \,
    \mathrm{d}x
  \end{align*}
  for $\xi(x): \mathbb R \rightarrow \mathbb C$ a $2\pi$-periodic function, and~$\hat \xi_u$ its Fourier coefficients such that
  \begin{align*}
    \xi(x)
    =
    \sum_{u \, = \, -\infty}^{\infty}
    \hat \xi_u \, \e^{\imag u x}
    \quad \text{ where } \quad
    \hat \xi_u
    =
    \frac{1}{2 \pi}
    \int_{-\pi}^{\pi}
    \xi(x) \, \e^{-\imag u x}
    \,
    \mathrm{d}x.
  \end{align*}

  For
  \begin{align*}
    \xi(x) =
    \left\{
      \begin{array}{cc}
        r' & x \in [-\pi / r', \pi / r') \\
        0 & \text{otherwise}
      \end{array}
    \right.
  \end{align*}
  it then holds that
  \begin{align*}
    \hat \xi_u
    &=
    \frac{1}{2 \pi}
    \int_{-\pi}^{\pi}
    \xi(x) \,
    \e^{-\imag u x}
    \,
    \mathrm{d}x
    =
    \frac{1}{2 \pi}
    \int_{-\pi / r'}^{\pi / r'}
    r' \, \e^{-\imag u x}
    \,
    \mathrm{d}x
    =
    \frac{r'}{2 \pi}
    \left[
      \frac{\imag \, \e^{-\imag u x}}{u}
    \right]_{-\pi / r'}^{\pi / r'}
    =
    \frac{\sin(\pi u / r')}{\pi u / r'}
  \end{align*}
  when $u \neq 0$, and furthermore
  \begin{align*}
    \hat \xi_0
    &=
    \frac{1}{2 \pi}
    \int_{-\pi}^{\pi}
    \xi(x)
    \,
    \mathrm{d}x
    =
    \frac{1}{2 \pi}
    \int_{-\pi / r'}^{\pi / r'}
    r'
    \,
    \mathrm{d}x
    =
    1
    =
    \lim_{u \rightarrow 0}
    \frac{\sin(\pi u / r')}{\pi u / r'}.
  \end{align*}

  It follows that
  \begin{align*}
    \sum_{u \, = \, -\infty}^{\infty}
    \frac{\sin^2(\pi u / r')}{(\pi u / r')^2}
    =
    \sum_{u \, = \, -\infty}^{\infty}
    |\, \hat \xi_u \,|^2
    &=
    \frac{1}{2 \pi}
    \int_{-\pi}^{\pi}
    \left|\, \xi(x) \,\right|^2
    \,
    \mathrm{d}x
    =
    \frac{1}{2 \pi}
    \int_{-\pi / r'}^{\pi / r'}
    \, (r')^2
    \,
    \mathrm{d}x
    =
    r',
  \end{align*}
  and so the claim follows.
\end{proof}

\subsubsection{Bounding the tails}
\begin{applemma}
  \label{lemma:fn-tails}
  It holds that
  \begin{align*}
    \sum_{\eta \, = \, -B_{\eta}}^{B_{\eta}} \:
    \sum_{\alpha_r' \, = \, -2^{m+\varsigma-\kappa_r-1}}^{2^{m+\varsigma-\kappa_r-1} - 1}
    2^{\kappa_r}
    f_{\eta}
    \left(
      \frac{2\pi}{2^{m+\varsigma}} \, 2^{\kappa_r} \alpha'_r
    \right)
    >
    1
    -
    \frac{2}{\pi^2}
    \frac{r}{2^{m}}
    \frac{1}{2^{\varsigma} (B_{\eta} + 1/2)}
    \left(
      1
      +
      \epsilon_{B_{\eta}}
    \right)
  \end{align*}
  for~$B_{\eta}$ and~$\varsigma$ non-negative integers, and for
  \begin{align*}
    \epsilon_{B_{\eta}} = \epsilon(2^{m+\varsigma-\kappa_r} (B_{\eta} + 1/2))
    \quad \text{ where } \quad
    \epsilon(x) = \frac{1}{2x} + \frac{1}{6x^2}.
  \end{align*}
\end{applemma}
\begin{proof}
  It holds that
  \begin{align*}
    T_{+}
    &=
    \sum_{\eta \, = \, -\infty}^{-B_{\eta} - 1} \:
    \sum_{\alpha_r' \, = \, -2^{m+\varsigma-\kappa_r-1}}^{2^{m+\varsigma-\kappa_r-1} - 1}
    2^{\kappa_r}
    f_{\eta}
    \left(
      \frac{2\pi}{2^{m+\varsigma}} \, 2^{\kappa_r} \alpha'_r
    \right) \\
    &=
    \frac{2^{\kappa_r}}{r}
    \sum_{\eta \, = \, -\infty}^{-B_{\eta} - 1} \:
    \sum_{\alpha_r' \, = \, -2^{m+\varsigma-\kappa_r-1}}^{2^{m+\varsigma-\kappa_r-1} - 1}
    \frac{\sin^2(\pi (2^{\kappa_r} \alpha'_r - 2^{m+\varsigma} \eta) / r)}{(\pi (2^{\kappa_r} \alpha'_r - 2^{m+\varsigma} \eta) / r)^2} \\
    &=
    \frac{2^{\kappa_r}}{r}
    \sum_{u \, = \, 2^{m+\varsigma-\kappa_r} (B_{\eta} + 1/2)}^{\infty}
    \frac{\sin^2(2^{\kappa_r} \pi u / r)}{(2^{\kappa_r} \pi u / r)^2}
    \le
    \frac{2^{\kappa_r}}{r}
    \frac{r^2}{(2^{\kappa_r} \pi)^2}
    \sum_{u \, = \, 2^{m+\varsigma-\kappa_r} (B_{\eta} + 1/2)}^{\infty}
    \frac{1}{u^2} \\
    &=
    \frac{r}{2^{\kappa_r} \pi^2}
    \sum_{u \, = \, 2^{m+\varsigma-\kappa_r} (B_{\eta} + 1/2)}^{\infty}
    \frac{1}{u^2}
    =
    \frac{r}{2^{\kappa_r} \pi^2}
    \, \psi'(2^{m+\varsigma-\kappa_r} (B_{\eta} + 1/2))
  \end{align*}
  for~$\psi'$ the trigamma function, and where we have used~\refeq{simplify-fn}.
  Analogously, it holds that
  \begin{align*}
    T_{-}
    &=
    \sum_{\eta \, = \, B_{\eta} + 1}^{\infty} \:
    \sum_{\alpha_r' \, = \, -2^{m+\varsigma-\kappa_r-1}}^{2^{m+\varsigma-\kappa_r-1} - 1}
    2^{\kappa_r}
    f_{\eta}
    \left(
      \frac{2\pi}{2^{m+\varsigma}} \, 2^{\kappa_r} \alpha'_r
    \right) \\
    &\le
    \frac{2^{\kappa_r}}{r}
    \sum_{u \, = \, -\infty}^{-2^{m+\varsigma-\kappa_r} (B_{\eta} + 1/2)}
    \frac{\sin^2(2^{\kappa_r} \pi u / r)}{(2^{\kappa_r} \pi u / r)^2}
    \le
    \frac{2^{\kappa_r}}{r}
    \frac{r^2}{(2^{\kappa_r} \pi)^2}
    \sum_{u \, = \, -\infty}^{-2^{m+\varsigma-\kappa_r} (B_{\eta} + 1/2)}
    \frac{1}{u^2} \\
    &=
    \frac{r}{2^{\kappa_r} \pi^2}
    \sum_{u \, = \, -\infty}^{-2^{m+\varsigma-\kappa_r} (B_{\eta} + 1/2)}
    \frac{1}{u^2}
    =
    \frac{r}{2^{\kappa_r} \pi^2}
    \, \psi'(2^{m+\varsigma-\kappa_r} (B_{\eta} + 1/2)).
  \end{align*}

  It follows from Lem.~\ref{lemma:fn-sum-to-one} that
  \begin{align*}
    1
    -
    T_{+}
    -
    T_{-}
    &=
    \sum_{\eta \, = \, -B_{\eta}}^{B_{\eta}} \:
    \sum_{\alpha_r' \, = \, -2^{m+\varsigma-\kappa_r-1}}^{2^{m+\varsigma-\kappa_r-1} - 1}
    2^{\kappa_r}
    f_{\eta}
    \left(
      \frac{2\pi}{2^{m+\varsigma}} \, 2^{\kappa_r} \alpha'_r
    \right) \\
    &\ge
    1
    -
    \frac{2r}{2^{\kappa_r} \pi^2}
    \, \psi'(2^{m+\varsigma-\kappa_r} (B_{\eta} + 1/2)) \\
    &>
    1 - \frac{2r}{2^{\kappa_r} \pi^2}
    \bigg(
      \frac{1}{2^{m+\varsigma-\kappa_r} (B_{\eta} + 1/2)}
      +
      \frac{1}{2 \cdot (2^{m+\varsigma-\kappa_r} (B_{\eta} + 1/2))^2} \\
      &\quad\quad\quad +
      \frac{1}{6 \cdot (2^{m+\varsigma-\kappa_r} (B_{\eta} + 1/2))^3}
    \bigg) \\
    &=
    1
    -
    \frac{2}{\pi^2}
    \frac{r}{2^{m}}
    \frac{1}{2^{\varsigma} (B_{\eta} + 1/2)}
    \left(
      1
      +
      \epsilon_{B_{\eta}}
    \right)
  \end{align*}
  where we have used Cl.~\ref{claim:bound-psi} to bound~$\psi'$, and so the lemma follows.
\end{proof}

\subsection{Invertibility in the classical post-processing}
\label{appendix:invertibility-classical-post-processing}
\begin{appclaim}
  \label{claim:r-minus-phi-r-bound}
  Suppose that no prime factor less than~$B_r$ divides~$r$, for $B_r \ge 2$ a bound.
  Then
  \begin{align*}
    r - \phi(r) \le \frac{r \log_2(r)}{B_r \log_2(B_r)}
  \end{align*}
  for~$\phi$ Euler's totient function.
\end{appclaim}
\begin{proof}
  It holds that $r - \phi(r)$ is the number of integers $u \in [0, r) \cap \mathbb Z$ not coprime to~$r$.

  There are at most $\log_2(r) / \log_2(B_r)$ distinct prime factors that divide~$r$.
  As~$u$ runs over all~$r$ integers in $[0, r) \cap \mathbb Z$, the probability that each distinct prime factor~$q$ that divides~$r$ also divides~$u$ is at most $1/q \le 1/B_r$.
  It follows that at most $r \log_2(r) / (B_r \log_2(B_r))$ values of~$u$ are divisible by at least one prime factor that also divides~$r$, and so the claim follows.
\end{proof}

\subsection{Comparison to Mosca's lower bound}
\label{appendix:comparison-mosca}

\corollarycomparisonmosca*
\begin{proof}
  Recall that by~\refeq{theorem:lower-bound} in Thm.~\ref{theorem:lower-bound} (which applies since~$2^{\kappa_r}$ divides $r < 2^m$, so $\kappa_r < m < m + \varsigma = \ell$), the probability of observing a $B_{\eta}$-$B_{\Delta}$-good pair is lower bounded by
  \begin{align*}
  &\max\left(
    0,
    1
    -
    \frac{2}{\pi^2}
    \frac{r}{2^{m}}
    \frac{1}{2^{\varsigma} (B_{\eta} + 1/2)}
    \left(
      1
      +
      \epsilon_{B_{\eta}}
    \right)
  \right) \cdot \\
  &\quad\quad
  \max\left(
    0,
    1
    -
    \frac{1}{\pi^2}
    \left(
      \frac{1}{B_{\Delta} + 1/2}
      (
        1
        +
        \epsilon_{B_{\Delta}}(1/2)
      )
      +
      \frac{1}{B_{\Delta} + 3/2}
      (
        1
        +
        \epsilon_{B_{\Delta}}(3/2)
      )
    \right)
  \right),
  \end{align*}
  for~$\varsigma$ a non-negative integer, for $B_{\eta} \in [0, \ceil{r/2}) \cap \mathbb Z$ and $B_{\Delta} \in [0, 2^{\ell-1}] \cap \mathbb Z$, and for
  \begin{align*}
    \epsilon_{B_{\Delta}}(x) = \epsilon(B_{\Delta} + x)
    \quad \text{ and } \quad
    \epsilon_{B_{\eta}} = \epsilon(2^{m+\varsigma-\kappa_r} (B_{\eta} + 1/2))
    \quad \text{ where } \quad
    \epsilon(x) = \frac{1}{2x} + \frac{1}{6x^2}.
  \end{align*}

  For $B_{\eta} = 0$ and $\varsigma = 2$, it holds that $\epsilon_{B_{\eta}} = \epsilon(2^{m+1-\kappa_r}) = \epsilon(2w)$ for $w = 2^{m-\kappa_r}$.
  As $r < 2^m$ and~$2^{\kappa_r}$ divides~$r$, it furthermore holds that $r \le 2^m - 2^{\kappa_r}$, and hence that $r / 2^m \le 1 - 1/w$.

  It follows for the first factor in~\refeq{theorem:lower-bound} that
  \begin{align*}
    1
    -
    \frac{2}{\pi^2}
    \frac{r}{2^{m}}
    \frac{1}{2^{\varsigma} (B_{\eta} + 1/2)}
    \left(
      1
      +
      \epsilon_{B_{\eta}}
    \right)
    &=
    1 - \frac{1}{\pi^2} \frac{r}{2^m}
    \left(
      1 + \epsilon(2w)
    \right) \\
    &\ge
    1 - \frac{1}{\pi^2}
    \left(
      1 - \frac{1}{w}
    \right)
    \left(
      1 + \frac{1}{4w} + \frac{1}{24 w^2}
    \right) \\
    &=
    1 - \frac{1}{\pi^2}
    \left(
      1 - \frac{3}{4w} - \frac{5}{24 w^2} - \frac{1}{24 w^3}
    \right)
    >
    1 - \frac{1}{\pi^2}.
  \end{align*}

  For $B_{\Delta} = 1$, it holds that $\epsilon_{B_{\Delta}}(1/2) = \epsilon(3/2) = 11/27$ and $\epsilon_{B_{\Delta}}(3/2) = \epsilon(5/2) = 17/75$, so it follows for the second factor in~\refeq{theorem:lower-bound} that
  \begin{align*}
    &1
    -
    \frac{1}{\pi^2}
    \left(
      \frac{1}{B_{\Delta} + 1/2}
      (
        1
        +
        \epsilon_{B_{\Delta}}(1/2)
      )
      +
      \frac{1}{B_{\Delta} + 3/2}
      (
        1
        +
        \epsilon_{B_{\Delta}}(3/2)
      )
    \right)
    = \\
    &\quad\quad 1 - \frac{1}{\pi^2}
    \left(
      \frac{2}{3} \cdot \frac{38}{27}
      +
      \frac{2}{5} \cdot \frac{92}{75}
    \right)
    =
    1 - \frac{14468}{10125} \frac{1}{\pi^2}.
  \end{align*}

  The lower bound given by~\refeq{theorem:lower-bound} is hence at least
  \begin{align*}
    \left(
      1 - \frac{1}{\pi^2}
    \right)
    \left(
      1 - \frac{14468}{10125} \frac{1}{\pi^2}
    \right)
    >
    0.7685,
  \end{align*}
  as both factors are positive, and so the corollary follows.
\end{proof}

\end{document}